\documentclass[12pt, a4paper, twoside, doublespace, spanish]{uns_tesis}

\begin{document}

\title{Caracterización Formal y Análisis Empírico de Mecanismos Incrementales de Búsqueda basados en Contexto}

\author{Carlos M. Lorenzetti}
\email{cml@cs.uns.edu.ar}
\department{Departamento de Ciencias e Ingenier\'{i}a de la Computaci\'{o}n}
\degree{Doctor en Ciencias de la Computaci\'{o}n}
\prevdegrees{Ingeniero en Sistemas de Computaci\'{o}n, Universidad Nacional del Sur (2005)}

\degreeyear{2011}
\degreebegin{30 de agosto de 2006}
\thesisdate{18 de marzo 2011}
\unilogo{uni}


\supervisor{\Dr}{Guillermo R. Simari}
\supervisor{\Dra}{Ana G. Maguitman}

\pagestyle{plain}

\changepage{-2cm}{-1cm}{}{}{}{1cm}{}{}{}  
\maketitle
\changepage{}{}{}{}{}{1cm}{}{}{}
\makepreface



\makeabstract

\begin{prefacepage}
\large
{\def\baselinestretch{1.2}\Large\bf Agradecimientos \par}
En primer lugar quiero agradecer a mis padres quienes siempre me alentaron para que continuara con mis estudios.
Luego, a mis directores de tesis y becas de postgrado, en especial a Ana quien me aceptó como becario, estando en el exterior, sin conocerme en persona. Ella es una directoria excelente que me enseñó lo que es la investigación y es un ejemplo a seguir.
También quisiera agradecer a los profesores investigadores del Departamento con los que he compartido grandes charlas durante los congresos y en los pasillos de la universidad. 
No quiero dejar de agradecer también a cada uno de los becarios del DCIC, y de otros lugares, con los que compartí estos años de beca doctoral. Cada uno contribuyó a que se creara un excelente ambiente de trabajo.
Finalmente, pero no menos importante, quiero agradecer a Rocío por su amor y compañía y su apoyo durante el doctorado y la escritura de esta tesis.

\par
\vfill
\end{prefacepage}
\changepage{2cm}{1cm}{}{}{}{-2cm}{}{}{}
\pagestyle{plain}
\cleardoublepage
\phantomsection
\addcontentsline{toc}{chapter}{\'{I}ndice general}
\begin{singlespace}
\tableofcontents
\end{singlespace}

\chapter*{Notación}
\phantomsection
\addcontentsline{toc}{chapter}{Notación}

\begin{tabular}{rl}
 $C$ & colección de documentos\\ 
 $t$ & número de términos en la colección $C$\\
 $N$ & número de documentos en la colección $C$\\ 
 $R$ & conjunto de documentos relevantes en la colección\\
 $\bar{R}$ & conjunto de documentos irrelevantes en la colección\\ 
 $A$ & conjunto de documentos recuperados por alguna consulta o pedido\\
 $\bar{A}$ & conjunto de documentos no recuperados por alguna consulta o pedido\\
 $k_i$ & término $i$-ésimo en un documento o en una lista de términos\\
 $d_j$ & documento $j$-ésimo en una lista de documentos\\
 $\overrightarrow{d_j}$ & vector asociado al documento $d_j$\\
 $w(k_i,d_j)$ & peso de un término $k_i$ en un documento $d_j$ \\
 $q$ & consulta formada por términos $k_i$, $i \in \{1 \dots n\}$\\
 $\overrightarrow{q}$ & vector asociado a la consulta $q$\\
 $\mathit{sim}(A, B)$ & similitud entre el elemento $A$ y el elemento $B$, expresada en forma general\\
 $\mathit{sim}_{\mathit{M1}}(A, B)$ & similitud entre el elemento $A$ y el elemento $B$, calculada con el método $\mathit{M1}$\\
\end{tabular}
\cleardoublepage
\pagestyle{headings}
\pagenumbering{arabic}
\chapter{Introducción}\label{chp:introduccion}

La Recuperación de Información (IR\footnote{del inglés, Information Retrieval.}) web es un área de investigación relativamente nueva, que se popularizó desde la aparición de la Internet a principios de los~'90s y trata de afrontar los desafíos de la IR en la Internet. La investigación de la IR con la ayuda de computadoras data de los~'50s, cuando el esfuerzo estaba enfocado en la resolución de problemas de IR en colecciones de documentos pequeñas, con consultas descriptivas, en un dominio acotado y con usuarios particulares.
%
Las características del nuevo entorno que resultó la World Wide Web (Web), hicieron que la tarea fuera algo diferente de la IR tradicional. La Web es un recurso prácticamente ilimitado, con información heterogénea, y con usuarios de todas las clases sociales, buscando información que satisfaga sus necesidades. Estos necesitan que la Web sea accesible a través de sistemas de recuperación de información efectivos y eficientes. El \textit{tamaño}, la \textit{heterogeneidad}, el \textit{dinamismo} y la \textit{estructura} de la Web, junto con la \textit{diversidad} en los comportamientos de búsqueda de los usuarios, son las principales características que hacen que la IR tradicional tenga grandes desafíos en la Internet.

Los motores de búsqueda comerciales, que son los sistemas de IR más populares, han resuelto parcialmente los desafíos con los que se enfrenta la IR en la Web, ofreciendo una herramienta para la búsqueda de información relevante. En efecto, los usuarios actuales esperan ser capaces de encontrar la información que buscan en la Web, de forma rápida y fácil.
%
La IR en la Web, sin embargo, continúa siendo un área con muchas cuestiones por resolver, probablemente con muchas aplicaciones por descubrir. 
En la actualidad sigue existiendo la necesidad de desarrollar métodos novedosos para facilitar el acceso eficiente a la información relevante en la Web.
Algunos problemas de investigación van desde comprender mejor las necesidades del usuario, al procesamiento de enormes cantidades de información para brindar mejores métodos de ordenamiento, que hagan uso de la estructura y las características de la Web.

\section{Motivación}\label{sec:intro:motivacion}
La omnipresencia de las computadoras personales, unida a la conectividad de la Internet han cambiado para siempre el rol de la información en la computación. Los recursos de información ya no están más relacionados con una única ubicación ni son accedidos sólo por profesionales.
Los sistemas de IR están disponibles para los usuarios de Internet cada día, desde el confort de su propia computadora personal.
Estos repositorios de información se acceden de la misma forma en la que se escriben artículos, se leen diarios y se navegan sitios de la Web.
Desafortunadamente, los sistemas de IR tradicionales resultaron difíciles de usar para usuarios nuevos, lo que impulsó el desarrollo de una gran cantidad de sistemas para buscar, filtrar y organizar la gran cantidad de información que se tenía disponible. Se desarrollaron sistemas de IR para aplicaciones que van desde la clasificación y organización de correo electrónico~\cite{maes94agents, cohen96learning}, el filtrado de noticias~\cite{lang95newsweeder}, sistemas para responder consultas basados en las FAQ\footnote{del inglés, Frequently Asked Questions, Preguntas frecuentes.} de Usenet\footnote{del inglés, USErs NETwork.}~\cite{hammond94casebased}, y la búsqueda en la Web~\cite{mcbryan94genvl,brin98anatomy}. También se han desarrollado algunas aplicaciones para organizar la información del usuario, como pueden ser archivos de notas, diarios y calendarios~\cite{jones86applied, lamming94forgetmenot}. 

Sin embargo, la mayoría de estos sistemas, que se han convertido en la piedra angular del acceso a la información, sólo se han concentrado en la generación de consultas para recuperar información por demanda, lo que significa que el usuario tiene que invocarlos explícitamente, interrumpiendo el proceso normal de navegación y esperando ocioso por los resultados de la búsqueda. Tales sistemas no pueden ayudar a un usuario cuando éste no está suficientemente familiarizado con el tema en cuestión, o desconoce el vocabulario exacto con el que debe formular las consultas para acceder a los recursos de interés.



Este escenario trae nuevos desafíos y oportunidades a los diseñadores de tales sistemas, tanto para crear sistemas accesibles como para aprovechar por completo este nuevo espacio de información oculta.
El crecimiento explosivo que ha tenido la Web y otras fuentes de información on-line han hecho crítica la necesidad de alguna clase de asistencia inteligente para el usuario que está buscando información relevante.
Al desarrollarse computadoras de escritorio cada vez más potentes, la mayor parte del tiempo de CPU de éstas se desperdicia esperando que el usuario presione la siguiente tecla, lea la siguiente página o se cargue el siguiente paquete de la red. No hay razón para que esos ciclos de CPU desperdiciados no puedan ser usados constructivamente para realizar búsquedas de información útil 
para el contexto actual del usuario. Por ejemplo, mientras un ingeniero lee un correo electrónico sobre un proyecto, un agente puede recordarle la planificación, los reportes de avance u otros recursos relacionados con ese proyecto. Cuando el ingeniero no lee más el correo y, por ejemplo, comienza a editar un archivo, el agente cambiaría automáticamente sus recomendaciones para adecuarse a la nueva tarea.


Para los diseñadores de interfaces de exploración de información también se presentan problemas interesantes, ya que la forma en la que un usuario genera una consulta depende de su conocimiento previo y de su entendimiento del tema. Algunas preguntas que surgen son: ¿cómo les presentamos a los usuarios las posibles acciones que pueden tomar teniendo en cuenta su entendimiento actual?, ¿cómo podemos ayudar a los usuarios a tener un mejor entendimiento de estas referencias?, y ¿cómo podemos ayudar a los usuarios a volver a sitios visitados con anterioridad en la exploración, una vez que se ganó una nueva perspectiva?

La motivación para las investigaciones presentadas en esta tesis es desarrollar una herramienta que ayude y asista al usuario de un sistema de IR en la tarea que está realizando, brindándole información relevante y basada en el contexto en el cual está trabajando. A continuación se presentan los objetivos específicos de la tesis.

\section{Objetivos}\label{sec:intro:objetivos}
Esta tesis tiene como principal objetivo proponer, investigar y evaluar nuevas técnicas semisupervisadas de IR orientadas a entender mejor las necesidades de los usuarios. Para abordar este objetivo, se plantearon las siguientes preguntas de investigación:
\begin{enumerate}
\item ¿Puede el contexto del usuario explotarse satisfactoriamente para acceder a material relevante en la Web?
\item ¿Puede un conjunto de términos específicos de un contexto ser refinado incrementalmente basándose en el análisis de los resultados de una búsqueda?
\item ¿Los términos específicos de un contexto aprendidos mediante métodos incrementales, son mejores para generar consultas comparados con aquellos encontrados por técnicas clásicas de IR o métodos clásicos de reformulación de consultas?
\end{enumerate}

Por lo tanto, los objetivos específicos de esta tesis son:
\begin{enumerate}
 \item Proponer un algoritmo semisupervisado capaz de aprender incrementalmente nuevos vocabularios con el propósito de mejorar consultas temáticas. El objetivo es que estas consultas reflejen la información contextual y así puedan recuperar efectivamente material relacionado semánticamente.
 \item Desarrollar una plataforma para evaluar las técnicas de IR propuestas, así como otras técnicas existentes. Dicha plataforma será especialmente apta para el análisis de buscadores temáticos y para incorporar métricas de evaluación novedosas basadas en las nociones de similitud semántica y relevancia parcial.
\end{enumerate}

\section{Contribuciones}\label{sec:intro:contribuciones}
Esta investigación propone una técnica de IR novedosa que incrementalmente aprende nuevos términos que pueden ayudar a reducir la distancia que existe 
entre el vocabulario empleado en las consultas formuladas por un usuario y el vocabulario utilizado para indexar los documentos relevantes para dicho usuario.
Es decir, las principales contribuciones de esta tesis son:
\begin{enumerate}
 \item Un \textit{Algoritmo semisupervisado} que utiliza una estrategia de recuperación incremental de documentos web para el ajuste de la importancia de los términos utilizados en la generación de consultas, de forma tal que éstos reflejen mejor su valor como descriptores y discriminadores del tópico del contexto del usuario. El vocabulario enriquecido de esta forma permite la generación de consultas para una búsqueda más efectiva.
 \item Una \textit{Plataforma de evaluación} de nuevos métodos y algoritmos desarrollados para la IR. Una  plataforma de evaluación es algo fundamental en el desarrollo de nuevos métodos en IR, permitiendo la comparación con las técnicas existentes. 
También se proponen nuevos métodos de evaluación sustentados en una métrica de \textit{similitud semántica} para la comparación de documentos.
\end{enumerate}
\newpage\section{Organización de la tesis}\label{sec:organizacion}
Esta tesis está organizada en 6 capítulos principales, seguidos del \hyperref[chp:figures]{índice de Figuras} y de las\linebreak
\hyperref[chp:biblio]{Referencias}.

El \autoref{chp2:state} describe los fundamentos de los sistemas de IR. Entre ellos, los modelos clásicos de representación de documentos más utilizados en el área, seguido de una explicación del proceso de formulación de una consulta, la etapa inicial de todo proceso de recuperación, en donde se incluye un análisis de los mecanismos de reformulación y optimización de consultas. Luego se analiza el concepto fundamental de la similitud, mostrando las métricas más difundidas. Finalmente se examina el potencial que tienen las ontologías en el proceso de evaluación de un sistema de IR y se define la noción de similitud semántica.

El \autoref{chp3:evaluacion} presenta la metodología de evaluación de los sistemas de IR, a la cual se la puede dividir en tres grandes componentes. Primero se hace un recorrido por la historia de las colecciones de prueba que se utilizan en la mayoría de las publicaciones del área. Luego se presenta otro componente necesario, como lo son los juicios de relevancia, que indicarán cuáles de los documentos recuperados por un sistema le son útiles a un usuario. Por último se enumeran las principales métricas de evaluación con las que es posible comparar un sistema con otros.

El \autoref{chp4:metodo_incremental} presenta las contribuciones teóricas de esta tesis. Se presenta el problema que tienen los sistemas actuales de IR para incorporar el contexto del usuario en las búsquedas, haciendo una revisión de la literatura existente en el tema. Luego se presenta la plataforma sobre la cual se basa el método incremental propuesto, para luego desarrollarlo de forma completa. El capítulo finaliza con los alcances y aplicaciones del método presentado.

El \autoref{chp5:evals} expone un análisis y una comparación de los resultados experimentales obtenidos. Se presenta la estructura de la plataforma de evaluación que es parte de las contribuciones de esta tesis, así como también las nuevas métricas de evaluación desarrolladas. 
Luego se muestran los resultados obtenidos, ilustrando la aplicación de la plataforma propuesta en la evaluación de distintos métodos de IR.

El \autoref{chp:conclusiones} establece las conclusiones de esta tesis y señala el trabajo de investigación a futuro.
\chapter{Fundamentos}\label{chp2:state}
\section{Modelos para Recuperación de Información}\label{sec:fundamentos}
Los modelos de IR clásicos consideran que un documento está representado por un conjunto representativo de palabras claves, llamadas \textit{términos índice}. Esta idea fue sugerida por Luhn en los~'50s~\cite{luhn57statistical}. Un término índice es una palabra simple dentro de un documento, cuya semántica nos ayuda a recordar los temas principales sobre los que trata el documento (de ahora en más nos referiremos a estas palabras simplemente como \textit{términos} de un documento, excepto que se indique lo contrario). Por lo tanto, se los utiliza tanto para indexar como para resumir el contenido de un documento. Se trata generalmente de sustantivos, porque ellos tienen un significado y su semántica es fácil de identificar y comprender. Otros tipos de términos, como son los adjetivos, los adverbios y los conectores no son tan útiles para realizar índices porque se utilizan como complementos de los sustantivos. Sin embargo es interesante considerar todas las palabras en un documento a la hora de realizar un índice. Algunos motores de búsqueda utilizan esta idea de indexado de \textit{texto completo}.

Dado un conjunto de términos de algún documento se puede notar que no todos son igualmente útiles a la hora de describir el documento. De hecho, hay algunos que son mucho más vagos que otros. No es un problema trivial decidir la importancia de un término como condensador del contenido de un documento. Más allá de esto, hay algunas propiedades de un término que son mensurables con facilidad y que son útiles para evaluar su potencialidad. Por ejemplo, consideremos una colección (o corpus) que contiene cien mil documentos. Una palabra que aparece en cada uno de los cien mil documentos es absolutamente inútil como término porque no nos dice nada acerca de cuáles documentos podrían interesarle a un usuario. Pero, otra palabra que aparezca en sólo cinco documentos sería más útil, porque reduce considerablemente el espacio de documentos que podría ser de interés para un usuario. Esto muestra que los distintos términos tienen una relevancia variable al usarlos para describir el contenido de los documentos.

Ahora vamos a definir la noción de peso de un término en un documento. Sea un término $k_i$, un documento $d_j$ y el \textit{peso} asociado a $(k_{i},d_{j})$, $w(k_i,d_j)\geq 0$. Este peso es una estimación de la importancia del término como descriptor del contenido semántico de un documento.
\begin{definition}\textnormal{\cite{salton71smart,baeza-yates99modern}}
 Sea $t$ el número de términos en un corpus $C$ y $k_{i}$ un término cualquiera. Entonces $K=\{ k_{1}, \dots, k_{t}\}$ es el conjunto de términos. Se asocia un peso\linebreak $w(k_i,d_j)>0$ a cada término $k_{i}\in d_{j}$ y $w(k_i,d_j)=0$ si $k_{i}\notin d_{j}$. Dado un documento $d_{j}$ se le asocia un vector de términos $\overrightarrow {d_j}$ que representa $\overrightarrow {d_j} = (w(k_1,d_j), w(k_2,d_j), \dots, w(k_t,d_j))$. Además, sea $g_i$ una función que devuelve el peso asociado con un término $k_i$ de cualquier vector de $t$ dimensiones, o sea, $g_i(\overrightarrow {d_j}) = w(k_i,d_j)$.
\label{def:doc_weights}
\end{definition}
Como mencionaremos más adelante, se asume que los pesos de los términos son \textit{mutuamente independientes}, lo que significa que el peso $w(k_i,d_j)$ asociado a $(k_{i}, d_{j})$ no está relacionado con el peso $w(k_{i+1},d_j)$ asociado al término $i+1$ del mismo documento. Se asume, por lo tanto, que la aparición de un término no está correlacionada con la ocurrencia de otro, lo que es claramente una simplificación del modelo. Un ejemplo que contradice esta simplificación son, p. ej. los términos \textit{red} y \textit{computadora}, que podrían aparecer en un documento relacionado con las redes de computadoras. En este documento seguramente aparecerá uno de estos términos muy cerca del otro. Luego, puede verse que estas palabras están correlacionadas y esta relación podría reflejarse en sus pesos. La independencia mutua es una gran simplificación, pero reduce los cálculos de los pesos y agiliza el cálculo del ranking de los documentos.

%
En un entorno en el que se quieren recuperar documentos, o en cualquier otro relacionado con coincidencia de patrones, en donde las entidades almacenadas (documentos) se comparan unas con otras o con nuevos patrones (pedidos de búsqueda), el mejor espacio de indexación es uno en el cual cada entidad está tan alejada como sea posible una de otra; en estas circunstancias la calidad de un sistema de indexado se puede expresar como una función de la densidad del espacio de objetos; en particular, el rendimiento en la recuperación puede correlacionarse inversamente con la densidad espacial~\cite{salton75vector}.

\begin{figure}[!ht]
\centerline{%
\includegraphics[scale=0.5]{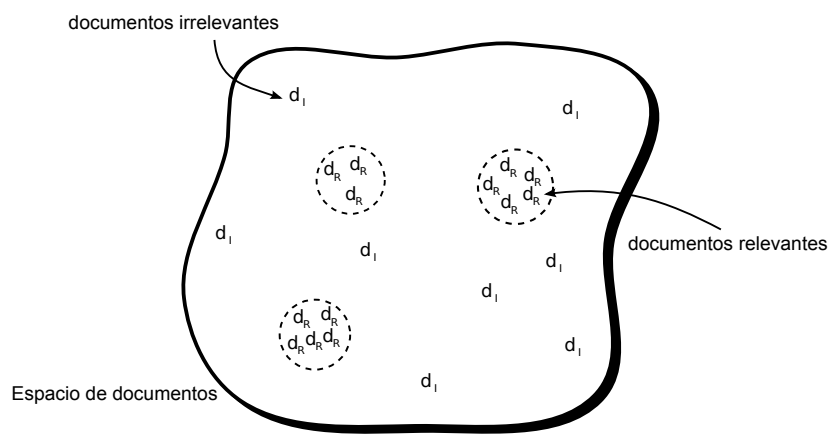}%
}
\caption{Espacio de documentos ideal.}
\label{fig:ideal_space}
\end{figure}

La pregunta que surge es si existe un espacio de documentos óptimo, esto es, uno que produzca un rendimiento óptimo en la etapa de recuperación. La configuración del espacio de documentos es una función de la forma en la que se asignan los términos y sus pesos a los documentos de una colección.
Si no se tiene ningún conocimiento especial sobre los documentos a indexar, se podría suponer que un espacio ideal de documentos es uno en donde los documentos más relevantes para ciertas consultas están agrupados, asegurando que se recuperen de forma conjunta al ingresar las mismas. De esta manera, aquellos documentos que no se quieren recuperar deberían estar muy separados en ese espacio. Esa situación se muestra en la \autoref{fig:ideal_space}, en donde la distancia entre dos documentos (representados con ``$d_i$'' si son documentos relevantes y con ``$d_r$'' si son documentos irrelevantes) está inversamente relacionada con la similitud entre los vectores correspondientes.
%
%
%

La configuración que muestra la \autoref{fig:ideal_space} representa la mejor situación posible, en la que los ítems que son relevantes e irrelevantes para varias consultas son separables, pero no existe una forma práctica de producir tal espacio, porque durante la etapa de indexado es muy difícil anticipar los juicios de relevancia de los usuarios a lo largo del tiempo. Esto quiere decir que la configuración óptima es muy difícil de generar si no se tiene un conocimiento \textit{a priori} de una colección de documentos dada.

Las definiciones que acabamos de dar nos ayudarán a discutir cuatro modelos clásicos de IR, el modelo \hyperref[sub:mod_bool]{Booleano}, el \hyperref[sub:mod_vectorial]{Vectorial}, el \hyperref[sub:mod_prob]{Probabilístico} y el \hyperref[sub:lsa]{Indexado de Semántica Latente}.

\subsection{Modelo Booleano}\label{sub:mod_bool}
El modelo Booleano es el modelo de recuperación más simple y se basa en la teoría de conjuntos y en el álgebra de Boole. El modelo Booleano proporciona una arquitectura fácil de comprender para un usuario común de un sistema de IR, ya que los conceptos de conjuntos son bastante intuitivos. Las consultas expresadas como expresiones de Boole tienen una semántica precisa. El modelo Booleano recibió una gran atención y fue adoptado por muchos sistemas bibliográficos comerciales debido, principalmente, a su inherente simplicidad y a su formalismo puro.

A pesar de esto, el modelo sufre de grandes desventajas. La primera es que su estrategia de recuperación se basa en un criterio de decisión binario; un documento \underline{es} relevante o \underline{no} lo es (la noción de relevancia se analiza en la \autoref{sub:relevancia}). No existe la noción de grado de relevancia, lo que evita que el modelo tenga un buen rendimiento en la recuperación de grandes volúmenes de información. Tal es así que este modelo es más un modelo de recuperación de \textit{datos} que uno de recuperación de \textit{información}. La segunda desventaja reside en la representación de las consultas, que como se dijo, tienen una semántica precisa pero no es tan simple transformar una idea de un usuario en una expresión Booleana. De hecho, la mayoría de los usuarios encuentran a este proceso muy dificultoso e incómodo y las expresiones Booleanas que formulan tienden a ser bastante simples.
A pesar de estos problemas, el modelo Booleano continúa siendo popular en bases de datos de documentos médicos~\cite{karimi09challenge} y legales~\cite{oard08trec} siendo un buen punto de partida para el estudio de los conceptos básicos de IR.

\begin{figure}[!ht]
\centerline{
  \includegraphics[scale=0.5]{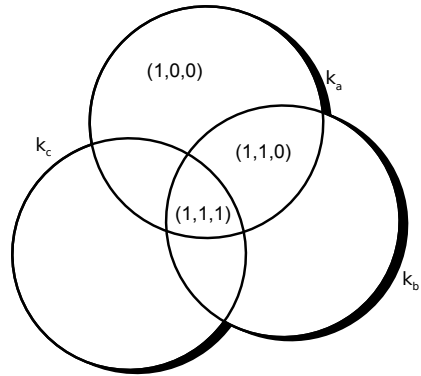}
}%
\caption{Componentes de una consulta booleana.} \label{fig:bool_sets}
\end{figure}

El modelo considera que los términos están \textit{presentes} o están \textit{ausentes} en un documento. Es por ello que los pesos de los términos son todos binarios, o sea, $w(k_i,d_j)\in\{0,1\}$. Una consulta $q$ puede estar compuesta de términos unidos entre sí por conectores lógicos: $not$, $and$ y $or$. Por lo tanto, una consulta es esencialmente una expresión Booleana convencional que puede ser representada como una disyunción de vectores conjuntivos, esto es, en la \textit{Forma Normal Disyuntiva} o DNF\footnote{del inglés, Disjunctive Normal Form.}. Por ejemplo, si consideramos la consulta $[q = k_a \wedge (k_b \vee \neg k_c)]$, esta puede escribirse (analizando su tabla de verdad) en la forma normal disyuntiva como $\overrightarrow {q_{\mathit{dnf}}} = (1, 1, 1) \vee (1, 1, 0) \vee (1, 0, 0)$, en donde cada uno de los componentes es un vector binario de pesos asociado con la tupla $(k_a, k_b, k_c)$. Estos vectores binarios de pesos se llaman componentes conjuntivos de $\overrightarrow {q_{\mathit{dnf}}}$. La \autoref{fig:bool_sets} muestra los tres componentes conjuntivos para el ejemplo.

\begin{definition}
\textnormal{\cite{lancaster73information, baeza-yates99modern}}
En el modelo Booleano, los pesos de los términos son todos valores binarios, esto es, $w(k_i,d_j)\in\{0, 1\}$. Una consulta $q$ es un expresión Booleana normal. Sea $\overrightarrow {q_{\mathit{dnf}}}$ la forma normal disyuntiva de la consulta $q$. Además, sea $\overrightarrow {q_{\mathit{cc}}}$ cualquiera de los componentes conjuntivos de $\overrightarrow {q_{\mathit{dnf}}}$. Luego, la similitud entre un documento $d_j$ y una consulta $q$ está definida como:
\begin{align*}
\mathit{sim}_{bool}(d_{j}, q) = \begin{cases}
  1\text{,} & \text{si }\exists\overrightarrow {q_{\mathit{cc}}} | (\overrightarrow {q_{\mathit{cc}}} \in \overrightarrow {q_{\mathit{dnf}}}) \wedge ({\forall _{{k_i}}},g_i(\overrightarrow{d_j}) = g_i(\overrightarrow{q_{cc}}))\text{,}\\
  0\text{,} & \text{si no.}
                         \end{cases}
\end{align*}
Si $\mathit{sim}(d_{j}, q) = 1$ entonces el modelo Booleano predice que el documento $d_j$ es \textnormal{relevante} para la consulta $q$ (puede que no lo sea). De otra forma, la predicción es que el documento es \textnormal{irrelevante}.
\label{def:bool_similarity}
\end{definition}
El modelo Booleano puede predecir que cada documento es \textit{relevante} o \textit{irrelevante}. No existe la noción de \textit{coincidencia parcial} con la consulta. Por ejemplo, sea un documento $d_{j}$ tal que\linebreak $\overrightarrow{d_{j}}=(0, 1, 0)$, entonces el documento incluye el término buscado $k_b$, pero se considera irrelevante para la consulta $[q = k_a \wedge (k_b \vee \neg k_c)]$.

Las principales \textit{ventajas} de este modelo son el formalismo que lo sustenta y su simplicidad. La principal \textit{desventaja} es que la coincidencia exacta de los términos puede acarrear que se recuperen muy pocos documentos o demasiados, dada la inexperiencia de los usuarios con este tipo de lógica. Puede verse que el modelo de indexado elegido tiene una gran influencia en el posterior proceso de recuperación, lo cual nos lleva al siguiente modelo.

\subsection{Modelo Vectorial}\label{sub:mod_vectorial}
\begin{figure}[!ht]
\centerline{%
\includegraphics[scale=0.5]{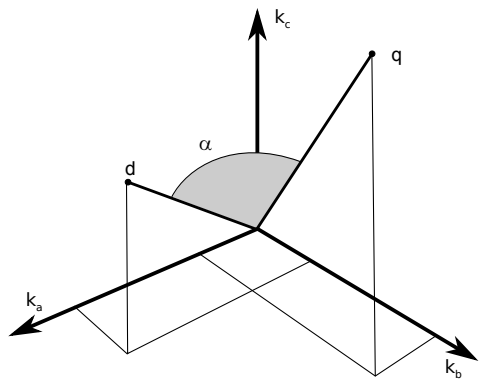}%
}%
\caption{Una consulta y un documento representados en un espacio vectorial.}
\label{fig:vect_sims}
\end{figure}

El modelo vectorial~\cite{salton68computer,salton71smart} tiene en cuenta que los pesos binarios son muy limitantes y propone una arquitectura en la cual es posible la coincidencia parcial. Su implementación se lleva a cabo asignando pesos \textit{no binarios} a los términos de las consultas y de los documentos. Por último, estos pesos se utilizan para calcular el \textit{grado de similitud} entre cada documento almacenado en el sistema y la consulta del usuario. 
El modelo tiene en cuenta aquellos documentos que coinciden parcialmente con la consulta, lo que permite 
que los documentos recuperados se puedan organizar
en orden decreciente a su grado de similitud. La principal consecuencia de esto es que el conjunto ordenado de documentos que responden a la consulta es mucho más preciso (en el sentido que hay una mayor coincidencia con las necesidades del usuario) que el conjunto recuperado por el modelo Booleano.  
\begin{definition}\label{def:vector_model}
\textnormal{\cite{salton71smart, salton75vector, baeza-yates99modern}}
En el modelo vectorial, los pesos de los términos\linebreak $w(k_i,d_j) \in d_j$ son valores positivos y no binarios. Los términos en las consultas también son ponderados, $w(k_i, q) \in q$, $w(k_i, q) \geq 0$. Luego, el vector consulta $\overrightarrow{q}$ se define como\linebreak $\overrightarrow{q}=(w(k_1,q), w(k_2,q), \dots, w(k_t,q))$ en donde $t$ es el número total de términos en el sistema. Al igual que el modelo anterior, un documento está representado por\linebreak $\overrightarrow{d_j}=(w(k_1,d_j), w(k_2,d_j), \dots, w(k_t,d_j))$.
\end{definition}
Esto nos permite la representación de un documento $d_{j}$ y de la consulta de un usuario $q$ como vectores en un espacio $t$-dimensional, como se muestra en la \autoref{fig:vect_sims}. El modelo propone evaluar el grado de similitud entre
un documento $d_j$ y la consulta $q$ como la correlación entre los dos vectores que los representan. Esta correlación puede medirse, por ejemplo, con el \textit{coseno del ángulo} entre dichos vectores, lo que es equivalente al producto escalar normalizado entre ambos vectores. Esto es:
\begin{align}
 \mathit{sim}_{cos}(d_j, q) &= \frac{\overrightarrow{d_j}.\overrightarrow{q}}{\abs{\overrightarrow{d_j}}.\abs{\overrightarrow{q}}} \label{eqn:cosine_similarity}\text{,} \\
&= \frac{\sum\nolimits_{i=1}^{t}{w(k_i,d_j) . w(k_i,q)}}{\sqrt {\sum\nolimits_{i=1}^t w(k_i,d_j)^{2}} .\sqrt {\sum\nolimits_{i=1}^t w(k_i,q)^{2}}}\text{,}\nonumber
\end{align}
en donde $\abs{\overrightarrow{d_j}}$ y $\abs{\overrightarrow{q}}$ son las normas del vector que representa al documento $d_j$ y del vector que representa a la consulta $q$ respectivamente. El factor escalar $\abs{\overrightarrow{q}}$ no afecta el orden de los documentos porque es constante en todos los documentos, mientras que $\abs{\overrightarrow{d_j}}$ es un factor de normalización sobre el espacio de documentos.

Dado que $w(k_i,d_j)\geq 0$ y $w(k_i,q)\geq 0$, entonces $\mathit{sim}(d_{j}, q)\in [0, +1]$, por lo que el modelo no trata de predecir si un documento es relevante o no, sino que los ordena de acuerdo a su \textit{grado de similitud} con la consulta. Puede recuperarse un documento incluso si sólo coincide \textit{parcialmente} con la consulta. Por ejemplo, se puede establecer un umbral sobre $\mathit{sim}(d_{j}, q)$ y recuperar todos los documentos con un grado de similitud por encima de ese límite. Para calcular el ranking es necesario especificar cómo obtener los pesos de los términos.

Los pesos de los términos pueden calcularse de distintas maneras. Aquí nos enfocaremos en las más eficientes y/o las utilizados a lo largo de esta tesis, para un análisis más detallado de las distintas técnicas de ponderación de términos puede consultarse~\cite{salton83introduction}. 

La idea está relacionada con los principios básicos de las técnicas de clasificación. Dada una colección de objetos $C$ y una descripción \textit{imprecisa} de un conjunto $R$, el objetivo de un algoritmo simple de clasificación podría ser separar esa colección en dos subconjuntos: el primero compuesto de objetos que están relacionados con $R$ y el segundo con aquellos que no lo están. En este contexto, una descripción imprecisa significa que no tenemos toda la información necesaria para decidir de forma exacta qué objetos están en $R$ y cuáles no. Un ejemplo de esto es un conjunto de autos $R$ que tienen un precio \textit{comparable} al de determinada marca y modelo. Dado que no está claro qué significa exactamente el término \textit{comparable}, no hay una descripción precisa (y única) del conjunto $R$. Otros algoritmos de clasificación más sofisticados podrían intentar separar a los objetos de la colección en varias clases de acuerdo a sus propiedades. Por ejemplo, los pacientes de un doctor especialista en cáncer pueden clasificarse en cinco clases: terminal, avanzado, metástasis, diagnosticado y saludable. Sin embargo, el problema no termina ahí porque es posible que las descripciones de estas clases sean imprecisas y no pueda decidirse en cuál clase debe asignarse a un paciente nuevo. Dado que en general, sólo debe decidirse si un documento es relevante o no, nos referiremos sólo a la primer clase de algoritmos (aquellos que sólo consideran dos clases).

Para ver al problema como uno de clasificación hacemos una analogía con los conjuntos explicados más arriba. Pensaremos a una colección de documentos como una colección de objetos $C$ y a la consulta del usuario como una especificación imprecisa de un conjunto de objetos $R$. Luego debe determinarse cuáles documentos están en $R$ y cuáles no. En un problema de clasificación deben resolverse dos cuestiones principales: primero deben determinarse las características que \textit{describen} mejor a los objetos de $R$, y segundo, deben determinarse las características que \textit{distinguen} mejor a los objetos de $R$ de los restantes objetos en $C$.
Los algoritmos de clasificación más exitosos son aquellos que balancean los efectos de ambas propiedades. 
A continuación veremos el método de ponderación de características (o términos) más conocido en el área de IR.

\subsubsection{TF-IDF}\label{sub:tf-idf}
La forma más natural de describir un documento es a través de los términos que lo componen, mientras más palabras utilicemos, más oportunidades tendremos de recuperar un documento. Esto sugiere que un factor que indique la \textit{frecuencia} de un término debe estar presente en una métrica de documentos o de consultas~\cite{salton71smart}.
Esta frecuencia se refiere usualmente como el \textit{factor TF}\footnote{del inglés, Term Frequency.} y proporciona una medida de qué tan bien describe un término al contenido de un documento.
Por otro lado, también debe incluirse alguna medida que tenga en cuenta el poder discriminatorio de un término, ya que por ejemplo, el término \textit{bebida} puede aparecer junto con términos como \textit{té}, \textit{café} o \textit{cacao} en documentos sobre té, café o chocolatada. Puede notarse que el primer término tiene un poder de distinción mucho menor que el de los últimos, por lo que aparecerá en muchos más documentos en una colección. Entonces debe considerarse un factor que ya no dependa de un documento en particular, sino de toda la colección~\cite{jones72statistical}.
A este factor se lo refiere usualmente como \textit{frecuencia inversa de documentos} o \textit{factor IDF}\footnote{del inglés, Inverse Document Frequency.}. La motivación detrás del uso de un factor IDF está dada porque aquellos términos que aparecen en muchos documentos, en general, no son muy útiles a la hora de distinguir a los documentos relevantes de aquellos que no lo son.
\begin{definition}
\textnormal{\cite{jones72statistical, baeza-yates99modern}}
Sea $N$ el número de documentos en la colección y $n_i$ el número de documentos en los que aparece el término $k_i$. Sea $\mathit{frec}(k_i,d_j)$ la frecuencia del término $k_i$ en el documento $d_j$, esto es la cantidad de ocurrencias. Entonces, la frecuencia normalizada del término $k_i$ en un documento $d_j$ viene dada por:
\begin{align}
 \label{eqn:tf}
\mathit{TF}(k_i,d_j) = \frac{\mathit{frec}(k_i,d_j)}{\sum\nolimits_{\forall h}{\mathit{frec}(k_h,d_j)}}\text{.}
\end{align}
Puede verse que si el término no aparece mencionado en el documento $d_j$ entonces $\mathit{TF}(k_i,d_j) = 0$. Por otro lado, la frecuencia inversa de documentos viene dada por:
\begin{align}
 \label{eqn:idf}
\mathit{IDF}(k_i) = log \frac{N}{n_i}\text{.}
\end{align}
Los esquemas más conocidos de ponderación de términos utilizan los pesos dados por la siguiente ecuación:
\begin{align}
 \label{eqn:tf-idf}
w(k_i,d_j) = \mathit{TF}(k_i, d_j) \times \mathit{IDF}(k_i)\text{,}
\end{align}
o por alguna variante de esta fórmula. Tales estrategias se llaman \textit{esquemas TF-IDF}.
 \label{def:tf-idf}
\end{definition}
Existen algunas variantes de la expresión anterior~\cite{salton88term} pero, en general, la misma proporciona un buen esquema de ponderación en muchas colecciones.

Las principales \textit{ventajas} del modelo vectorial son: 
(1) su esquema de ponderación de términos (TF-IDF) mejora el rendimiento en la recuperación de documentos; 
(2) su estrategia de coincidencia parcial permite la recuperación de documentos que \textit{aproximan} a la información pedida por la consulta; y 
(3) su fórmula de similitud por coseno ordena los documentos de acuerdo a su similitud con la consulta. La \textit{desventaja} teórica que existe es la independencia mutua que se asumió más arriba (la \autoref{eqn:tf-idf} \textit{no} tiene en cuenta las dependencias entre los términos). Sin embargo, en la práctica, tener en cuenta esta dependencia podría volverse en contra ya que, debido a la localidad de muchas dependencias, su utilización indiscriminada en todos los documentos de la colección podría, de hecho, \textit{perjudicar} el rendimiento global.
 
A pesar de su simplicidad, este modelo es una estrategia de ranking muy utilizada en las colecciones de documentos generales. 
Es un modelo que sigue siendo muy popular dado que es rápido y simple.
El rendimiento obtenido en sus conjuntos ordenados de respuestas es difícil de mejorar, a menos que se incorpore dentro de la arquitectura del modelo vectorial, alguna estrategia de expansión de consultas o de realimentación de relevancia (ver \autoref{sub:feedback}). 

%
\subsection{Modelo Probabilístico}\label{sub:mod_prob}
El modelo probabilístico~\cite{robertson76relevance} se basa en una arquitectura probabilística. Dada una consulta de usuario, existe un conjunto de documentos que contiene sólo a los documentos relevantes y a ningún otro. Este conjunto de documentos es el conjunto de respuestas \textit{ideal}. Si tuviéramos la descripción de este conjunto ideal no tendríamos problemas para recuperar estos documentos. Luego, podemos pensar que el proceso de generación de consultas es el proceso de especificación de las propiedades del conjunto ideal (lo cual es, una vez más, análogo a interpretarlo como un problema de clasificación). Pero el problema es que no sabemos cuáles son estas propiedades exactamente. Todo lo que sabemos es que existen términos cuya semántica podría utilizarse para caracterizar esas propiedades. Inicialmente es necesario realizar un esfuerzo por adivinar cuáles podrían ser las propiedades, ya que no las conocemos al momento de realizar la consulta. Estas suposiciones iniciales nos permiten generar una descripción probabilística preliminar del conjunto ideal, la cual se utiliza para recuperar un primer conjunto de documentos. Luego se inicia una interacción con el usuario con el fin de mejorar la descripción probabilística. Tal proceso sería de la siguiente manera.

El usuario analiza los documentos recuperados y decide cuáles son relevantes y cuáles no lo son (aunque, en realidad, no se los analiza a todos, sino a los que se encuentran más arriba en una lista ordenada). Entonces el sistema utiliza esta información para refinar la descripción del conjunto ideal. Repitiendo este proceso se espera que la descripción evolucione y se acerque más a la descripción real. Veamos ahora la siguiente suposición sobre la que se basa el modelo probabilístico.

\paragraph{Suposición (Principio Probabilístico)} Dada una consulta de usuario $q$ y un documento en la colección $d_j$, el modelo probabilístico trata de estimar la probabilidad de que el usuario encuentre interesante al documento $d_j$ (esto es, relevante). El modelo supone que esta probabilidad de relevancia depende solamente de la consulta y del documento. Además, asume que existe un subconjunto de la colección que el usuario acepta como el conjunto que responde a $q$. Ese conjunto \textit{ideal} se lo etiqueta como $R$ y debería maximizar la probabilidad de relevancia para el usuario. Se predice que los documentos que contiene $R$ son \textit{relevantes} a la consulta y que los que no están contenidos son \textit{irrelevantes}.\\


Para cada documento $d_j$, el modelo asigna como medida de similitud con una consulta $q$ dada, la relación $P(d_j \text{ \textit{es-relevante-para} } q)/P(d_j \text{ \textit{no-es-relevante-para} } q)$, que mide la correspondencia entre que un documento sea relevante y que sea irrelevante para una consulta. Tomando esta relación como un orden, se minimiza la probabilidad de un juicio erróneo~\cite{vanrijsbergen79information,fuhr92probabilistic}.

\begin{definition}
\textnormal{\cite{robertson76relevance, baeza-yates99modern}}
En el modelo probabilístico, los pesos de los términos son valores binarios, esto es, $w(k_i,d_j)\in\{0,1\}$ y $w(k_i,q)\in\{0,1\}$. Una consulta $q$ es un subconjunto de términos. Sea $R$ el conjunto conocido de documentos relevantes (o que se suponen relevantes). Sea $\bar{R}$ el complemento de $R$, esto es, el conjunto de documentos irrelevantes. Sea $P(R|\overrightarrow{d_j})$ la probabilidad de que un documento $d_j$ sea relevante a una consulta $q$ y sea $P(\bar{R}|\overrightarrow{d_j})$ la probabilidad de que ese mismo documento no sea relevante para la misma consulta. Entonces la similitud de un documento $d_j$ con la consulta $q$ se define como la relación:
\begin{align}
\label{eqn:sim-probab}
\mathit{sim}_{prob}(d_j, q) = \frac{P(R|\overrightarrow{d_j})}{P(\bar{R}|\overrightarrow{d_j})}\text{.}
\end{align}
\label{def:sim-probab}
\end{definition}
\vspace{-15mm}
Aplicando la regla de Bayes,
\begin{align*}
\mathit{sim}_{prob}(d_j, q) &= \frac{P(\overrightarrow{d_j}|R)\times P(R)}{P(\overrightarrow{d_j}|\bar{R})\times P(\bar{R})}\text{.}
\end{align*}
En donde,
\vspace{-3mm}
\begin{itemize}
 \item $P(\overrightarrow{d_j}|R)$ representa la probabilidad de elegir aleatoriamente a $d_j$ del conjunto de documentos relevantes $R$.
 \item $P(R)$ representa la probabilidad de elegir aleatoriamente un documento de la colección y que éste sea relevante.
 \item Las probabilidades sobre $\bar{R}$ son análogas y complementarias.
\end{itemize}

Dado que $P(R)$ y $P(\bar{R})$ son constantes para todos los documentos de la colección, puede escribirse:
\vspace{-2mm}
\begin{align}
\mathit{sim}_{prob}(d_j, q) &\sim \frac{P(\overrightarrow{d_j}|R)}{P(\overrightarrow{d_j}|\bar{R})}\text{,} \nonumber
\intertext{y asumiendo la independencia de los términos:
\vspace{-2mm}}
\mathit{sim}_{prob}(d_j, q) &\sim \frac{{(\prod\nolimits_{{g_i}(\overrightarrow{d_j}) = 1} {P({k_i}|R)} ) \times (\prod\nolimits_{{g_i}(\overrightarrow{d_j}) = 0} {P(\bar{k_i}|R)} )}}{{(\prod\nolimits_{{g_i}(\overrightarrow{d_j}) = 1} {P({k_i}|\bar{R})} ) \times (\prod\nolimits_{{g_i}(\overrightarrow{d_j}) = 0} {P(\bar{k_i}|\bar{R})} )}}\text{.}\nonumber
\intertext{\vspace{-3mm}
En donde,
\begin{itemize}
 \item $P({k_i}|R)$ representa la probabilidad de que un término $k_i$ \textit{esté} presente en un documento que se selecciona aleatoriamente de $R$,
 \item $P(\bar{k_i}|R)$ representa la probabilidad de que un término $k_i$ \textit{no esté} presente en un documento que se selecciona aleatoriamente de $R$,
 \item Las probabilidades asociadas a $\bar{R}$ tienen significados análogos a estos.
\end{itemize}
Si aplicamos logaritmos, recordando que $P({k_i}|R) + P(\bar{k_i}|R) = 1$ e ignorando los factores que son constantes para todos los documentos en el contexto de una misma consulta, podemos escribir:}
 \mathit{sim}_{prob}(d_j, q) &\sim \sum\limits_{i=1}^{t}{%
w(k_i,q)\times w(k_i,d_j)\times%
\left(%
 \log \frac{P(k_i|R)}{1-P(k_i|R)} + %
 \log \frac{1-P(k_i|\bar{R})}{P(k_i|\bar{R})}%
\right)}\text{,}
  \end{align}
que representa la expresión fundamental del cálculo de ranking en el modelo probabilístico.

Recordando que en un principio no se conoce al conjunto $R$, es necesario encontrar un método para realizar los cálculos iniciales de las probabilidades $P(k_i|{R})$ y $P(k_i|\bar{R})$, existiendo varias alternativas para el cómputo.

Al comenzar el proceso, o sea, inmediatamente después de haber especificado la consulta, aún no se han recuperado documentos. Entonces se tienen que hacer algunas suposiciones para simplificar los cálculos: (a) asumir que $P(k_i|{R})$ es constante para todos los términos (p. ej., igual a $0.5$) y (b) asumir que la distribución de los términos a través de los documentos irrelevantes puede aproximarse por la distribución de términos en toda la colección de documentos. Numéricamente esto es:
\begin{align*}
 P(k_i|{R}) &= 0.5 &
 P(k_i|\bar{R}) &= \frac{n_i}{N}\text{,}
\intertext{donde $n_i$ representa la cantidad de documentos que contienen el término $k_i$ y $N$ el total de documentos en el corpus, respectivamente. Dadas estas suposiciones iniciales, se pueden recuperar documentos que contienen los términos de la consulta y brindan un ranking probabilístico inicial. Luego, se mejora el ranking de la siguiente manera.\newline
\indent Se define un subconjunto, $V$, de los documentos recuperados inicialmente, por ejemplo con los $r$ mejores documentos del ranking, siendo $r$ un umbral definido previamente. Además, sean $V_i \subset V$ aquellos documentos que contienen el término $k_i$. Para mejorar el ranking probabilístico se necesita mejorar $P(k_i|{R})$ y $P(k_i|\bar{R})$, y esto puede realizarse con las siguientes suposiciones: (a) se puede aproximar $P(k_i|{R})$ con la distribución de $k_i$ en los documentos que ya se recuperaron, y (b) se puede aproximar $P(k_i|\bar{R})$ considerando que todos los documentos que todavía no se han recuperado son irrelevantes. Numéricamente, esto es:}
P(k_i|{R}) &= \frac{\abs{V_i}}{\abs{V}} &
P(k_i|\bar{R}) &= \frac{n_i - \abs{V_i}}{N-\abs{V}}\text{.}
\intertext{Repitiendo este proceso recursivamente se van mejorando las probabilidades sin ninguna asistencia del usuario (contrario a la idea original). Sin embargo, puede utilizarse la ayuda del usuario en la selección de los documentos del subconjunto $V$.}
\intertext{Las últimas fórmulas de probabilidad tienen algunos problemas prácticos, como por ejemplo, cuando $V=1$ o $V_i=0$. Para resolverlos se pueden agregar algunos factores de ajuste:}
P(k_i|{R}) &= \frac{\abs{V_i}+0.5}{|V|+1} &
P(k_i|\bar{R}) &= \frac{n_i - |V_i|+0.5}{N-|V|+1}\text{.}
\intertext{Una alternativa al uso de una constante igual a $0.5$, que no siempre es satisfactoria, es utilizar ${n_i}/N$ como factor de ajuste.}
P(k_i|{R}) &= \frac{|V_i|+\frac{n_i}{N}}{|V|+1} &
P(k_i|\bar{R}) &= \frac{n_i - |V_i|+\frac{n_i}{N}}{N-|V|+1}\text{.}
\end{align*}

La principal \textit{ventaja} del modelo probabilístico es
que los documentos son ordenados en orden decreciente de acuerdo a su \textit{probabilidad} de ser relevante. Las \textit{desventajas} incluyen: (1) es necesario realizar una separación de los documentos recuperados inicialmente en relevantes e irrelevantes; (2) el método \textit{no} tiene en cuenta la frecuencia de los términos en los documentos, o sea, los pesos son binarios; y (3) supone la independencia de los términos. Sin embargo, como se dijo en el modelo vectorial, en la práctica no está claro si ésta es una mala suposición.
\subsection{Modelo de Indexado de Semántica Latente}\label{sub:lsa}
Al resumir el contenido de los documentos y las consultas por medio de términos se puede producir un bajo rendimiento del proceso de recuperación por dos motivos. El primero es la aparición de documentos irrelevantes en el conjunto de resultados, mientras que el segundo es que los documentos relevantes, que no fueron indexados con ninguno de los términos de la consulta, nunca son recuperados. La razón principal para la aparición de estos efectos es la imprecisión inherente al proceso de recuperación que, como se dijo, está basado en conjuntos de términos clave.

Las ideas en un texto están relacionadas más a los conceptos que describen que a los términos que se utilizan para su descripción. Luego, el proceso de buscar las coincidencias de los documentos con una consulta dada podría basarse en coincidencia de conceptos en lugar de coincidencia de términos. Esto permitiría la recuperación de documentos aun si no estuvieran indexados con los términos de la consulta. El Indexado de Semántica Latente (LSI\footnote{del inglés, Latent Semantic Indexing.}) es un método que hace uso de estas ideas.

La idea principal en el \textit{Modelo de Indexado de Semántica Latente}~\cite{furnas88information, deerwester90indexing} es mapear cada vector que representa a un documento y cada vector que representa a una consulta, a un espacio de menos dimensiones que está asociado a conceptos, con la suposición de que el proceso de recuperación en ese espacio reducido puede ser superior a la obtenida en el espacio de términos. Veamos a continuación algunas definiciones.
\begin{definition}
\textnormal{\cite{furnas88information}}
 Sea $t$ el número de términos índice en la colección y $N$ el número total de documentos. Se define $\overrightarrow{M}=(M_{ij})$ como una matriz de asociación entre términos-documentos con $t$ filas y $N$ columnas. En cada elemento $M_{ij}$ de esta matriz se asigna un peso $w_{i,j}$ asociado al par término-documento $[k_{i},d_{j}]$, $w_{i,j}=w(k_i,d_j)$. Este peso podría calcularse usando TF-IDF.
\end{definition}
\begin{figure}[!ht]
\begin{align*}
 \overrightarrow{M} &= \bordermatrix{
      & d_1 & d_2 & \cdots & d_N \cr
  k_1 & w_{1,1} & w_{1,2} & \cdots & w_{1,N} \cr
  k_2 & w_{2,1} & w_{2,2} & \cdots & w_{2,N} \cr
  \vdots & \vdots & \vdots & \ddots & \vdots \cr
  k_t & w_{t,1} & w_{t,2} & \cdots & w_{t,N} \cr
}
\end{align*}
\caption{Matriz de asociación.} \label{fig:m_svd}
\end{figure}

El proceso de indexado de semántica latente plantea descomponer a la matriz de asociación $\overrightarrow{M}$ en tres componentes, utilizando descomposición en valores singulares, de la siguiente manera:
\begin{align*}
 \overrightarrow{M}=\overrightarrow{K}\overrightarrow{S}\overrightarrow{D^{T}}\text{.}
\end{align*}
La matriz $\overrightarrow{K}$ es la matriz de autovectores que se deriva de la matriz de correlación término-término, dada por $\overrightarrow{M}\overrightarrow{M^{T}}$, ortonormal por columnas. La matriz $\overrightarrow{D^{T}}$ es la matriz de autovectores que se deriva de la matriz traspuesta de correlación documento-documento, dada por $\overrightarrow{M^{T}}\overrightarrow{M}$, ortonormal por columnas. La matriz $\overrightarrow{S}$ contiene valores singulares en orden decreciente, es diagonal y de tamaño $r\times r$, en donde $r\leq\min(t,N)$ es el rango de $\overrightarrow{M}$. En la \autoref{fig:svd} se muestra de forma esquemática una matriz de $t\times N$ componentes.
\begin{figure}[!ht]
\centerline{
$\begin{matrix} 
 & \text{documentos} & & & & \\
\side{\hspace{-7mm}términos}
&
\begin{bmatrix} 
w_{1,1} & \dots & w_{1,N} \\
\\
\vdots & \ddots & \vdots \\
\\
w_{t,1} & \dots & w_{t,N} \\
\end{bmatrix}
&
=
&
\begin{bmatrix} 
\begin{bmatrix} \, \\ \, \\ \mathbf{u}_1 \\ \, \\ \,\end{bmatrix} 
\dots
\begin{bmatrix} \, \\ \, \\ \mathbf{u}_r \\ \, \\ \, \end{bmatrix}
\end{bmatrix}
&
\begin{bmatrix} 
\sigma_1 & \dots & 0 \\
\vdots & \ddots & \vdots \\
0 & \dots & \sigma_r \\
\end{bmatrix}
&
\begin{bmatrix} 
\begin{bmatrix} & & \textbf{v}_1 & & \end{bmatrix} \\
\vdots \\
\begin{bmatrix} & & \textbf{v}_r & & \end{bmatrix}
\end{bmatrix} \\
& M & =  & K & S & D^T \\
\end{matrix}$
}
\caption{Descomposición en valores singulares de la matriz $\vec{M}$.} \label{fig:svd}
\end{figure}

Si se descartan los valores singulares más pequeños y sólo se consideran los $s$ valores singulares más grandes de $\overrightarrow{S}$, junto con las columnas correspondientes de $\overrightarrow{K}$ y las filas correspondientes de $\overrightarrow{D^{T}}$, se obtiene una matriz $\overrightarrow{M_s}$ de rango $s$ que es la más cercana a la matriz $M$ original en el sentido de los cuadrados mínimos. Esta matriz está dada por:
\begin{align*}
 \overrightarrow{M_s}=\overrightarrow{K_s}\overrightarrow{S_s}\overrightarrow{D^{T}_s}\text{,}
\end{align*}
en donde $s$, $s<r$, es la dimensión del espacio reducido de conceptos. En la \autoref{fig:svd_reduced} se muestra el modelo reducido. Al elegir el valor de $s$ se quiere que sea lo suficientemente grande como para preservar toda la estructura presente en los datos originales, pero lo suficientemente pequeño como para descartar los errores de muestreo o los detalles irrelevantes, que están presentes en las representaciones basadas en términos índice. Como puede verse al favorecerse una característica se empeora la otra.
\begin{figure}[!ht]
 \centerline{$
\begin{matrix} 
 &  & & & & \text{documentos} \\
\begin{bmatrix} 
\begin{bmatrix} \, \\ \, \\ \mathbf{u}_1 \\ \, \\ \,\end{bmatrix} 
\dots
\begin{bmatrix} \, \\ \, \\ \mathbf{u}_r \\ \, \\ \, \end{bmatrix}
\end{bmatrix}
&
\begin{bmatrix} 
\sigma_1 & \dots & 0 & 0 \\
\vdots & \ddots & \vdots & \multirow{2}*{\vdots} \\
0 & \dots & \sigma_s & \\
0&\multicolumn{2}{c}{\dots} & 0\\
\end{bmatrix}
&
\begin{bmatrix} 
\begin{bmatrix} & & \textbf{v}_1 & & \end{bmatrix} \\
\vdots \\
\begin{bmatrix} & & \textbf{v}_r & & \end{bmatrix}
\end{bmatrix}
&
=
&
\side{\hspace{-7mm}términos}
&
\begin{bmatrix} 
w'_{1,1} & \dots & w'_{1,N} \\
\\
\vdots & \ddots & \vdots \\
\\
w'_{t,1} & \dots & w'_{t,N} \\
\end{bmatrix}
\\
 K & S' & D^T & = & & M_s \\
\end{matrix}$
}
\caption{Descomposición en valores singulares reducida de la matriz $\vec{M}$.} \label{fig:svd_reduced}
\end{figure}

La relación entre dos documentos en el espacio reducido de dimensión $s$ puede calcularse a partir de la matriz $\overrightarrow{M^{T}_s}\overrightarrow{M_s}$ dada por:
\begin{align*}
 \overrightarrow{M^{T}_s}\overrightarrow{M_s}	&= (\overrightarrow{K_s}\overrightarrow{S_s}\overrightarrow{D^{T}_s})^{T}\overrightarrow{K_s}\overrightarrow{S_s}\overrightarrow{D^{T}_s}\text{,} && \text{por definición de $\overrightarrow{M}$,}\\
			&= \overrightarrow{D_s}\overrightarrow{S_s}\overrightarrow{K^{T}_s}\overrightarrow{K_s}\overrightarrow{S_s}\overrightarrow{D^{T}_s}\text{,} && \text{por propiedad de traspuesta,}\\
			&= \overrightarrow{D_s}\overrightarrow{S_s}\overrightarrow{S_s}\overrightarrow{D^{T}_s}\text{,} && \text{por $\overrightarrow{K}$ ortonormal por columnas,}\\
			&= (\overrightarrow{D_s}\overrightarrow{S_s})(\overrightarrow{D_s}\overrightarrow{S_s})^{T}\text{,}
\end{align*}
en donde el elemento $(j,k)$ de la matriz cuantifica la relación entre los documentos $d_j$ y $d_k$.

Para ordenar los documentos respecto de alguna consulta de un usuario se modela la consulta como un \textit{pseudodocumento} en la matriz de términos-documentos original $\overrightarrow{M}$. Si la consulta es el documento 0 de la matriz original, entonces la primera columna de la matriz $\overrightarrow{M^{T}_s}\overrightarrow{M_s}$ contiene el ranking de todos los documentos respecto de la consulta.

Las matrices que se utilizan en el modelo de indexado de semántica latente son de rango $s$, $s<<t$ y $s<<N$, por lo que es un sistema de indexación eficiente de los documentos de una colección, permitiendo además la eliminación de ruido y de redundancia. Además, proporciona un nuevo enfoque al problema de la recuperación de información, basado en la teoría de descomposición en valores singulares.

Una de las limitaciones del LSI es determinar el valor óptimo de $s$.
En la mayoría de las aplicaciones $s$ es mucho menor que el número de términos en el índice. Sin embargo, aún no existe una teoría o método para predecir el valor óptimo. Se ha conjeturado que la dimensión óptima es algo intrínseco al dominio de los documentos indexados y, por lo tanto, debe ser determinado empíricamente~\cite{landauer08lsa}. 
%
%
Los resultados experimentales muestran que la determinación de la dimensión óptima es un problema complejo. En 1991, Susan Dumais, una de las inventoras del LSI, reportó que $s=100$ funcionaba bien para una colección de aproximadamente 1000 documentos, pero también notó que ``el número necesario para capturar la estructura de otra colección probablemente dependerá de su tamaño''~\cite{dumais91improving}. Desde ese momento, otros trabajos han especulado que mientras más cantidad de conceptos contenga una colección, más grande deberá ser el valor de $s$ necesario para representarlos adecuadamente. En~\cite{bradford08empirical} se analizan los efectos de variar esta dimensión, determinando cómo afecta esto al rendimiento del proceso de recuperación y al costo computacional. Los resultados mostraron que un valor de $s$ entre 300 y 500 es un valor apropiado para obtener un rendimiento estable en colecciones de entre diez mil y varios millones de documentos, por lo que el factor de crecimiento es bastante lento.

A continuación se detallará el proceso de formulación y optimización de consultas. También se explicarán las nociones de relevancia y similitud que serán utilizadas en los capítulos siguientes.
\section{Formulación de consultas}\label{sec:query_formulation}
El proceso de formulación de una consulta es complejo y está condicionado a quien lo inicia, a su conocimiento sobre el contenido del repositorio, a su conocimiento acerca de los procesos de indexación y búsqueda, a su familiaridad con el tópico que está buscando, a sus preferencias personales con respecto al vocabulario y a otros aspectos. De hecho, el usuario debe tomar una decisión estadística basándose en su experiencia personal para obtener los resultados que quiere. En cualquier sistema de recuperación la probabilidad a priori de que una consulta satisfaga las necesidades de un usuario varía en gran medida. Por ejemplo, un usuario que conoce de la existencia de un documento dentro de un repositorio, puede formular una consulta que seguramente será exitosa, ya que por ejemplo puede crear una consulta que sea idéntica al documento en cuestión. Como otro caso extremo está el usuario que necesita información sobre un tópico que desconoce por completo. Claramente las probabilidades de que este usuario genere una consulta que recupere el mejor conjunto de documentos son muy bajas.

Por lo tanto en la práctica, las probabilidades de que un sistema satisfaga las necesidades de un usuario son muy variables y, consecuentemente, es importante considerar técnicas que reduzcan esta varianza. Esto puede llevarse a cabo desde dos puntos de vista diferentes.
Uno de ellos es desde el punto de vista del usuario, optimizando las consultas para que se minimicen los costos de recuperación, los costos de optimización, y el costo de la información. Por otro lado, también es 
de interés separar los defectos del proceso de indexación de los defectos producidos por consultas mal formuladas. Es así que los sistemas de IR pueden evaluarse respecto de un conjunto particular de consultas predefinidas, o
también es posible definir una consulta \textit{óptima} para un determinado pedido de información, y de esta forma evaluar mejor el poder de una técnica dada. A continuación se estudiarán algunas técnicas en este sentido.

\subsection{Reformulación y Optimización de la consulta}
Para definir un pedido óptimo es necesario trabajar con una representación explícita del proceso de coincidencia pedido-documento. Una función de coincidencia muy utilizada es la correlación por coseno del pedido con los documentos en el repositorio (\autoref{eqn:cosine_similarity}). Como se mencionó en la \autoref{def:doc_weights}, los pesos asignados a los términos de un documento son positivos, por lo que esta correlación induce un orden en los elementos del repositorio que es equivalente a su distancia angular con el vector de la consulta (esto es, $0 \le \mathit{sim}(d_{j}, q) \le 1$ corresponde a una separación angular que va desde 90 a 0 grados sexagesimales).

Dado cualquier pedido (consulta) $q$, asumimos la existencia de un subconjunto $R$ ($R \subset C$) del conjunto de documentos contenidos en el repositorio $C$. Este conjunto es el conjunto de documentos relevantes para el pedido $q$ y debe estar especificado fuera del contexto del sistema de recuperación.

Habiendo definido $R$, un pedido \textit{ideal} puede ser definido como uno que induce un orden de elementos de $C$ tal que todos los miembros de $R$ están posicionados más altos (o sea que tienen una correlación más alta) que todos los otros elementos de $C$.

El pedido ideal es 
incierto porque la relevancia es 
una apreciación subjetiva
que, en teoría, la determina el usuario. En tal caso puede decirse que el proceso de indexación no se define desde el punto de vista de un usuario, porque no permite hacer las distinciones que hace un usuario. Por supuesto esto será la norma en lugar de la excepción, ya que la indexación está diseñada para reducir la cantidad de información almacenada en lugar de preservarla. Por esta razón, un pedido óptimo no ambiguo se define como una función de $R$, $C$ y una transformación en el índice la cual es única para cada subconjunto único no vacío $R$ de $C$.

Un pedido \textit{óptimo} correspondiente a un subconjunto $R$ dado de un repositorio $C$, bajo una transformación del índice $T$, es aquel pedido que maximiza la diferencia entre la media de las correlaciones de los documentos relevantes (miembros de $R$) y la media de las correlaciones de los documentos no relevantes (miembros de $C$ que están en $\bar{R}$).
 
En términos matemáticos, el vector de un pedido \textit{óptimo} $\overrightarrow{q_{opt}}$ correspondiente a un conjunto $R \subset C$ está definido como el vector $\overrightarrow{q}$ para el cual:
\begin{align*}
A &= \frac{1}{|R|}\sum\limits_{d_j \in R } {\mathit{sim}(d_{j}, q)}  - \frac{1}{|\bar{R}|}\sum\limits_{d_j  \in \bar{R} } {\mathit{sim} (d_{j}, q)}\text{,} 
\intertext{es máximo, en donde $\mathit{sim}(d_{j}, q)$ es alguna medida de la calidad de los documentos recuperados. Si deseamos tener en cuenta solamente a los pedidos que tienen componentes no negativos en la consulta, entonces el problema se modifica para maximizar $A$ restringido a $w(k_i, q) \ge 0$.\newline
\indent Suponiendo que utilizamos como medida de la calidad de los documentos a la similitud por coseno, substituimos $\mathit{sim}(d_{j}, q)$ de acuerdo a la \autoref{eqn:cosine_similarity} y utilizando la notación de vectores, la ecuación anterior queda:}
A &= \frac{1}{|R|}\left( {\frac{\overrightarrow{q}}{{\abs{ \overrightarrow{q} }}}} \right) \cdot \sum\limits_{\overrightarrow{d_j}  \in R } {\frac{{\overrightarrow{d_j} }}{{\abs{ {\overrightarrow{d_j} } }}}}  - \frac{1}{|\bar{R}|}\left( {\frac{\overrightarrow{q}}{{\abs{ \overrightarrow{q} }}}} \right) \cdot \sum\limits_{\overrightarrow{d_j} \in \bar{R} } {\frac{{\overrightarrow{d_j} }}{{\abs{ {\overrightarrow{d_j} } }}}}\text{,}
\intertext{o}
A &= \frac{\overrightarrow{q}}{\abs{\overrightarrow{q}}}\left[ {\frac{1}{|R|}\sum\limits_{\overrightarrow{d_j}\in R}{\frac{\overrightarrow{d_j}}{{\abs{\overrightarrow{d_j}}}}} - \frac{1}{|\bar{R}|}\sum\limits_{\overrightarrow{d_j} \in \bar{R}}{\frac{\overrightarrow{d_j}}{\abs{\overrightarrow{d_j}}}}} \right]\text{,}\\
&= \hat{q} \cdot \overrightarrow{v}\text{.}
\end{align*}
En la ecuación de arriba $\hat{q}$ es el vector unitario en la dirección de $\overrightarrow{q}$ y $\overrightarrow{v}$ es un vector igual a la parte encerrada entre corchetes de la ecuación. Claramente, por definición de producto escalar, el $\overrightarrow{q_{opt}}$ es uno en la dirección $\overrightarrow{v}$, o sea $\overrightarrow{q_{opt}} = k\overrightarrow{v}$ (en donde $k$ es un escalar arbitrario), o
\begin{align}
\overrightarrow{q_{opt}}  &= \frac{1}{|R|}\sum\limits_{\overrightarrow{d_j}  \in R } {\frac{{\overrightarrow{d_j} }}{{\abs{ {\overrightarrow{d_j} } }}}}  - \frac{1}{|\bar{R}|}\sum\limits_{\overrightarrow{d_j}  \in \bar{R} } {\frac{\overrightarrow{d_j} }{\abs{ {\overrightarrow{d_j}}}}}\text{.}
\intertext{Además, es trivial demostrar que $A$ restringido a $w(k_{i},q) \ge 0$ (donde $i$ varía sobre todas las coordenadas de $\overrightarrow{q}$) se maximiza con el vector compuesto por:}
{w(k_i,q'_{opt})}  &= %
\begin{cases}
  {w(k_i,q_{opt})}\text{,} & \text{para } {w(k_i,q_{opt})}  \ge 0\text{,} \\
  0\text{,}         & \text{para } {w(k_i,q_{opt})} < 0\text{.}
\end{cases} \nonumber
\end{align}
Por lo tanto, con estas suposiciones, existe una consulta óptima no ambigua que se corresponde con cualquier subconjunto no vacío $R$ de $C$. Durante el proceso de evaluación de los sistemas de recuperación automáticos, esta representación de un pedido óptimo: (1) brinda la capacidad de aislar el proceso de indexado de la varianza debido al proceso de formulación de la consulta; (2) mide la habilidad de la transformación del índice para distinguir un conjunto particular de documentos de todos los otros en el repositorio, en donde este conjunto particular se asume que 
está compuesto por todos los documentos que fueron juzgados como relevantes para algún tópico en particular.
%
\subsection{Realimentación de relevancia}\label{sub:feedback}
\begin{figure}[!ht]
\centerline{%
   \includegraphics[scale=0.5]{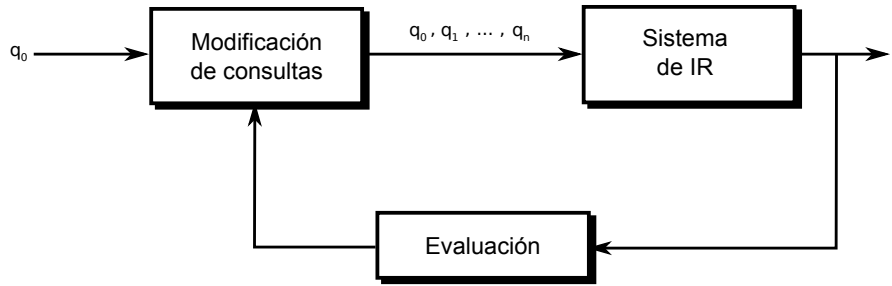}%
}%
\caption{Realimentación de relevancia.}
\label{fig:control_realimentation}
\end{figure}

La generación de una consulta óptima, que se corresponda con un conjunto particular de documentos, tiene una consecuencia directa en la operación de recuperación de información, dado que el conjunto de documentos en cuestión es el objeto de la búsqueda. Por lo tanto, no hay una forma directa de crear una consulta óptima, ya que si se tiene esa habilidad se elimina la necesidad de la búsqueda. Este tipo de circularidad sugiere una gran analogía con la realimentación en la teoría de control y fue estudiada por Rocchio en los~'70s~\cite{rocchio71relevance}. Puede verse más claramente la analogía con un sistema de realimentación secuencial si consideramos una secuencia de operaciones de recuperación comenzando con una $q_0$, la cual es luego modificada basándose en la salida que produce el sistema de recuperación (utilizando $q_0$ como entrada), de forma que la consulta modificada $q_1$ esté más cerca de la consulta óptima para ese usuario. Si dejamos que el usuario especifique cuáles de los documentos recuperados son relevantes (los recuperados con $q_0$) y cuáles no, tendremos la señal de error del sistema de IR. Esto puede apreciarse gráficamente en la \autoref{fig:control_realimentation}. Basándonos en las señales de error y la original es posible producir una consulta modificada de forma tal que la salida esté más cerca de lo que el usuario desea o, de forma que la consulta esté en efecto más cerca de la consulta óptima para ese usuario. 

Un escenario típico para este proceso involucra los siguientes pasos:
\begin{enumerate}
\renewcommand{\labelenumi}{\bfseries{Paso \arabic{enumi}:}}
\setlength{\leftskip}{5mm}
\item Se formula la consulta.
\item El sistema devuelve un conjunto inicial de resultados.
\item \label{item:assessment} Se calcula la relevancia de los resultados obtenidos.
\item El sistema determina una mejor representación de las necesidades de información basándose en esta realimentación.
\item El sistema devuelve un conjunto revisado de resultados.
\end{enumerate}

Dependiendo del nivel de automatización del Paso~\ref{item:assessment} podemos distinguir tres formas de realimentación que explicaremos más adelante. La efectividad de este proceso dependerá de: (a) qué tan buena es la consulta inicial, ya que los resultados obtenidos deben contener un número razonable de documentos relevantes al tópico que se está buscado; (b) qué tanto vocabulario comparten los documentos relevantes encontrados con los documentos que se están buscando; y (c) qué tan rápido el proceso iterativo converge hacia el óptimo~\cite{harman92relevance}.
%


\subsubsection{Realimentación Supervisada}\label{sub:feedback:sup}
La realimentación de relevancia supervisada necesita una realimentación \textit{explícita}, la cual se obtiene típicamente de usuarios que indican la importancia de cada documento~\cite{rocchio71relevance,xu08active}.
Dada la consulta original $q_0$ y dados los resultados que ella obtuvo como una lista ordenada de acuerdo a su relación con la consulta, el usuario examina esta lista y especifica cuáles son relevantes y cuáles no. Se asume que faltarán algunos resultados relevantes en la lista, por lo que sólo se modificará la consulta con la información que proporciona el usuario. La consulta modificada está compuesta por la $q_0$ más un vector de optimización basado en la información de realimentación. La suposición fundamental de estos sistemas es que el vector resultante será una mejor aproximación que $q_0$ a la consulta óptima y producirá mejores resultados cuando se lo envíe al sistema.

Luego buscamos una relación de la forma:
\begin{align}
q_1 &= f(q_0, R, \bar{R}), \nonumber \\
\intertext{en donde $q_0$ es la consulta original, $R$ es el subconjunto de documentos que el usuario estima que son relevantes y $\bar{R}$ es el subconjunto que estima que no son relevantes. La forma que sugiere esto es:}
\overrightarrow{q_1}  &= \overrightarrow{q_0}  + \frac{1}{|R|}\sum\limits_{\overrightarrow{d_j} \in R}{\overrightarrow{d_{j}}}  - \frac{1}{|\bar{R}|}\sum\limits_{\overrightarrow{d_j} \in \bar{R}}{\overrightarrow{d_{j}}}\text{,} \nonumber\\ 
\intertext{en donde todos los vectores son unitarios. Luego $\overrightarrow{q_1}$ es el vector que resulta de la suma del vector original más el vector diferencial entre los miembros del conjunto $R$ y los del conjunto $\bar{R}$. \newline
\indent La ecuación anterior puede ser rescrita en la forma:}
\overrightarrow{q_1}  &= {|R|}{|\bar{R}|} \overrightarrow{q_0}  + {|\bar{R}|} \sum\limits_{\overrightarrow{d_j} \in R}{\overrightarrow{d_{j}}} - |R| \sum\limits_{\overrightarrow{d_j}\in \bar{R}}{\overrightarrow{d_{j}}}\text{.}
\label{eqn:rocchio} \\
\intertext{Los componentes de $\overrightarrow{q_1}$ pueden restringirse a valores positivos haciendo:}
{w(k_i,q'_1)} &=%
    \begin{cases}
      {w(k_i,q_1)}\text{,} & \text{para }{w(k_i,q_1)}\ge 0\text{,} \\
      0\text{,} & \text{para }{w(k_i,q_1)}  < 0\text{.}
    \end{cases}  \nonumber
\end{align}
Esto representa la relación básica para la modificación de una consulta utilizando \textit{realimentación de relevancia}~\cite{rocchio71relevance}. Esta relación puede modificarse de distintas maneras imponiendo restricciones adicionales. Por ejemplo, el peso de la consulta original ($|R||\bar{R}|$) puede ser una función de la cantidad de realimentación, de forma tal que con una realimentación alta la consulta original tenga un efecto menor en la consulta final, y, por el contrario, un efecto alto si la realimentación es baja.

El proceso descripto puede hacerse iterativamente y luego, de forma general, puede escribirse:
\begin{align*}
q_{i+1} =f(q_i, R_i, \bar{R}_i)\text{,}
\end{align*}
en donde $q_i$ es la $i$-ésima consulta de la secuencia y $R_i$ y $\bar{R}_i$ son los subconjuntos identificados como relevantes e irrelevantes respectivamente, obtenidos como resultado de enviar $q_i$ al sistema de IR.

\begin{figure}[!ht]
\centerline{
  \includegraphics[scale=0.5]{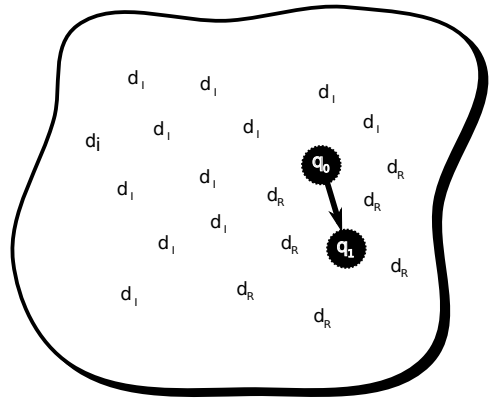}%
}%
\caption{Modificación de una consulta basada en la realimentación del usuario.} \label{fig:relevance_regions}
\end{figure}

La consulta original del usuario se utiliza para identificar una región en el espacio del índice, la cual debería contener documentos relevantes. Pero, dado que no existe un conocimiento detallado acerca de las características de los documentos en el repositorio, es poco probable que esa región sea óptima. Luego de que el usuario identifica los documentos relevantes en la región, el sistema tiene suficiente información para intentar generar una consulta mejorada que se centra en los documentos relevantes y, a su vez, maximiza la distancia a los documentos no relevantes. Esto sólo es posible si los conjuntos de documentos relevantes e irrelevantes son diferenciables. Una interpretación gráfica puede verse en la \autoref{fig:relevance_regions}.

Una desventaja que se ha encontrado en este método es que requiere que los resultados obtenidos por la primer consulta sean lo suficientemente buenos como para poder encontrar documentos relevantes que guíen al algoritmo hacia la región correcta del espacio de documentos. Esto puede ser un gran problema si el usuario no tiene el suficiente conocimiento del tema que está buscando. También, para que la \autoref{eqn:rocchio} funcione correctamente, es necesario que la distribución de términos en los documentos relevantes sea similar y, además, distinta de la existente en los irrelevantes.

Una variante del método de Rocchio es la presentada en~\cite{jordan04extending}, en donde se compara la consulta con el contexto del usuario. Los autores argumentan que la versión original del algoritmo de Rocchio trabaja solamente basándose en un modelo de usuario que se construye cada vez que se realiza una consulta, y toda información que pudo haberse obtenido en consultas previas se pierde. Es por esto que se utiliza esta información para construir perfiles múltiples del usuario que se preservan a lo largo del tiempo.

\subsubsection{Realimentación sin supervisión}
\label{sub:feedback:nosup}
Esta variante aplica realimentación de relevancia de forma \textit{ciega}, sin la asistencia de un usuario que indique cuáles documentos son relevantes y cuáles no. Existen dos tipos básicos de realimentación ciega~\cite{he06comparing}. La primera clase típicamente asume que los $k$ mejores resultados del proceso de búsqueda son relevantes~\cite{evans93design, buckley95new}.
El segundo tipo se ha investigado como una forma alternativa a la realimentación ciega. La expansión de consultas se realiza agregando términos de los $k$ documentos más relevantes recuperados con la consulta original pero, en esta variante, de una colección \textit{diferente}~\cite{kwok98improving, he06comparing}. Esta colección se suele llamar ``colección de expansión'', 
la cual debe ser razonablemente similar a la colección de ``búsqueda'', de otra manera sólo se introduciría ruido. Esta segunda fuente de información incrementa el tamaño total de la colección y aumenta también la probabilidad de tener documentos relevantes en posiciones altas del ranking. Sin embargo, las mejoras obtenidas no han mostrado ser estadísticamente superiores a la primera versión.

Otra forma de expansión de consultas se basa en la utilización de diccionarios de sinónimos\footnote{del inglés, thesaurus. En español la palabra tesauro está en desuso de acuerdo a la RAE.}. Esta técnica expande la consulta utilizando palabras o frases que tienen un significado similar a las que se encuentran en la consulta. Esto aumenta las probabilidades de tener coincidencias con el vocabulario utilizado en los documentos relevantes que se están buscando. Sin embargo, no existen resultados concluyentes acerca de las mejoras en el rendimiento introducidas por el método, incluso si las palabras son elegidas por los usuarios~\cite{voorhees94query}.

La creación del diccionario también puede realizarse a partir de los documentos recuperados por la consulta original. En~\cite{attar77local} los $k$ mejores documentos recuperados se utilizan como una fuente de información sobre la que se construye el diccionario de forma automática. Los términos en esos documentos se agrupan con un algoritmo de clustering y se los trata como cuasi sinónimos. En~\cite{croft79using} se utiliza esta misma fuente información pero para re-estimar la importancia de los términos en el conjunto de documentos relevantes. Luego, basándose en esta información y sin agregar términos nuevos, se modifican los pesos de los términos de la consulta original. Los experimentos realizados han mostrado mejoras en el rendimiento, pero sólo sobre una colección pequeña. Una alternativa es analizar la colección de documentos de forma completa para generar un diccionario más efectivo~\cite{jones71automatic}.

Una generalización exitosa del método de Rocchio es el mecanismo de la \textit{Divergencia de la Aleatoriedad} (DFR\footnote{del inglés, Divergence From Randomness.}) con estadísticas de Bose-Einstein (\mbox{Bo1-DFR})~\cite{amati02probabilistic}. Para aplicar este modelo necesitamos primero asignar un peso a los términos basándonos en su grado de información. Este se estima con la divergencia entre su distribución en los documentos mejor clasificados y en una distribución aleatoria, de la siguiente manera:
\begin{align*}
 w(k_i)=\mathit{frec}(k_i) . log_2\frac{1+ P_n}{P_n}+log_2(1+P_n)\text{,}
\end{align*}
en donde $\mathit{frec}(k_i)$ es la frecuencia del término $k_i$ en los documentos mejor clasificados y $P_n$ es el porcentaje de documentos en la colección que contienen $k_i$. Finalmente, la consulta se expande mezclando los términos más informativos con los términos originales de la consulta.

\subsubsection{Realimentación Semisupervisada}%
\label{sub:feedback:semisup}%
En esta última forma de realimentación tampoco es necesaria la intervención del usuario, ya que la relevancia es \textit{inferida} por el sistema. Una forma de hacer esto es monitorear el comportamiento del usuario.
En~\cite{sufyan07websearch} se propone el uso de una métrica de \textit{calidad de búsqueda} que mide la ``satisfacción'' que un usuario obtiene con determinados documentos. La puntuación que el usuario le asigna a los resultados de un motor de búsqueda se tiene en cuenta para mejorar los resultados futuros. Esta realimentación se obtiene de forma \textit{implícita}, monitoreando las acciones que el usuario realiza normalmente sobre los resultados de una búsqueda, infiriéndose de ellas la realimentación. Esto es una ventaja respecto de la realimentación supervisada que, como se mostrará en la \autoref{sec:antecedentes}, es una práctica muy demandante.
%
En particular, en~\cite{sufyan07websearch} se identifican siete acciones, o características, que es posible monitorear: 
(a) el orden en que se visitan los documentos; 
(b) el tiempo que se examina un documento;
(c) si se imprime;
(d) se guarda;
(e) se agrega a \textit{Favoritos\footnote{del inglés, bookmarks o favourites.}}; o
(f) se envía por correo electrónico un documento; y
(g) el número de palabras que se copian y pegan en otra aplicación.
Estas características se utilizan para obtener el nivel de importancia que el usuario le asigna a un documento para una consulta dada. Este proceso continúa en el tiempo y va mejorando el conocimiento que tiene el sistema sobre las necesidades del usuario y, por lo tanto, le ayudará a obtener resultados cada vez más acordes a él.

Algunas de estas características ya habían sido utilizadas en otros trabajos previos, una clasificación puede encontrarse en~\cite{kelly03implicit}. En~\cite{joachims02optimizing} se utiliza la información histórica de los clicks de los usuarios, sobre los resultados de un motor de búsqueda, para entrenar un clasificador y así mejorar el ranking de varios motores web. 
%
En~\cite{white04simulated} se hizo también una simulación para determinar la efectividad de distintos modelos de realimentación basados en los clicks de los usuarios. Los modelos analizados fueron votación binaria, varios modelos probabilísticos y un modelo aleatorio. Los documentos de los resultados se presentaron con distintas granularidades, desde solamente el título y algunas sentencias hasta el texto completo, y dependiendo de la granularidad elegida por el usuario, el sistema aprendía qué objetos eran más relevantes para una consulta.
En~\cite{sugiyama04adaptive} se utilizan perfiles de usuario colaborativos basados en la historia de navegación de los usuarios para mejorar los resultados de una búsqueda. 

En~\cite{kelly04display} se utilizó el tiempo que emplea un usuario para explorar un documento como una forma de realimentación, mostrando que esta es una característica muy subjetiva y que deben aprenderse las preferencias de cada usuario en particular para lograr los mejores resultados.
%
%
En~\cite{amati04query} se introduce la noción de \textit{robustez}, ya que, aunque en muchos casos la expansión de consultas tiene éxito, en otros empeora la calidad de los documentos recuperados. Si se mide solamente el rendimiento global de un sistema, se obtiene una imagen del comportamiento promedio, pero no es posible analizar la varianza de los resultados y se pueden estar obviando resultados muy pobres en algunas consultas.
Los autores notaron que las consultas que presentaban un peor rendimiento,
aun sin la utilización de expansión de consultas, empeoraban al habilitar esta característica. Por lo tanto introducen, en la conferencia TREC (\autoref{sub:trec}), una métrica que predice cuándo la expansión resultará beneficiosa y cuándo no, logrando mejoras significativas sobre el método estándar.

\section{Similitud}\label{sec:similitud}
La idea de definir un valor numérico para la similitud (o su inversa, la distancia) entre dos objetos, cada uno caracterizado por un conjunto común de atributos,
es uno de los problemas fundamentales en IR y en otras áreas. 
Por ejemplo, en matemática se utilizan los métodos geométricos para medir la similitud en estudios de congruencia y homotecia, así como también en campos como la trigonometría~\cite{alt88congruence}. Los métodos topológicos se aplican en campos como la semántica espacial~\cite{kuipers00spatial}. La teoría de grafos se usa ampliamente para evaluar las similitudes cladísticas en taxonomías~\cite{thorley98information}. La teoría de grafos difusa también ha desarrollado sus propias métricas de similitud, que tienen aplicación en las más diversas áreas, como administración~\cite{dubois03fuzzy}, medicina~\cite{adlassnig86fuzzy} y meteorología~\cite{mcbratney85applications}. Medir la similitud de secuencias de pares de proteínas es un problema muy importante en la biología molecular~\cite{lipman85rapid}.

La similitud también ha jugado un papel fundamental en experimentos psicológicos, por ejemplo, en experimentos con personas a las cuales se les pregunta acerca de la similitud de pares de objetos. En estos estudios se utilizan una gran variedad de técnicas, pero la más común es preguntar si los objetos son similares o no, o indicar en una escala el nivel de similitud que la persona piensa que tienen los objetos. El concepto de similitud también juega un rol fundamental en el modelado de tareas psicológicas, especialmente en teorías de reconocimiento, identificación y clasificación de objetos, como ser los gustos por ciertas marcas comerciales. Una suposición común es que cuando se está evaluando un producto, las personas se imaginan un producto ideal y luego juzgan la similitud de un producto nuevo con respecto a éste~\cite{coombs64theory}.

Debido a la gran diversidad de áreas de aplicación que se mencionó más arriba, y a una falta de comunicación entre ellas, hubo mucho esfuerzo duplicado y, en general, los coeficientes más utilizados se reinventaron varias veces, existiendo diversos nombres para las mismas métricas.
Algunos coeficientes son métricas de la distancia, o de la disimilitud entre los objetos, por lo que tienen un valor nulo para los objetos idénticos, mientras que otros miden directamente la similitud, y tienen un valor máximo para los objetos idénticos.
Si la métrica tiene un rango $[0, 1]$ puede transformarse un coeficiente de similitud en un coeficiente de distancia complementario simplemente restándose el valor de similitud con la unidad, o calculando una inversa~\cite{boriah08similarity}. En algunos casos una similitud y su complemento se desarrollaron de manera independiente y tienen nombres distintos.

Con una métrica de similitud pueden compararse dos documentos, dos consultas o un documento con una consulta. Entonces, esta medida puede ser una función que calcule el grado de similitud entre cualquier par de textos. También es posible crear un \textit{orden} de los documentos recuperados por un sistema de IR a partir de esta medida. Esto es algo fundamental en la evaluación de muchos sistemas de IR. Dado que aún no existe la mejor métrica, existe un gran número de métricas de similitud propuestas en la literatura. Recopilar una lista completa es una tarea imposible, por lo que aquí nos concentraremos en las métricas más utilizadas o las más comunes en IR, y en particular en las utilizadas en esta tesis. Una recopilación de coeficientes de similitud puede encontrarse en~\cite{ellis93measuring}.

Algunos métodos para resolver el problema del cálculo de la similitud incluyen: varios tipos de \textit{distancias de edición}, o de Levenshtein~\cite{levenshtein65binary}, entre dos términos. Esta distancia se define como la cantidad mínima de cambios necesarios en un término para transformarlo en el otro, en donde las operaciones permitidas se eligen de un conjunto fijo, como ser inserción o sustitución de una letra. También puede mencionarse la distancia de \textit{Hamming}, entre dos cadenas de caracteres de la misma longitud. Se basa  en~\cite{hamming50error} y es igual a la cantidad de posiciones en las que difieren ambas cadenas. Asimismo existen distancias \textit{universales}, o independientes de un modelo hipotético, como la \textit{distancia de información}~\cite{bennett98information}; y métodos basados en algoritmos de compresión universales, para la estimación de la entropía relativa de pares de secuencias de símbolos~\cite{ziv93measure}, o divergencia de Kullback-Leibler~\cite{kullback51information}. En un espacio multidimensional pueden definirse distancias entre pares de vectores en ese espacio, tales como la distancia \textit{Euclídea} y la distancia de \textit{Manhattan}. La primera se calcula como la norma Euclídea del vector diferencia de ambos vectores, y la última como la suma de las distancias en cada dimensión. La distancia de Hamming puede verse como la distancia de Manhattan entre vectores de bits~\cite{esposito02classical}. En el Modelo Vectorial propuesto por Salton y visto en la \autoref{sub:mod_vectorial}, la medida de similitud más común es la \textit{similitud por coseno} (\autoref{eqn:cosine_similarity}). A continuación se mencionarán algunas métricas muy utilizadas en el área y más adelante se detallará el concepto de \textit{similitud semántica}.

\subsection{Coeficiente de Jaccard}
El coeficiente de similitud de Jaccard o \textit{índice de Jaccard} mide la similitud entre dos conjuntos de muestras y fue concebido con la intención de comparar los tipos de flores presentes en un ecosistema de la cuenca de un río, con los tipos presentes en las regiones aledañas~\cite{jaccard01etude}. Se define como la relación entre el tamaño de la intersección de ambos conjuntos y el tamaño de la unión:
\begin{align}\label{eqn:jaccard_sim}
 \mathit{sim}_J(A,B) = \frac{\abs{A \cap B}}{\abs{A \cup B}}.
\end{align}

La \textit{distancia de Jaccard} mide la disimilitud entre dos conjuntos de muestras y se define como el complemento del coeficiente de Jaccard:
\begin{align*}
 J_{\delta}(A,B) = 1 - \mathit{sim}_J(A,B) =  \frac{ \abs{A \cup B} - \abs{A \cap B} }{\abs{A \cup B}}.
\end{align*}
Muchos años después, Lee R. Dice, también en el campo de la Biología, propone una métrica relacionada con el coeficiente de Jaccard, conocida como \textit{coeficiente de Dice}, para medir el nivel de relación que tienen dos especies animales comparada con una relación fortuita esperada~\cite{dice45measures}:
\vspace{-3mm}
\begin{align*}
\mathit{sim}_D(A,B)	&= \frac{ 2\abs{A \cap B}}{\abs{A} + \abs{B}}.\\
\intertext{\vspace{-3mm}Puede encontrarse la relación antes mencionada de la siguiente manera:\vspace{-3mm}}
 \mathit{sim}_D(A,B)	&= {\frac{ 2\abs{A \cap B}}{\abs{A} + \abs{B} - \abs{A \cap B} + \abs{A \cap B}}},\\
			&= {\frac{ 2\abs{A \cap B}}{\abs{A \cup B} + \abs{A \cap B}}},\\
			&= {\frac{ 2\abs{A \cap B}}{\abs{A \cup B} + \abs{A \cap B}}}{\frac{\abs{A \cup B}}{\abs{A \cup B} }},\\
			&= {\frac{ 2\abs{A \cap B}}{\abs{A \cup B} }}{\frac{\abs{A \cup B}}{\abs{A \cup B} + \abs{A \cap B}}},\\
			&= {\raise0.7ex\hbox{${{{2\mathit{sim}_J(A, B)}}}$} \!\mathord{\left/
			    {\vphantom {{2\mathit{sim}_J(A, B)} {\frac{\abs{A \cup B} + \abs{A \cap B}}{\abs{A \cup B}}}}}\right.\kern-\nulldelimiterspace}
			    \!\lower0.7ex\hbox{${{\frac{\abs{A \cup B} + \abs{A \cap B}}{\abs{A \cup B}}}}$}},\\
			&= {\raise0.7ex\hbox{${{{2\mathit{sim}_J(A, B)}}}$} \!\mathord{\left/
			    {\vphantom {{2\mathit{sim}_J(A, B)} {{1+{\frac{\abs{A \cap B}}{\abs{A \cup B}}}}}}}\right.\kern-\nulldelimiterspace}
			    \!\lower0.7ex\hbox{${1+{\frac{\abs{A \cap B}}{\abs{A \cup B}}}}$}},\\
			&= {\frac{2\mathit{sim}_J(A, B)}{1+\mathit{sim}_J(A, B)}}. 
\end{align*}\vspace{-15mm}

Para aplicar la \autoref{eqn:jaccard_sim} para medir el grado de similitud entre un documento y una consulta debemos llevar o extender esta ecuación, que está expresada en función de \textit{conjuntos} de términos, a una expresión en función de \textit{vectores} de términos. Esta forma \textit{extendida} del coeficiente de Jaccard también se conoce con el nombre de \textit{coeficiente de Tanimoto}~\cite{tanimoto58elementary}:
\vspace{-3mm}
\begin{align*}
 \mathit{sim}_J(A,B)	&= \frac{\abs{A \cap B}}{\abs{A \cup B}}\text{,}\\
			&= \frac{\abs{A \cap B}}{\abs{A}+\abs{B}-\abs{A \cap B}}\text{.}
\intertext{Expresada de esta forma es trivial ver que:\vspace{-3mm}}
\mathit{sim}_J(d_{j}, q)	&= \frac{\overrightarrow{d_j}.\overrightarrow{q}}{\abs{\overrightarrow{d_j}}+\abs{\overrightarrow{q}}-\overrightarrow{d_j}.\overrightarrow{q}}\text{,}\\
				&= \frac{\sum\nolimits_{i=1}^{t}{w(k_i,d_j) . w(k_i,q)}}{\sqrt {\sum\nolimits_{i=1}^t w(k_i,d_j)^{2}} + \sqrt {\sum\nolimits_{i=1}^t w(k_i,q)^{2}} - \sum\nolimits_{i=1}^{t}{w(k_i,d_j) . w(k_i,q)}}\text{.}\\
\end{align*}

\subsection{Similitud Okapi}\label{sub:okapi}
La medida de similitud Okapi~\cite{singhal95document} es una de las métricas más populares en el campo de la IR tradicional y en varias conferencias de área~\cite{hawking99acsys,robertson99okapi}. A diferencia de la Similitud por coseno del Modelo Vectorial, el método Okapi además de considerar la frecuencia de los términos de la consulta, también tiene en cuenta la longitud promedio de los documentos en la colección completa y la longitud del documento evaluado. En este método, la similitud entre una consulta $q$ y un documento $d_j$, puede describirse como el producto escalar del vector de la consulta $\overrightarrow{q}$ y el vector que describe el documento $\overrightarrow{d_j}$:
\begin{align*}
 sim_{o}(d_j, q) = \overrightarrow{q} . \overrightarrow{d_j} = \sum\limits_{i = 1}^{m}{w(k_i,q) . w_{o}(k_i,d_j)}\text{,}
\end{align*}
en donde $w(k_i,q)$ es la frecuencia de $i$-ésimo término de la consulta $q$ y $w_{o}(k_i,d_j)$ es el peso del documento de acuerdo a la siguiente expresión:
\begin{align}\label{eqn:okapi_weight}
w_{o}(k_i,d_j)  = \frac{\mathit{frec}(k_i,d_j) .\log ((N - n_i  + 0.5)/(n_i  + 0.5))}{2.(0.25 + 0.75.l(d_j)/\mathit{proml}) + \mathit{frec}(k_i,d_j) }\text{,}
\end{align}
en donde $\mathit{frec}(k_i,d_j)$ es la frecuencia del $i$-ésimo término en el documento $d_j$, $N$ es el número de documentos en la colección, $n_i$ es el número de documento en la colección que contienen el término $k_i$ de la consulta, $l(d_j)$ es la longitud del documento (en bytes) y $\mathit{proml}$ es la longitud promedio de los documentos en la colección (en bytes).

La principal razón para tener en cuenta la longitud de los documentos en el cálculo del peso de los términos es que los documentos largos tienen más términos que los documentos cortos, por lo que los primeros tienen más probabilidades de ser recuperados que los segundos. La normalización que se incluye en la \autoref{eqn:okapi_weight} es una forma de reducir esta ventaja. Para aplicar esta métrica a la Web es necesario estimar el parámetro $\mathit{proml}$~\cite{li01relevance}.
\newpage
\subsection{Rango de Densidad de Cobertura}\label{sub:cover}
Existen otros métodos, como el \textit{Rango de Densidad de Cobertura} (CDR\footnote{del inglés, Cover Density Range.}), que en lugar de calcular la relevancia basándose en la aparición de los términos, se basan en la ocurrencia de \textit{frases}~\cite{clarke00relevanceCDR}. En CDR, los resultados de las consultas  se ordenan en dos pasos:
\begin{enumerate}
 \item Los documentos que poseen uno o más términos de la consulta $q$ se ordenan por su nivel de \textit{coordinación}, o sea que aquellos que tengan más términos coincidentes estarán mejor posicionados. De esta manera los documentos quedarán agrupados y su ranking queda establecido por el grupo al que pertenecen. Los documentos con un nivel cero son descartados. Esto produce $n$ conjuntos, en donde $n$ es igual a la cantidad de términos de la consulta $q$.
 \item Los documentos dentro de cada grupo se ordenan para producir un orden general. Este se basa en la proximidad y en la densidad de los términos de la consulta dentro de cada documento. Cada documento $d_h$ se convierte en una lista ordenada de términos $k_1, k_2, \dots, k_m$ con sus posiciones $1, \dots, m$, en donde $m$ es la cantidad de términos del documento. Luego se calculan las \textit{coberturas}, que consisten de pares ordenados de la forma $(i, j)$ tales que $i < j$ y $k_i, k_j \in q$ que sean lo más cortos posibles, es decir, que no existe otra cobertura $({i'}, {j'})$ tal que $i < {i'} < {j'} < j$. Finalmente se define el conjunto cobertura $\omega(d_h) = \{({i_1}, {j_1}), ({i_2}, {j_2}), \dots,({i_c}, {j_c})\}$, en donde $c$ representa la cantidad de coberturas encontradas, y se calcula un puntaje para cada documento, $S(\omega)$, de la siguiente manera:
\begin{align}\label{eqn:coverdensity}
S(\omega ) &= \sum\limits_{z = 1}^{c}{I(\omega_z)} = \sum\limits_{z = 1}^{c}{I({i_z}, {j_z})}\text{,}\\
\intertext{en donde $I(i,j)$ está definido de la siguiente manera:}
I(i, j) &= %
  \begin{cases}
    \frac{\lambda}{{j - i + 1}}\text{,} & \text{si }(j - i  + 1) > \lambda \text{,}  \\
    1\text{,} & \text{en otro caso.}
  \end{cases}\label{eqn:cdr_Impact}
\end{align}
En la fórmula de arriba $\lambda$ es una constante, en particular en~\cite{clarke95shortest} se utilizó $\lambda = 16$, obteniendo buenos resultados para consultas booleanas.
La \autoref{eqn:cdr_Impact} le asigna un peso de 1 a las coberturas cuya longitudes sean iguales o menores a $\lambda$ y, a las mayores, les asigna un peso menor a 1 que es proporcional a la inversa del intervalo entre los términos.
\end{enumerate}
La \autoref{eqn:coverdensity} puede interpretarse como una función de similitud, aunque su rango va a depender de $\omega$ y del $\lambda$ elegido. Un ejemplo puede encontrarse en~\cite{clarke00relevanceCDR}. El beneficio de este método es que no sólo tiene en cuenta el número de términos distintos dentro de un documento, sino qué tan cerca están unos de otros. Esto puede resultar acorde a las expectativas de un usuario que, en general, busca que las palabras ingresadas estén cercanas en los documentos recuperados.

\subsection{Método de Puntaje de Tres niveles}\label{sub:three}
El \textit{Método de Puntaje de Tres niveles} (TLS\footnote{del inglés, Three-Level Scoring.}) está diseñado para tener en cuenta las expectativas del usuario acerca de los resultados de la búsqueda y su modelo se basa en algunos métodos manuales existentes. Muchos métodos manuales usan el siguiente criterio para asignar puntajes de relevancia~\cite{chu96search, leighton99first}:
\begin{enumerate}
 \item Los enlaces relevantes, aquellos que están relacionados con las necesidades de información de una consulta, o los que tienen enlaces a otras páginas que pueden ser útiles para la consulta, obtienen 2 puntos.
 \item Los enlaces que son ligeramente relevantes, aquellos que sólo están un poco relacionados con una consulta, aquellos con una definición muy corta que es útil, o que contienen soluciones técnicas a un problema que es relevante para la consulta, o contienen alguna descripción de un trabajo relacionado, obtienen 1 punto.
  \item Los siguientes enlaces obtienen 0 puntos:

    \begin{itemize}
    \item Enlaces duplicados. Este tipo no tiene en cuenta a los sitios espejo\footnote{en inglés estos sitios se denominan \textit{mirrors}.}.
    \item Los enlaces inactivos, que son aquellos que producen algún tipo de mensaje de error. Por ejemplo, error de ``archivo no encontrado'', ``prohibido'' o ``el servidor no responde''.
    \item Enlaces irrelevantes, o sea aquellos que contienen información irrelevante para la consulta.
    \end{itemize}

\end{enumerate}
Este método calcula la relevancia de una página Web con una consulta de la siguiente manera:
\begin{itemize}
 \item Dada una frase de consulta $q$ con $n$ términos y una página Web $d_j$, se calcula un primer puntaje para esa página como:
  \begin{align*}
   A(d_j, q) = \frac{f_n . k^{n-1} + f_{n-1} . k^{n-2} + \dots + f_1}{k^{n-1}}\text{,}
  \end{align*}
  en donde $k$ es una constante y $f_i$, $1\leq i \leq n$ es el número de subfrases de longitud $i$, o sea, que contienen $i$ términos de $q$ en $d_j$. El orden de los términos en cada subfrase tiene que ser el mismo que en la consulta.
 \item Se convierte $A(d_j, q)$ en un puntaje de similitud de tres niveles por medio de un umbral, de forma de asignar 2 puntos a los documentos relevantes, 1 punto a los parcialmente relevantes y 0 puntos a los irrelevantes:
  \begin{align*}
   \mathit{sim}_{TLS}(d_j, q) = %
    \begin{cases}
      2\text{,} & \text{si }A(d_j, q) \geq \Theta\text{,}\\
      1\text{,} & \text{si }\Theta > A(d_j, q) \geq \alpha\Theta\text{,}\\
      0\text{,} & \text{si }A(d_j, q) < \alpha\Theta\text{.}\\
    \end{cases}
  \end{align*}
  en donde $\Theta$ es una constante de umbral para considerar a un documento relevante, y $\alpha$ es un valor entre $0$ y $1$ que representa el porcentaje de $\Theta$ que se pide para que un documento sea parcialmente relevante.
\end{itemize}
Entre las ventajas de este método se pueden mencionar que califica mejor a los documentos que tengan más términos coincidentes, y que también considera el orden de aparición de esos términos en el documento, ya que el orden puede cambiar el significado de una frase~\cite{li02improvement}.

A continuación se muestra un ejemplo ilustrativo simple. En la consulta ``procesamiento paralelo distribuido'', hay 1 frase con 3 términos (procesamiento paralelo distribuido), 3 subfrases con 2 términos (procesamiento paralelo, paralelo distribuido, procesamiento distribuido) y 3 subfrases con 1 término (procesamiento, paralelo, distribuido). Supongamos que existe una página Web tal que, el número de apariciones de la frase exacta de la consulta es $f_3=2$, el número total de apariciones de subfrases con dos términos es $f_2=7$, y que número total de apariciones de subfrases con un término es $f_1=18$. Si ajustamos los tres parámetros del algoritmo a $\Theta=1$, $\alpha=0.1$, y $k=10$, obtenemos $ A=(2{.}10^2+7{.}10+18)/10^2=2.88$, y $sim_{TLS}=2$.

\section{Similitud semántica}\label{sec:similitud_semantica}

El desarrollo de mecanismos de búsqueda web está fuertemente condicionado a resolver los siguientes interrogantes: (1) ¿cómo encontrar páginas relevantes? y, (2) dado un conjunto de páginas web potencialmente relacionadas, ¿cómo ordenarlas de acuerdo a su relevancia? Para evaluar la efectividad de un mecanismo de búsqueda web en estos aspectos se necesitan medidas de \textit{similitud semántica}.
La similitud semántica, al igual que en caso de la similitud, puede utilizarse para comparar objetos en las áreas más diversas. En esta tesis en particular está relacionada con el cálculo de la similitud entre documentos, que a pesar de estar descriptos con vocabularios diferentes y por lo tanto ser léxicamente diferentes, son similares conceptualmente~\cite{hliaoutakis06information}.

En los mecanismos tradicionales las similitudes, o juicios de relevancia, son proporcionadas por los usuarios de forma manual, lo cual como se mencionó anteriormente, es muy difícil de obtener. Más importante aún es el problema de la escalabilidad, ya que en grandes colecciones de datos como la Web es imposible cubrir exhaustivamente todos los tópicos existentes. 
El Open Directory Project (ODP\footnote{http://dmoz.org}) es un gran directorio de la Web editado por personas, y utilizado por cientos de portales y sitios de búsqueda. El ODP clasifica millones de URLs en una ontología temática. Las ontologías ayudan a darle sentido a un conjunto de objetos y, con esta información, pueden derivarse relaciones semánticas entre esos objetos. Por lo tanto, el ODP es una fuente muy útil de donde se pueden obtener medidas de similitud semántica entre páginas web.

Una ontología es un tipo especial de red. El problema de evaluar la similitud semántica en una red tiene una larga historia en la teoría psicológica~\cite{tversky77features}. Más recientemente, la similitud semántica se ha vuelto algo fundamental en la representación de conocimiento, en donde este tipo especial de redes se utilizan para describir objetos y sus relaciones~\cite{gruber93translation}.

\begin{figure}[!ht]%
\centerline{%
 \includegraphics[scale=0.3]{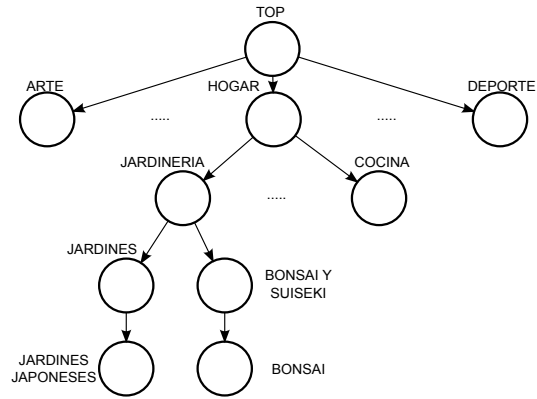}%
}%
\caption{Porción de una taxonomía de tópicos.}%
\label{fig:taxonomy}
\end{figure}

Existen muchas propuestas para medir la similitud semántica por medio del cálculo de distancia entre nodos en una red. La mayoría se basan en la premisa de que mientras más relacionados semánticamente estén dos objetos, más cerca deberán estar en la red. Sin embargo, tal como ha sido discutido por varios autores, surgen problemas al utilizar medidas basadas en distancia en redes en las cuales los enlaces no representan distancias uniformes~\cite{resnik95using}.
Un ejemplo de esto puede verse en la porción del ODP que se muestra en la \autoref{fig:taxonomy}, en donde el tópico \textit{Jardines Japoneses}  se encuentra a la misma distancia
del tópico \textit{Cocina} que del tópico \textit{Bonsai}, pero es claro que la relación semántica con el último es más fuerte que con el primero. La diferencia radica en el ancestro común más próximo a ambos pares, que en el caso de ``Jardines Japoneses'' y ``Bonsai'' es \textit{Jardinería} que es \underline{más específico} que \textit{Hogar}, el ancestro común con el otro tópico.

En las ontologías algunos enlaces conectan categorías muy densas y generales, mientras que otros a categorías muy específicas. Para resolver este problema algunas propuestas estiman la similitud semántica basándose en la noción de contenido de información~\cite{resnik95using,lin98information}. La similitud entre dos objetos se asocia con sus similitudes y con sus diferencias. Dado un conjunto de objetos en una taxonomía ``es-un'', la similitud entre dos objetos puede estimarse encontrando el ancestro común más cercano.

Las ontologías a veces son consideradas como taxonomías ``es-un'', pero no están limitadas a esta forma. Por ejemplo, la ontología ODP es más compleja que un árbol simple. Algunas categorías tienen múltiples criterios para clasificar sus subcategorías. La categoría ``Negocios'', por ejemplo, está subdividida en tipos de organizaciones (cooperativas, negocios pequeños, grandes compañías), así como también en áreas (automotores, cuidado de la salud, telecomunicaciones). Además, el ODP tiene varios tipos de enlaces de referencias cruzadas entre categorías, por lo cual cada nodo puede tener varios nodos padre e incluso pueden existir ciclos.

Las medidas de similitud semántica que se basan en árboles son muy estudiadas~\cite{ganesan03exploting}, sin embargo el diseño de medidas bien fundamentadas para objetos almacenados en nodos de grafos arbitrarios es un problema abierto.

\subsection{Similitud basada en Árboles}
Lin~\cite{lin98information} ha investigado una definición de similitud basada en la teoría de la información que es aplicable en tanto el dominio tenga un modelo probabilístico. Esta propuesta puede utilizarse para derivar una medida de similitud semántica entre tópicos en una taxonomía ``es-un''.

De acuerdo a esta propuesta, la similitud semántica se define como una función del significado que comparten los tópicos y el significado de cada tópico de forma individual. En una taxonomía, el significado que se comparte puede hallarse buscando el ancestro común más cercano a los dos tópicos. Una vez que se tiene identificada esta clasificación común se puede medir el significado como la cantidad de información que se necesita para alcanzar ese estado. Asimismo, el significado de cada tópico se mide como la cantidad de información que se necesita para describirlo en forma completa.

En el campo de la teoría de la información~\cite{cover91elements}, el contenido de información de una clase o tópico $\tau$ se mide como el negativo del logaritmo de una probabilidad $- \log \Pr[\tau]$.

\begin{figure}[!ht]%
\centerline{%
 \includegraphics[scale=0.3]{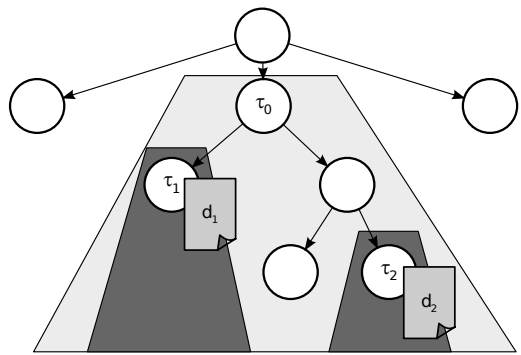}%
}%
\caption{Ejemplo de similitud semántica basada en un árbol.}%
\label{fig:tree_based_sim}
\end{figure}

\begin{definition}[\textnormal{adaptada de~\cite{lin98information}}]\label{def:sim_sem_arboles}
Para medir la similitud semántica entre dos tópicos $\tau_1$ y $\tau_2$ en una taxonomía $T$ se utiliza la relación entre el significado común y los significados individuales de cada tópico, de la siguiente manera:
\begin{align}\label{eqn:sim_sem_arboles}
\mathit{sim}_s^T (\tau_{1}, \tau_2) = \frac{{2.\log \Pr [\tau_0 (\tau_{1}, \tau_2 )]}}{{\log \Pr [\tau_1 ] + \log \Pr [\tau_2 ]}}\text{,}
\end{align}
en donde $\tau_0 (\tau_{1}, \tau_2 )$ es el tópico ancestro común más cercano en el árbol para los tópicos $\tau_1$ y $\tau_2$, y $\Pr [\tau]$ representa la probabilidad a priori de que cualquier página sea clasificada bajo el tópico $\tau$.
\end{definition}
Dado un documento $d_j$ que está clasificado en un tópico de la taxonomía, utilizamos $\mathfrak{T}(d_j)$ para referirnos al nodo del tópico que contiene a $d_j$. Dados dos documentos $d_1$ y $d_2$ en una taxonomía de tópicos, la similitud semántica entre ellos se estima como $\mathit{sim}_s^T (\mathfrak{T}(d_1) ,\mathfrak{T}(d_2) )$. Para simplificar la notación usaremos $\mathit{sim}_s^T (d_1 ,d_2 )$ en lugar de la expresión anterior. De aquí en adelante nos referiremos a $\mathit{sim}_s^T$ como la medida de similitud semántica basada en árboles. Un ejemplo de esta medida se muestra en la \autoref{fig:tree_based_sim}. Allí los documentos $d_1$ y $d_2$ están contenidos en los tópicos $\tau_1$ y $\tau_2$ respectivamente, mientras que el tópico $\tau_0$ es su ancestro común más cercano. En la práctica $\Pr [\tau]$ puede calcularse, para cada tópico $\tau$ en la ontología ODP, contando la fracción de páginas que están almacenadas en el subárbol cuya raíz es el nodo $\tau$ (que llamaremos $subtree(\tau)$) respecto de todas las páginas del árbol.

Esta medida tiene varias propiedades interesantes y una sólida justificación teórica. Es la extensión directa de la medida de similitud de teoría de información~\cite{lin98information}, diseñada para compensar el desbalanceo del árbol en cuanto a los términos de su topología y en cuanto al tamaño relativo de sus nodos. En un árbol balanceado perfectamente $\mathit{sim}_s^T$ se corresponde con la medida clásica de distancia en árboles~\cite{kleinberg99approximation}.

En~\cite{menczer04combining,menczer04correlated,menczer05mapping} se calculó la medida $\mathit{sim}_s^T$ para todos los pares de páginas de una muestra por capas de alrededor de 150.000 páginas del ODP. Para cada uno de los $3.8 \times 10^9 $ pares resultantes también se calcularon las medidas de similitud del texto y de los enlaces, y se hizo una correlación entre éstas y la similitud semántica. Un resultado interesante es que estas correlaciones fueron un poco débiles sobre todos los pares, pero se fortalecieron en páginas dentro de algunas categorías de los niveles altos, como ``Noticias'' y ``Referencia''.

\begin{figure}[!ht]%
\centerline{%
 \includegraphics[scale=0.4]{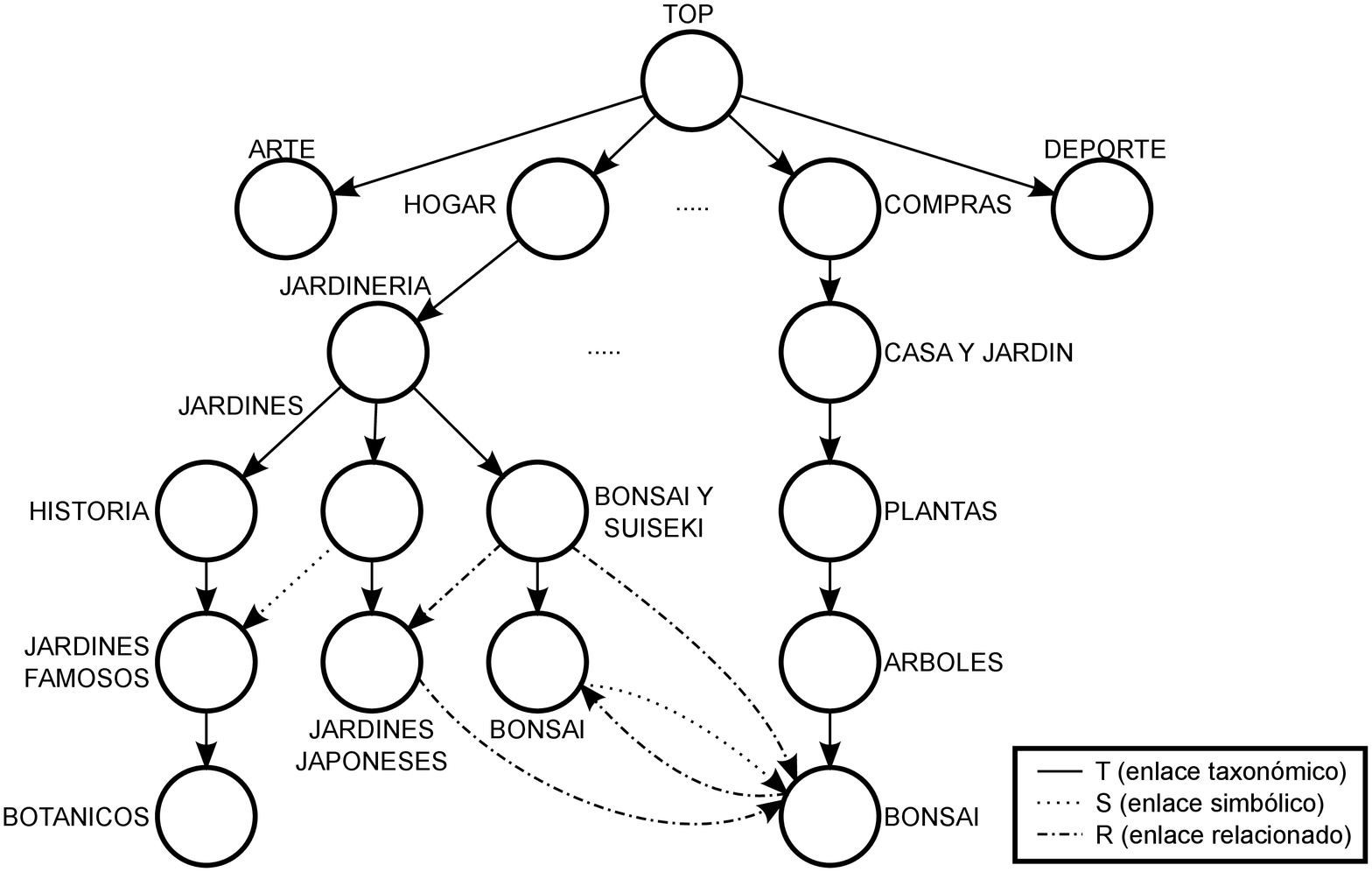}%
}%
\caption{Porción de una ontología de tópicos.}%
\label{fig:taxonomy_odp}
\end{figure}

Dado que $\mathit{sim}_s^T$ sólo está definida en término de los componentes jerárquicos del ODP no es buena a la hora de capturar muchas de las relaciones semánticas inducidas por otros componentes (enlaces simbólicos y relacionados). Es por esto que la similitud semántica basada en árboles entre las páginas que pertenecen a los tópicos de los niveles superiores es cero, incluso si los tópicos están claramente relacionados. Esto muestra un escenario poco confiable al considerar todos los tópicos de la ontología, como sería por ejemplo, la \autoref{fig:taxonomy_odp}.

En la figura puede verse que si sólo se consideran los enlaces de la taxonomía, la similitud semántica entre los tópicos \textit{Hogar/.../Bonsai} y \textit{Compras/.../Bonsai} sería cero, lo cual claramente no refleja la fuerte relación que existe entre ambos. Es por esto que es necesario definir una métrica que contemple todos los tipos de relaciones que encontramos en las ontologías.

\subsection{Similitud basada en Grafos}
Si generalizamos la medida de similitud semántica para que tenga en cuenta grafos arbitrarios, necesitamos definir una medida $\mathit{sim}_s^G$ que saque provecho de los componentes jerárquicos y no jerárquicos de una ontología.

Un grafo de una ontología temática es un grafo de nodos que representan tópicos. Cada nodo contiene objetos que representan documentos (páginas). Está constituido por un componente \textit{jerárquico} (árbol) compuesto por enlaces ``es-un'' y un componente \textit{no jerárquico} compuesto por enlaces cruzados de distintos tipos.

\begin{figure}[!ht]%
\centerline{%
 \includegraphics[scale=0.4]{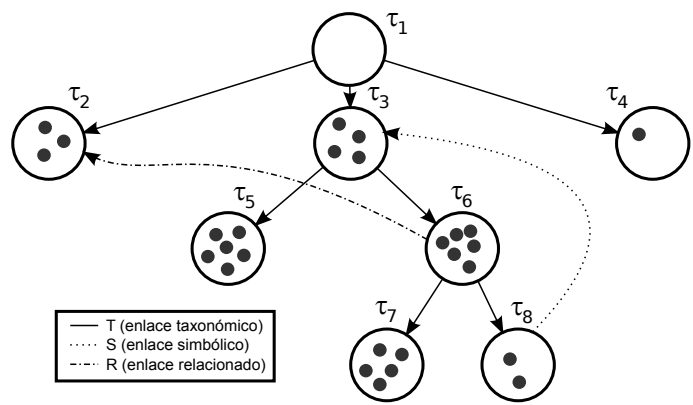}%
}%
\caption{Ejemplo de una ontología.}%
\label{fig:ontologia}
\end{figure}

Por ejemplo, la ontología ODP es un grafo dirigido $G = (V,E)$ en donde:
\begin{itemize}
\item $V$ es un conjunto de nodos que representan los tópicos que contienen documentos;
\item $E$ es un conjunto de aristas entre los nodos de $V$, divididos en tres subconjuntos $T$, $S$ y $R$ tales que:
	\begin{itemize}
	\item $T$ es el componente jerárquico de la ontología,
	\item $S$ es el componente no jerárquico compuesto de enlaces cruzados ``simbólicos'',
	\item $R$ es el componente no jerárquico compuesto de enlaces cruzados ``relacionados''.
	\end{itemize}
\end{itemize}
La \autoref{fig:ontologia} muestra un ejemplo simple de un grafo G definido por los conjuntos\linebreak $V = \{\tau_1, \tau_2, \tau_3, \tau_4, \tau_5, \tau_6, \tau_7, \tau_8\}$, $T = \{ (\tau_1, \tau_2 ), (\tau_1, \tau_3), (\tau_1, \tau_4), (\tau_3, \tau_5), (\tau_3, \tau_6), (\tau_6, \tau_7), (\tau_6, \tau_8) \}$, $S = \{ (\tau_8, \tau_3) \}$, y $R = \{ (\tau_6, \tau_2) \}$. Además, cada nodo $\tau \in V$ contiene un conjunto de objetos. El número de objetos almacenados en un nodo $\tau$ se representa como $\abs{\tau}$ (p. ej., $\abs{ \tau_3 } = 4$).

La extensión de la métrica $\mathit{sim}_s^T$ para considerar grafos genera dos interrogantes. La primera, ¿cómo buscar el ancestro común más específico de un par de tópicos en un grafo?; y la segunda, ¿cómo extender la definición de subárbol con raíz en un tópico dado?


Una diferencia importante entre las taxonomías y las ontologías, como el grafo ODP, es que en las primeras las aristas son todas del mismo tipo (enlaces ``es-un''), mientras que en las segundas pueden existir distintos tipos (p. ej., ``es-un'', ``simbólico'', ``relacionado''). Los distintos tipos de aristas tienen diferentes significados y deberían ser utilizados coherentemente. Una forma de distinguir el rol de las aristas es asignarles pesos y variarlos de acuerdo a su tipo. El peso de una arista $w_{ij}  \in [0, 1]$ entre los tópicos $\tau_i$ y $\tau_j$ puede interpretarse como una medida explícita del grado de pertenencia de $\tau_j$ a la familia del tópico cuya raíz es $\tau_i$. Los enlaces ``es-un'' y los ``simbólicos'', en ODP, tienen el mismo peso ya que, como no se permiten los enlaces repetidos, estos últimos son la única manera de representar membresías múltiples. Por otro lado, los enlaces ``relacionados'' suponen una relación más débil.

Se asume que $w_{ij} > 0$ si y sólo si existe algún tipo de arista entre los tópicos $\tau_i$ y $\tau_j$. Sin embargo, para estimar la pertenencia a un tópico deben considerarse relaciones transitivas entre las aristas. 
\begin{definition}[\textnormal{\cite{maguitman05algorithmic}}]
Se define $\cone{i}$ a la familia de tópicos $\tau_j$ tales que $i = j$ o existe un camino $(e_1, \dots, e_n)$ que satisface:
\begin{enumerate}
\item $e_1 = (\tau_i, \tau_k)$ para algún $\tau_k \in V$;
\item $e_n = (\tau_k, \tau_j)$ para algún $\tau_k \in V$;
\item $e_k \in T\cup S\cup R$ para $k = 1\dots n$; y
\item $e_k \in S\cup R$ a lo sumo para un $k$.
\end{enumerate}
\end{definition}

Las condiciones anteriores expresan que $\tau_j \in \cone{i}$ si hay un camino directo desde $\tau_i$ a $\tau_j$ en el grafo $G$, en donde a lo sumo una arista de $S$ o de $R$ está contenida en ese camino. Esta última condición es necesaria por tres motivos. 
Primero, si se permiten más aristas, el cómputo no podría llevarse a cabo, ya que en la práctica cualquier tópico podría pertenecer a cualquier otro. 
Segundo, se perdería robustez, porque algunos enlaces cruzados producirían grandes cambios en todas las membresías. 
Finalmente y más importante, haría que la clasificación de un tópico pierda sentido al mezclar todos los tópicos unos con otros. Los distintos tópicos $\tau_j$ ahora están asociados a lo que denominamos el \textit{cono} $\cone{i}$ de un tópico $\tau_i$ con diferentes grados de pertenencia.

La estructura del grafo se puede representar como matrices de adyacencia. La matriz $\mathbf{T}$ representa la estructura jerárquica de la ontología y es equivalente a las aristas de $T$ a las que se le agregan 1s en la diagonal. Los componentes no jerárquicos de la ontología se representan con la matriz $\mathbf{S}$ y con la matriz $\mathbf{R}$, como se muestra a continuación.
\begin{align*}
\begin{matrix}
\mathbf{T}_{ij}  = \left\{ {\begin{array}{*{20}l}
   1 		& \mbox{si }i = j\mbox{,} \\
   \alpha	& \mbox{si }i \ne j\mbox{ e }(i, j)\in {T}\mbox{,}\\
   0		& \mbox{si no.} \\
\end{array}} \right. &
\mathbf{S}_{ij}  = \left\{ {\begin{array}{*{20}l}
   \beta	& \mbox{si }(i, j)\in {S}\mbox{,}\\
   0		& \mbox{si no.} \\
\end{array}} \right. &
\mathbf{R}_{ij}  = \left\{ {\begin{array}{*{20}l}
   \gamma	& \mbox{si }(i, j)\in {R}\mbox{,}\\
   0		& \mbox{si no.} \\
\end{array}} \right.
\end{matrix}
\end{align*}
\begin{definition}[\textnormal{\cite{maguitman05algorithmic}}]
Se define la operación sobre matrices $\vee$ como:
\begin{align*}
 [\mathbf{A}\vee\mathbf{B}]_{ij} = \mathop {\max }(\mathbf{A}_{ij}, \mathbf{B}_{ij})\text{.}
\end{align*}
\end{definition}
Se calcula la matriz $\mathbf{G} = \mathbf{T}\vee \mathbf{S}\vee \mathbf{R}$, que es la matriz de adyacencia del grafo $G$ a la que se le agregaron 1s en la diagonal.
\begin{definition}[\textnormal{\cite{maguitman05algorithmic}}]
Sea la función de composición difusa MaxProduct definida en~\cite{kandel86fuzzy} como:
\begin{align*}
[\mathbf{A}\varodot\mathbf{B}]_{ij}=\mathop {\max }\limits_{k}(\mathbf{A}_{ik}, \mathbf{B}_{kj}).
\end{align*}
Entonces se define:
\begin{align*}
\mathbf{T}^{(1)} &= \mathbf{T}\text{,} \\
\mathbf{T}^{(r+1)} &=\mathbf{T}^{(1)}\varodot \mathbf{T}^{(r)}, \\
\intertext{y la clausura de $\mathbf{T}$, $\mathbf{T}^{+}$, como:}
\mathbf{T}^{+} &= \lim_{r\to\infty}\mathbf{T}^{(r)}. 
\end{align*}
En esta matriz, $\mathbf{T}^{+}_{ij} = 1$ si $\tau_{j}\in subtree(\tau_i)$ y si no es cero. 
\end{definition}
Finalmente se calcula la matriz $\mathbf{W}$ de la siguiente manera:
\begin{align*}
\mathbf{W} = \mathbf{T}^{+}\varodot\mathbf{G}\varodot\mathbf{T}^{+}. 
\end{align*}
Cada elemento $\mathbf{W}_{ij}$ puede interpretarse como un valor de pertenencia difusa del tópico $\tau_{j}$ al cono $\cone{i}$ ~y luego se puede ver a $\mathbf{W}$ como la \textit{matriz de pertenencia difusa} de $G$.

A modo de ejemplo consideremos a la ontología que se mostró en la \autoref{fig:ontologia}. En este caso las matrices $\mathbf{T}$, $\mathbf{G}$, $\mathbf{T}^{+}$ y $\mathbf{W}$ están definidas de la siguiente manera:
\begin{spacing}{0.95}
\begin{align*}
\allowbreak
\mathbf{T} &=  
\bordermatrix{ 
      & \tau_1 & \tau_2 & \tau_3 & \tau_4 & \tau_5 & \tau_6 & \tau_7 & \tau_8 \cr
  \tau_1 & 1 & 1 & 1 & 1 & 0 & 0 & 0 & 0 \cr
  \tau_2 & 0 & 1 & 0 & 0 & 0 & 0 & 0 & 0 \cr
  \tau_3 & 0 & 0 & 1 & 0 & 1 & 1 & 0 & 0 \cr
  \tau_4 & 0 & 0 & 0 & 1 & 0 & 0 & 0 & 0 \cr
  \tau_5 & 0 & 0 & 0 & 0 & 1 & 0 & 0 & 0 \cr
  \tau_6 & 0 & 0 & 0 & 0 & 0 & 1 & 1 & 1 \cr
  \tau_7 & 0 & 0 & 0 & 0 & 0 & 0 & 1 & 0 \cr
  \tau_8 & 0 & 0 & 0 & 0 & 0 & 0 & 0 & 1 \cr
}\\
\mathbf{G} & =  
\bordermatrix{ 
      & \tau_1 & \tau_2 & \tau_3 & \tau_4 & \tau_5 & \tau_6 & \tau_7 & \tau_8 \cr
  \tau_1 & 1 & 1 & 1 & 1 & 0 & 0 & 0 & 0 \cr
  \tau_2 & 0 & 1 & 0 & 0 & 0 & 0 & 0 & 0 \cr
  \tau_3 & 0 & 0 & 1 & 0 & 1 & 1 & 0 & 0 \cr
  \tau_4 & 0 & 0 & 0 & 1 & 0 & 0 & 0 & 0 \cr
  \tau_5 & 0 & 0 & 0 & 0 & 1 & 0 & 0 & 0 \cr
  \tau_6 & 0 & .5 & 0 & 0 & 0 & 1 & 1 & 1 \cr
  \tau_7 & 0 & 0 & 0 & 0 & 0 & 0 & 1 & 0 \cr
  \tau_8 & 0 & 0 & 1 & 0 & 0 & 0 & 0 & 1 \cr
}\\
\mathbf{T}^+  &=  
\bordermatrix{ 
              & \tau_1 & \tau_2 & \tau_3 & \tau_4 & \tau_5 & \tau_6 & \tau_7 & \tau_8 \cr
 subtree(\tau_1) & 1 & 1 & 1 & 1 & 1 & 1 & 1 & 1 \cr
 subtree(\tau_2) & 0 & 1 & 0 & 0 & 0 & 0 & 0 & 0 \cr
 subtree(\tau_3) & 0 & 0 & 1 & 0 & 1 & 1 & 1 & 1 \cr
 subtree(\tau_4) & 0 & 0 & 0 & 1 & 0 & 0 & 0 & 0 \cr
 subtree(\tau_5) & 0 & 0 & 0 & 0 & 1 & 0 & 0 & 0 \cr
 subtree(\tau_6) & 0 & 0 & 0 & 0 & 0 & 1 & 1 & 1 \cr
 subtree(\tau_7) & 0 & 0 & 0 & 0 & 0 & 0 & 1 & 0 \cr
 subtree(\tau_8) & 0 & 0 & 0 & 0 & 0 & 0 & 0 & 1 \cr
}\\
\mathbf{W}  &=  
\bordermatrix{ 
           & \tau_1 & \tau_2 & \tau_3 & \tau_4 & \tau_5 & \tau_6 & \tau_7 & \tau_8 \cr
  \cone{1} & 1 & 1 & 1 & 1 & 1 & 1 & 1 & 1 \cr
  \cone{2} & 0 & 1 & 0 & 0 & 0 & 0 & 0 & 0 \cr
  \cone{3} & 0 & .5 & 1 & 0 & 1 & 1 & 1 & 1 \cr
  \cone{4} & 0 & 0 & 0 & 1 & 0 & 0 & 0 & 0 \cr
  \cone{5} & 0 & 0 & 0 & 0 & 1 & 0 & 0 & 0 \cr
  \cone{6} & 0 & .5 & 1 & 0 & 1 & 1 & 1 & 1 \cr
  \cone{7} & 0 & 0 & 0 & 0 & 0 & 0 & 1 & 0 \cr
  \cone{8} & 0 & 0 & 1 & 0 & 1 & 1 & 1 & 1 \cr
} 
\end{align*}
\end{spacing}
Entonces la similitud semántica entre dos tópicos $\tau_1$ y $\tau_2$ en un grafo de una ontología puede calcularse de la siguiente manera.
\begin{definition}[\textnormal{\cite{maguitman05algorithmic}}]\label{def:semantic_similarity}
 Sean dos tópicos $\tau_1$ y $\tau_2$ en un grafo dirigido $G=(V,E)$, con $V$ el conjunto de nodos que representan a los tópicos que contienen documentos y $E$ el conjunto de aristas. ${\abs{ U }}$ representa el número de documentos en el grafo. 
Sea $\Pr [\tau_k ]$ la probabilidad a priori de que un documento cualquiera sea clasificado en un tópico $\tau_k$ y se calcula como:
\begin{align*}
\Pr [\tau_k ] = \frac{{\sum\nolimits_{\tau_j  \in V} {(\mathbf{W}_{kj}  \cdot \abs{ {\tau_j } })} }}{{\abs{ U }}}.
\end{align*}
Sea $\Pr [\tau_i |\tau_k ]$ la probabilidad posterior de que cualquier documento será clasificado en un tópico $\tau_i$ luego de ser clasificado en $\tau_k$, y se calcula:
\begin{align*}
\Pr [\tau_i |\tau_k ] = \frac{{\sum\nolimits_{\tau_j \in V} {(\min (\mathbf{W}_{ij} ,\mathbf{W}_{kj} ) \cdot \abs{ \tau_j})} }}{{\sum\nolimits_{\tau_j  \in V} {(\mathbf{W}_{kj}  \cdot \abs{ {\tau_j } })} }}.
\end{align*}
Quedando finalmente,	
\begin{align}
\mathit{sim}_S^G (\tau_1 ,\tau_2 ) = \mathop {\max }\limits_{\tau_k \in V} \frac{{2 \cdot \min (\mathbf{W}_{k1} ,\mathbf{W}_{k2} ) \cdot \log \Pr [\tau_k ]}}{{\log (\Pr [\tau_1 |\tau_k ] \cdot \Pr [\tau_k ]) + \log (\Pr [\tau_2 |\tau_k ] \cdot \Pr [\tau_k ])}}.
\label{eqn:sim_sem_grafos}
\end{align}
\end{definition}
La definición $\mathit{sim}_S^G$ es una generalización de $\mathit{sim}_S^T$, ya que si se da el caso de que $G$ sea un árbol ($S = R = \emptyset$), entonces $\cone{i}$ es igual a $subtree(\tau_i)$, el subárbol del tópico cuya raíz es $\tau_i$, y todos los tópicos $\tau\in subtree(\tau_i)$ pertenecen a $\cone{i}$ con un grado igual a 1. Si $\tau_k$ es un ancestro de $\tau_1$ y $\tau_2$ en una taxonomía, entonces $\min (\mathbf{W}_{k1} ,\mathbf{W}_{k2} ) = 1$ y $\Pr [\tau_i |\tau_k ] \cdot \Pr [\tau_k ] = \Pr [\tau_i ]$ para $i = 1, 2$.

\section{Resumen}
En este capítulo se definieron los fundamentos de los sistemas de IR que serán utilizados en los capítulos siguientes de esta tesis. Entre ellos, los modelos clásicos de representación en IR, como son el modelo Booleano, el Vectorial, el Probabilístico y el Indexado por Semántica Latente. Luego se explicó el proceso de generación de una consulta, los mecanismos más usados para reformulación de consultas y el concepto de realimentación de relevancia. Finalmente se mostraron dos conceptos fundamentales para el desarrollo de esta tesis, la similitud y la similitud semántica. Se analizaron las métricas más comunes en el área y se definió la noción de similitud semántica basada en árboles, mostrándose sus limitaciones. En vista de esto se definió la similitud semántica basada en grafos, una métrica que tiene en cuenta los distintos tipos de enlaces presentes en una ontología, y que es capaz de medir más fielmente las relaciones existentes entre tópicos relacionados conceptualmente.

\chapter{M\'{e}todos de evaluaci\'{o}n}\label{chp3:evaluacion}
La recuperación de información se ha vuelto una disciplina altamente empírica, por lo que para demostrar que un sistema con una técnica novedosa es superior a otro, necesita pasar por un riguroso proceso de evaluación sobre colecciones de documentos representativas. 
En este capítulo discutiremos las medidas de efectividad de los sistemas de IR y las colecciones que más se utilizan para este propósito.
\section{Colecciones de prueba}\label{sec:test_col}
Para medir de forma estándar la efectividad de un sistema de IR necesitamos una colección de prueba en común y de acceso público. Las colecciones de prueba son una herramienta muy útil a la hora de medir y comparar algoritmos en IR. Consisten de corpus creados y compartidos por la comunidad de IR para promover una plataforma de prueba sobre la que se pueda medir la efectividad de los sistemas de recuperación. Una colección ideal está compuesta por:
\begin{itemize}
  \item un conjunto de documentos;
  \item un conjunto de consultas; y,
  \item una lista de documentos que son \textit{relevantes} para esas consultas.  
\end{itemize}

Hacia fines de los años~'40, uno de los principales problemas era la accesibilidad de los trabajos científicos que se habían desarrollado motivados por investigaciones desarrolladas durante la Segunda Guerra Mundial.
En aquel momento había dos tipos convencionales de índices y dos técnicas de indexado. Un índice podía existir en la forma de un catálogo de tarjetas, como se encontraban en la mayoría de las bibliotecas, o alternativamente en forma impresa, por ejemplo, en compendios anuales de resúmenes de una revista. En cuanto a las técnicas de indexado, en Europa existía una tendencia al uso de sistemas de clasificación, mientras que en los Estados Unidos la práctica más usual era la utilización de los catálogos alfabéticos de temas.

El modelo de clasificación bibliotecario era el dominante, y bajo él, un documento era clasificado/indexado por un ser humano. El resultado era una descripción corta del documento en un lenguaje más o menos formal.
Cuando comenzaron a aumentar el número de publicaciones, en la forma de reportes técnicos, los índices y las técnicas de indexado fueron puestas a prueba y
en los~'50s hubo varios intentos por desprenderse de los sistemas convencionales.

Los puntos de discusión en esa época eran: la forma particular del lenguaje de indexación, el tipo de análisis que llevaba a la construcción del mismo y su posterior aplicación en un documento y la granularidad, especificidad y divisibilidad de la representación. Otros tipos de discusiones eran más bien filosóficas, ya que los bibliotecarios trataban de imponer su visión del conocimiento humano y su forma de representarlo. El \textit{lenguaje natural}, hasta ese momento, era considerado como algo separado del sistema porque la clasificación formal trataba de alguna forma de evitar sus falencias. 

En este contexto, las experimentaciones empíricas eran una idea radical. Había una resistencia a ver a los sistemas de IR desde un punto de vista funcional, además de las dificultades que había para formular e implementar tales funcionalidades.
\subsection{Cranfield}\label{sub:cranf1}
Algunas investigaciones continuaron su curso estimuladas por
\cite{mcninch48conference} y algunos pequeños experimentos alrededor del mundo. El interés sería retomado en la Conferencia Internacional sobre Información Científica de 1958~\cite{icsi58proceedings}, pero para ese tiempo Cyril Cleverdon, bibliotecario del Colegio Cranfield de Aeronáutica, tenía sus propias ideas y comenzó el primero de dos proyectos Cranfield que fue publicado en 1962~\cite{cleverdon62report}. Este proyecto derribó las divisiones filosóficas que dominaban el área al someter a una competencia a cuatro esquemas de indexación. Cada uno de éstos eran ejemplos contrapuestos sobre cómo debía organizarse la información. El trabajo comparó la eficiencia de cada uno respecto del otro indexando 16000 documentos sobre Ingeniería Aeronáutica. Esos cuatro sistemas son descriptos a continuación.
\subsubsection[Clasificación Decimal Universal]{Clasificación Decimal Universal (UDC\footnote{del ingles, Universal Decimal Classification})}
Este sistema de clasificación bibliotecario fue desarrollado a finales del siglo XIX. Utiliza números arábigos y, como su nombre lo indica, se basa en el sistema decimal. Es un sistema jerárquico en donde el conocimiento se divide en diez clases, cada una de éstas se subdivide en partes y cada subdivisión se vuelve a subdividir hasta alcanzar el nivel de detalle deseado. Cada número se supone es una fracción decimal cuyo punto inicial no se escribe, pero para facilitar la lectura se agrega un punto cada tres dígitos. Por ejemplo, después de 00 `\textit{Fundamentos de la ciencia y de la cultura}' viene las subdivisiones 001 a 009; debajo de 004 `\textit{Ciencia y tecnología de los ordenadores}' están sus subdivisiones 004.1 a 004.9; debajo de 004.4 están todas sus subdivisiones antes que 004.5, por lo que después de 004.49 viene 004.5; y así siguiendo. Una ventaja de este sistema es que puede ser extendido indefinidamente, y cuando se introducen nuevas subdivisiones no hay que modificar las existentes. 
\subsubsection{Catálogo alfabético de temas}\label{subsub:cat_alp_tem}
Es un sistema de clasificación en donde se describen los documentos por medio de términos que indican sobre qué trata el contenido del mismo. El proceso de indexado comienza con algún tipo de análisis sobre el tema del documento. Un indexador (una persona) debe identificar los términos que mejor expresan el contenido mediante la extracción de palabras que aparecen en el documento o utilizando palabras de algún vocabulario predeterminado. La calidad de los índices dependerá en gran medida de la experticia que tenga cada bibliotecario en el área.
\subsubsection{Esquema de clasificación facetado}
Esta técnica analiza los temas y los divide en características. Permite que un objeto tenga múltiples clasificaciones, y habilita a un usuario a navegar a través de la información por distintos caminos, que se corresponden con los distintos ordenamientos de las características. Estas ideas se contraponen con los sistemas jerárquicos como el UDC, en donde sólo hay un único orden predeterminado. Este sistema se originó a partir de la Clasificación por dos puntos\footnote{del inglés, Colon Classification.}, publicada por primera vez en~1933~\cite{ranganathan33colon}. Se le da este nombre porque utiliza signos de puntuación (como ``:'', ``;'') para separar las facetas. El sistema representa los objetos a partir de 42 clases principales, divididas en 5 categorías fundamentales: \textit{personalidad}, \textit{material}, \textit{energía}, \textit{espacio} y \textit{tiempo}. Por ejemplo, supongamos que tenemos un libro titulado ``\textit{La administración de la Educación Elemental en el Reino Unido en los~'50s}'' ~\cite{garfield84tribute}. Para este tópico, ``educación'' es la clase principal y se representa con una letra `T'. La característica distintiva, o personalidad, es ``Elemental'', representada por el número `15'. Al no existir un material se introduce directamente un signo ``:''. La acción que ocurre respecto de la clase principal, o energía, es ``administración'', representada por un número `8'. El espacio, o ubicación geográfica, viene precedida de un signo ``.'', en este caso ``Reino Unido'', representado con un número `56'. Luego se continúa con una comilla simple que indica el comienzo del campo tiempo, en este caso 1950, que se representa con `N5'. Finalmente tenemos, T15:8.56'N5.
\subsubsection{Sistema de unitérminos de indexado coordinado}
Este sistema está basado en la implementación de \textit{unitérminos}\footnote{del inglés, uniterms.} y la aplicación de lógica booleana (\autoref{sub:mod_bool}). Los unitérminos son términos individuales que son seleccionados por los indexadores para representar las distintas características de los documentos. Mortimer Taube, un bibliotecario que trabajaba para el gobierno de los Estados Unidos, analizó cerca 40000 encabezados temáticos en un gran catálogo de tarjetas y encontró que tales encabezados eran combinaciones de, ``sólo'', 7000 palabras distintas. Entonces propuso utilizar estas palabras como términos de indexado que pueden ser \textit{coordinados}, o interconectados, en la etapa de búsqueda~\cite{cleverdon91significance}. Los unitérminos pueden considerarse, en cierto sentido, como las palabras clave de hoy en día ya que ambos están derivados del documento original y no se hace ningún control del vocabulario. Típicamente se utilizan varios unitérminos para representar un único documento, como con las palabras clave.\\

Para la realización de cada índice se emplearon distintas personas expertas en cada sistema, los cuales fueron implementados con tarjetas. La evaluación se llevó a cabo con 1200 consultas pero, para evitar los juicios de relevancia, las consultas fueron creadas a partir de algún documento dentro de la colección. Se consideró que evaluar la relevancia de cada documento para cada consulta sería una tarea excesivamente larga. Algunos trabajos anteriores habían intentado obtener juicios de relevancia conciliando las opiniones de distintos jueces, pero lo encontraron también muy difícil. Los sistemas fueron comparados entonces de acuerdo a si recuperaban o no el documento ``fuente''. Sin embargo, este método fue severamente criticado por tres razones principales:
\begin{itemize}
  \item Las consultas podrían considerarse irreales;
  \item La recuperación o no del documento ``fuente'' no es un buen test;
  \item Sólo se está evaluando la cobertura y no la precisión (\autoref{sub:prec_and_rec}).
\end{itemize}
Los resultados mostraron que los cuatro sistemas alcanzaron una eficiencia en el rango del 74\% al 82\%, con el sistema de \textit{unitérminos} en primer lugar y la clasificación facetada en el último. Se hizo un análisis detallado sobre las causas de los errores en la localización del documento deseado y la mayoría se debieron a errores humanos en alguna de las etapas del proceso de indexación o recuperación. 
\subsection{Cranfield 2}\label{sub:cranf2}
Las discuciones que originó Cranfield 1 llevaron a la creación de un segundo proyecto. La transición a la versión 2 fue un gran salto en cuanto al contenido y a la metodología~\cite{cleverdon66factors}. En cuanto al contenido, la atención continuó en los ``lenguajes de indexación'': lenguajes artificiales creados para permitir representar de manera formal a los documentos y a las consultas. Se estudió meticulosamente el proceso de construcción de un lenguaje evaluándose 8 que utilizaban términos simples (unitérminos), 15 basados en conceptos y 6 basados en términos fijos (diccionarios). 
Sin embargo, el logro más importante de Cranfield 2 fue el de definir la noción de \textit{metodología de experimentación} en IR. Las ideas básicas sobre recolectar los documentos y las consultas se habían heredado de la primera versión del proyecto, pero el cambio más grande ocurrió con la noción de \textit{respuesta correcta}. 
Los documentos elegidos para responder a cada una de las consultas se eligieron con una variedad de métodos, incluyendo la búsqueda manual y una forma de indexado basado en citas. La intención fue la completitud, es decir, descubrir todos (o casi todos) los documentos relevantes dentro de la colección.
La experiencia les había demostrado que no eran esenciales demasiados documentos en la colección, pero si lo era tener todos los juicios de relevancia de todos los documentos para todas las consultas. En esta oportunidad se les enviaron cartas a aproximadamente 200 autores de artículos recientes, para que describieran en forma de pregunta el problema sobre el que trataban sus trabajos, y que agregaran preguntas adicionales que surgieron en el curso de la investigación. Entonces también se les indicó que le dieran una puntuación de 1 a 5 a cada una de las referencias que citaron en sus trabajos. El resultado fueron 1400 documentos y 279 consultas. Cada consulta fue analizada para cada documento por un grupo de estudiantes y los documentos que se consideraron relevantes fueron reenviados al autor de la pregunta para el juicio final.

El mejor rendimiento fue alcanzado por los lenguajes que utilizaban \textit{términos simples}. Aunque hoy en día casi todos los motores de búsqueda se basan en búsquedas por términos del lenguaje natural, esta conclusión fue impactante en aquel entonces. Hay que destacar que, incluso en este proyecto, los experimentos fueron realizados sin la ayuda de computadoras.

Por último, en los comienzos del proyecto, se tenía la idea de encontrar que alguno de los índices utilizados mejoraría la cobertura o la precisión, llegando a la conclusión de que ninguno era particularmente mejor que otro. También se observó que, en general, tiende a existir una relación inversa entre la cobertura y la precisión.

\subsection{SMART}\label{sub:smart}
El sistema SMART~\cite{salton71smart} es un sistema de los años~'60s con el que se realizaron los experimentos más importantes de la época en IR utilizando computadoras. Liderado por Gerard Salton, se desarrolló mayoritariamente en la Universidad de Cornell hasta su finalización en los años~'90s. Muchas de las ideas que se desarrollaron allí se utilizan ampliamente hoy en día en los motores de búsqueda. En particular,  pueden destacarse: la utilización de métodos completamente automáticos basados en el contenido de los documentos; la noción de función de ordenamiento (para medir el grado de relevancia de un documento) y su posterior ubicación en una lista que se muestra al usuario. Es importante mencionar que esto último, que estuvo incluido en SMART desde su comienzo, no apareció en ningún sistema comercial hasta finales de los~'80s y terminó de triunfar recién a mediados de los~'90s con los motores de búsqueda web.

El sistema utilizó un modelo de representación vectorial (\autoref{sub:mod_vectorial}) y el sistema de pesos TF-IDF ideado por Spärck Jones~\cite{jones72statistical}. La idea de realimentación de relevancia\linebreak (\autoref{sub:feedback}) también fue introducida en el sistema por Rocchio~\cite{rocchio71relevance}.

Los primeros experimentos se realizaron sobre pequeñas colecciones de prueba (decenas o cientos de documentos) creadas por ellos mismos o reutilizadas de otros experimentos. En particular utilizaron la colección Cranfield cuando estuvo disponible para su acceso electrónico. En principios de los~'80s comenzaron a utilizar, y a su vez a construir, algunas colecciones más grandes. Hay que tener en cuenta que en el año 1973 una sola corrida sobre la colección Cranfield 2 tomaba 10 minutos a un costo monetario altísimo.

\subsection{MEDLARS}\label{sub:medlars}
El Servicio de Búsqueda por Demanda (MEDLARS\footnote{del inglés, Medical Literature Analysis and Retrieval System.}) fue uno de los primeros sistemas de recuperación bibliográfica de literatura médica computarizado. Desde 1879 la Biblioteca Nacional de Medicina (NLM\footnote{del inglés, National Library of Medicine}) editaba un índice bibliográfico anual, pero no fue hasta los~'40s que divisaron nuevas tecnologías que acelerarían el proceso. En enero de 1959 comenzaron a editar un índice mensual llamado \textit{Index Medicus} con la ayuda de un proceso mecanizado de tarjetas. En 1960 comenzaron a analizar la posibilidad de utilizar una computadora, pero dados los costos que implicaba, el proyecto fue largamente estudiado. Con su uso se deberían reducir los tiempos de producción de las ediciones mensuales del índice y la búsqueda y recuperación de bibliografía. El sistema de indexado utilizado fue el Catálogo Alfabético de Temas, descripto en la \autoref{sub:cranf1}, que para el área de medicina se llama MeSH\footnote{del inglés, Medical Subject Headings.}. Por otro lado, para el almacenamiento de la información se utilizaron cintas magnéticas de acceso secuencial. El sistema estuvo funcional para 1964.

El aporte más grande de este proyecto estuvo en las \textit{búsquedas por demanda}. Un investigador que estuviera interesado en buscar determinada bibliografía debía contactar al experto en búsquedas que trabajaba en el NLM personalmente o por carta. Las consultas eran creadas exclusivamente sobre los términos de indexado del lenguaje MeSH y utilizaban lógica booleana (\autoref{sub:mod_bool}). El experto creaba las consultas en base a su interacción con el usuario, lo cual requería un gran entrenamiento. La salida era un conjunto de referencias impresas sin ningún orden en particular.
Como puede notarse, no había una interacción directa entre los usuarios y la computadora, por otro lado, el proceso era caro y el tiempo de obtención de los resultados oscilaba entre 3 a 6 semanas. Es por eso que en los~'70s se implementó un nuevo sistema de formulación de consultas on-line llamado MEDLINE (MEDLARS onLINE) que se mantiene hasta nuestros días.

\subsection{TREC}\label{sub:trec}
Desde comienzos de los años~'60 y por un período de 20 años, Karen Spärck Jones condujo una serie de experimentos en clustering y ponderación de términos. A diferencia de Cleverdon, e incluso de otros investigadores de su época, no construyó su propia colección de prueba. Sus primeros experimentos se basaron únicamente en la versión electrónica de Cranfield 2~\cite{jones71automatic} y fue una de las primeras en darse cuenta de los beneficios y de las dificultades de reutilizar colecciones de prueba construidas por otros investigadores. Durante ese período inventó una forma de ponderar términos basada en el número de documentos en los que aparecen (IDF,~\cite{jones72statistical}). Luego el equipo de SMART combinó esta métrica con la frecuencia del término dentro del documento (TF) para crear una de las medidas que más influyó en los algoritmos de ponderación de términos y de ordenamiento de documentos de la siguiente generación (\autoref{eqn:tf-idf}). Desde mediados de los~'70s hasta entrados los~'80s lideró los esfuerzos para depurar los estándares básicos y para mejorar los métodos y los datos de experimentación en IR~\cite{jones75report}. También editó un libro con varios autores del área~\cite{jones81information}, que fue la única fuente coherente de consulta a la hora de planificar y ejecutar un experimento en IR hasta, quizás, la edición del libro sobre las experiencias de TREC en 2005~\cite{voorhees05trec}. Otro resultado de su trabajo fue el diseño de una colección de prueba estándar, ya que hasta la época sólo existían colecciones que habían sido creadas para llevar a cabo un experimento específico, pero que eran reutilizadas en otros experimentos. El estudio incluyó un minucioso análisis sobre los aspectos más importantes de la preparación de los datos, como por ejemplo, la técnica de selección para obtener los juicios de relevancia sobre una gran colección de datos~\cite{jones75report}. Contando con pocos recursos para sus proyectos el desarrollo se detuvo, hasta que a comienzos de los~'90 se puso en marcha el proyecto TREC.

La Conferencia de Recuperación de Información (TREC\footnote{del inglés, Text REtrieval Conference.}) es un conjunto de workshops anuales enfocados en las distintas áreas de IR. El proceso en su conjunto está dirigido por el Instituto Nacional de Estándares y Tecnología de los Estados Unidos (NIST\footnote{del inglés, US National Institute of Standards and Technology.}), pero cada workshop en particular (track) es en general organizado por los grupos de investigación participantes. En la~\autoref{tbl:trec_evol} se muestran los distintos tracks y su evolución a lo largo de los años~\cite{voorhees05trec,voorhees04trec2004,voorhees05trec2005,voorhees06trec2006,voorhees07trec2007,voorhees08trec2008,voorhees09trec2009}. Como se dijo más arriba no existía un esfuerzo conjunto de parte de los grupos de investigación por utilizar los mismos datos de prueba o, por comparar los resultados entre sistemas de IR.
La intención de estas comparaciones no era mostrar que un sistema era mejor que otro, sino poder utilizar una variedad mucho mayor de técnicas, sobre todo al momento de proponer nuevas. Por otro lado, también hacía falta una colección de un tamaño mucho más realista, ya que como se dijo, la colección Cranfield sólo alcanzaba a los miles de documentos. Había que ser capaz de demostrar que las técnicas eran aplicables a grandes colecciones de documentos. Es por esto que en 1990 se le encomienda al NIST construir una colección de cerca de un millón de documentos. 

La primera conferencia se llevó a cabo en 1992 sobre la base de una colección de aproximadamente 2 GB de artículos de diarios y documentos gubernamentales. El track central de las primeras conferencias se llamó track \textit{ad~hoc}, el cual medía la capacidad de un sistema para recuperar listas precisas y completas de documentos \textit{ordenados} (éste fue el principal cambio respecto de Cranfield). Estas listas de resultados se generan en respuesta a determinadas \textit{consultas} que fueron creadas conjuntamente con la colección y suele asumirse que son temáticas (documentos \textit{acerca de} X). En TREC se las llama `tópicos'. Un `tópico' tiene un título corto (el cual puede usarse como consulta) e información adicional acerca del conocimiento que supuestamente se necesita. Esta descripción contiene reglas explícitas para juzgar la relevancia de los documentos. Un ejemplo de título y descripción son: ``Boeing 747'' y ``Aerolíneas que utilizan actualmente aviones Boeing 747''.

El proceso de creación de un nuevo track incluye la producción de la colección de documentos, los tópicos y el conjunto de juicios de relevancia (sobre esto se hablará en particular más adelante). Luego la colección se distribuye entre los participantes, cada uno la indexa y corre experimentos sobre sus sistemas de búsqueda. Algunos resultados de estos experimentos son enviados al NIST en donde se evalúa su relevancia. Finalmente se realiza la conferencia, los organizadores escriben un reporte sobre los resultados de cada track y los equipos participantes escriben reportes técnicos sobre sus sistemas. También pueden llegar a escribirse ediciones especiales de alguna revista con los resultados de una conferencia~\cite{harman95trec2, voorhees00trec6} o de un track en particular~\cite{hersh01trecinteractive, robertson02trec}.

\begin{table}[ht!]
\begin{center}
{\sffamily
\scriptsize
\begin{tabular}{r|xxxxxxxxxxxxxxxxxx|l}
 Enfoque & \side{'92} & \side{'93} & \side{'94} & \side{'95} & \side{'96} & \side{'97} & \side{'98} & \side{'99} & \side{'00} & \side{'01} & \side{'02} & \side{'03} & \side{'04} & \side{'05} & \side{'06} & \side{'07} & \side{'08} & \side{'09} & Track\\
\hline
\multirow{3}*{\minitab[r]{Texto\\ estático}} & \multicolgray{8} &  &  &  &  &  &  &  &  &  &  & ad~hoc \\ 
 &  &  &  &  &  &  &  &  &  &  &  & \multicolgray{3} &  &  &  &  & robust \\ 
 &  &  &  &  &  &  &  &  &  &  &  &  &  &  &  & \multicolgray{3} & million query \\
\hline
\multirow{2}*{\minitab[r]{streamed\\ text}} & \multicolgray{6} &  &  &  &  &  &  &  &  &  &  &  &  & routing \\ 
 &  &  &  & \multicolgray{8} &  &  &  &  &  &  &  &  filtering \\
\hline
\multirow{4}*{\minitab[r]{con usuarios}} &  &  & \multicolgray{9} &  &  &  &  &  &  &  & interactive \\ 
&  &  &  &  &  & \multicolgray{2} &  &  &  &  &  &  &  &  &  &  &  & high precision \\ 
&  &  &  &  &  &  &  &  &  &  &  & \multicolgray{3} &  &  &  & & HARD \\ 
&  &  &  &  &  &  &  &  &  &  &  &  &  &  &  &  & \multicolgray{2}  & relevance feedback \\
\hline
\multirow{3}*{\minitab[r]{No sólo\\ inglés}} &  &  & \multicolgray{3} &  &  &  &  &  &  &  &  &  &  &  &  &  & spanish \\ 
&  &  &  &  & \multicolgray{2} &  &  &  &  &  &  &  &  &  &  &  &  & chinese \\ 
&  &  &  &  &  & \multicolgray{6} &  &  &  &  &  &  &  & cross-language \\
\hline
\multirow{3}*{\minitab[r]{Más allá del\\ texto}} &  &  &  & \multicolgray{2} &  &  &  &  &  &  &  &  &  &  &  &  &  & confusion \\ 
&  &  &  &  &  & \multicolgray{4} &  &  &  &  &  &  &  &  &  &  speech \\ 
&  &  &  &  &  &  &  &  &  & \multicolgray{2} &  &  &  &  &  &  &  &  video \\
\hline
\multirow{4}*{\minitab[r]{Web,\\ tamaño}} &  &  &  &  &  &  & \multicolgray{2} &  &  &  &  &  &  &  &  &  &  & very large corpus \\ 
&  &  &  &  &  &  &  & \multicolgray{6} &  &  &  &  & \multicolgray{1} & web \\ 
&  &  &  &  &  &  &  &  &  &  &  &  & \multicolgray{3} &  &  &  &  terabyte \\ 
&  &  &  &  &  &  &  &  &  &  &  &  &  & \multicolgray{4} &  &  enterprise \\
\hline
\multirow{3}*{\minitab[r]{Respuestas}} &  &  &  &  &  &  &  & \multicolgray{9} &  &  &  question answering \\ 
&  &  &  &  &  &  &  &  &  &  & \multicolgray{3} &  &  &  &  &  &  novelty \\ 
&  &  &  &  &  &  &  &  &  &  &  &  &  &  &  &  &  & \multicolgray{1} & entity \\
\hline
\multirow{3}*{\minitab[r]{Dominio}} &  &  &  &  &  &  &  &  &  &  &  & \multicolgray{5} &  &  &  genome \\ 
&  &  &  &  &  &  &  &  &  &  &  &  &  &  & \multicolgray{4} & legal \\ 
&  &  &  &  &  &  &  &  &  &  &  &  &  &  &  &  &  & \multicolgray{1} & chemical \\
\hline
\multirow{2}*{\minitab[r]{Documentos\\ personales}} &  &  &  &  &  &  &  &  &  &  &  &  &  & \multicolgray{3} &  &  &  spam \\
&  &  &  &  &  &  &  &  &  &  &  &  &  &  & \multicolgray{4} & blog \\
\hline
\end{tabular}}

\caption{Evolución de los tracks en la Conferencia TREC.}
\label{tbl:trec_evol}
\end{center}
\end{table}

\section{Juicios de relevancia manuales}\label{sec:juicios_man}
Crear colecciones de prueba presenta muchos desafíos metodológicos, pero el mayor incumbe al conjunto de documentos a los que se le evaluará la \textit{relevancia}. Una colección ideal está compuesta de un conjunto de documentos, un conjunto de consultas y una lista que indica qué documentos son relevantes para esas consultas. Los documentos y las consultas suelen ser relativamente fáciles de obtener, sin embargo, los juicios de relevancia son algo costoso. Las colecciones de los~'60s, '70s y principios de los~'80s eran pequeñas: no más de 3 MB de texto, por lo que los juicios se realizaban de manera exhaustiva, examinando cada documento para cada consulta. Sin embargo, si su tamaño crece hasta tamaños más realistas se necesita un esfuerzo considerable, evitando que cada juez pueda analizar todos los documentos de manera completa.

Ya en los~'70s Spärck Jones y van Rijsbergen notaron que examinar cada documento de manera exhaustiva no era escalable y propusieron un método \textit{no exhaustivo}. En el reporte de 1975~\cite{jones75report} y en los siguientes~\cite{jones77report, gilbert79bases} describieron la construcción de una colección de prueba `ideal'. El uso del \textit{pooling} fue propuesto como una forma eficiente de encontrar los documentos relevantes dentro de una gran colección de documentos. Se asumió que dentro de este pool, o fondo común, se encontrarían casi todos los documentos relevantes. Para crear los juicios de relevancia se tomaría una muestra aleatoria del pool de documentos.

Como se describe en~\cite{salton83extended}, este método se utilizó para recopilar los juicios de relevancia de la colección Inspec de 5.5 MB. Cada consulta se procesó con siete métodos distintos y luego todos los documentos recuperados fueron mezclados, eliminando los repetidos. El pool resultante fue examinado por los asesores de relevancia.

TREC utiliza este sistema y es uno de los más utilizados. Aquí sólo se evalúan los $k$ mejores documentos enviados por cada uno de los $n$ grupos participantes. Si $k$ y $n$ son grandes se considera que el conjunto de documentos relevantes juzgados es representativo del conjunto ideal, y luego útil para ser usado en la evaluación de los resultados de un proceso de recuperación. Por ejemplo, en los experimentos del track ad~hoc de TREC-6~\cite{harman97trec6} se utilizaron valores de $k=100$ y $n=30$, lo que implicó, para las 50 consultas temáticas, unos 60000 juicios. El número total de juicios requeridos si se utilizaba un examen exhaustivo hubiera sido de 500000.

Este método tiene críticas y hay estudios acerca de la completitud de este subconjunto de documentos y sobre la consistencia en los juicios de relevancia entre los distintos jueces~\cite{voorhees98variations,zobel98reliable}. Se mostró que es necesario tener: suficientes \textit{tópicos} para lograr una buena estabilidad al promediar los juicios de todos los jueces, suficientes \textit{documentos} en los subconjuntos de documentos para asegurar que se encuentren los más relevantes, y suficientes \textit{participantes} para tener variedad en el conjunto final.
 
\section{Juicios de relevancia automáticos}\label{sec:juicios_aut}
\subsection{Motivación}
La forma no exhaustiva de evaluación que se presentó en la sección anterior también tiene problemas a la hora de escalar o cambiar la colección de documentos, por lo que se han desarrollado métodos que intentan reducir los problemas de escalabilidad. Una forma es utilizar colecciones o consultas que contengan \textit{información semántica} que ayude a la evaluación de la relevancia. Ciertas colecciones de noticias incluyen etiquetas de categorías codificadas manualmente. Con ellas pueden ser entrenados y probados sistemas de clasificación de manera automática, para luego ser utilizados en la evaluación de algoritmos de ranking y métodos de representación de texto~\cite{lewis92evaluation}. Otra posibilidad es crear consultas para las cuales sólo existan \textit{pocos documentos relevantes} en la colección. En~\cite{sheridan97cross} se construyó una colección de noticias en la cual las consultas se referían a eventos con una fecha de inicio, lo que hizo que los evaluadores pudieran limitarse sólo a los documentos a partir de esa fecha. Las consultas pueden limitarse más aún hasta que sólo exista un documento relevante; esta clase de búsqueda se llama de \textit{ítem conocido} y fueron usadas en las primeras ediciones de TREC~\cite{garofolo97spoken}. A pesar de estos intentos la mayor parte de las colecciones disponibles siguen el método de pooling. Es por esto que~\cite{zobel98reliable} se interesó en maximizar el número de documentos relevantes encontrados y notó, como podría ser lógico pensar, que la cantidad de documentos relevantes para cada consulta no es constante y que esto afecta la eficiencia del proceso de análisis de relevancia. Teniendo en cuenta la contribución de cada participante al pool de consultas, en~\cite{cormack98efficient} notaron que ciertos sistemas contribuían con más documentos relevantes que otros, por lo que propusieron un sistema de pooling que juzga primero a los documentos de los mejores sistemas. Este método fue utilizado hasta alcanzar el 50\% de los documentos examinados en TREC-6 y fue correlacionado con éste alcanzando una distancia de Kendall\footnote{la distancia $\tau$ es una función que cuantifica el número mínimo de intercambios entre los elementos de un  ranking para convertirlo en el otro.}~\cite{kendall38newmeasure} de $\tau  = 0.999$. Reduciendo los documentos al 10\% de los examinados en TREC la correlación se redujo muy levemente, alcanzando un valor de $\tau = 0.990$. Estas ideas también fueron implementadas luego en~\cite{carterette05incremental}. Puede verse que ambos sistemas son extremadamente similares y, por lo tanto, que la forma de hacer pooling en TREC es altamente ineficiente. Esta conclusión se basa en~\cite{voorhees98variations} y en~\cite{voorhees01evaluation} en donde se muestra que una correlación de al menos $0.9$ implica que ambos sistemas son equivalentes.

Por otro lado también se propuso una nueva forma de hacer pooling que incluye una combinación de búsqueda, evaluación y reformulación de consultas que dieron en llamar ISJ\footnote{del inglés, Interactive Searching and Judging.}~\cite{cormack98efficient}. Para cada consulta los jueces debían realizar la búsqueda con consultas que refinarían de acuerdo a sus criterios personales hasta que no encontraran más resultados relevantes. La correlación encontrada al realizar la comparación fue de $0.89$, muy cercana al límite antes mencionado. Los autores encontraron que esta correlación ligeramente inferior pudo deberse a las diferencias de opiniones de los jueces de ISJ y los de TREC en cuanto a qué documentos eran relevantes y cuáles no. En el track Filtering de TREC-2002~\cite{soboroff03building}, con el objetivo de realizar un proceso de evaluación más exhaustivo, se implementó un sistema de pooling que utilizaba refinamiento de consultas pero de forma automática. Se crearon conjuntos iniciales de documentos que se les dieron a evaluar a los jueces. A partir de estos juicios se realimentó esta información en los sistemas de recuperación y se obtuvieron nuevos documentos para evaluar. Este proceso se realizó durante cinco iteraciones o hasta que no se encontraran más documentos relevantes. 

Analizando los problemas de crear continuamente nuevas colecciones de documentos, se propuso mantener los conjuntos de TREC existentes para limitar el impacto de estos cambios, ya que el mantenimiento es más sencillo que la búsqueda de nuevos tópicos y la realización de nuevos juicios de relevancia~\cite{soboroff06dynamic}. Finalmente en~\cite{sanderson04forming} se analizaron métodos para producir colecciones de prueba de forma automática, sin utilizar pooling, encontrando altas correlaciones con las colecciones de TREC. Estos sistemas de evaluación automática son muy utilizados en dominios en los que la evaluación manual requeriría un esfuerzo prohibitivo~\cite{goldstein05intrinsic}.

Las técnicas de evaluación automática en IR podrían clasificarse en dos grupos: \textit{Inferencia automática} y \textit{Fuentes externas} de relevancia.

\subsection{Inferencia automática}
Las técnicas dentro de este grupo infieren los juicios de relevancia a partir de los documentos recuperados. Algunos métodos muestrean aleatoriamente los conjuntos de documentos recuperados basándose en estadísticas conocidas acerca de la distribución típica de los documentos relevantes, tomándolos como \textit{pseudorelevantes}. Este tipo de técnica es muy utilizada en conjunto con la \textit{Realimentación de relevancia sin supervisión}, vista en el \autoref{chp2:state}, en donde la forma más básica de seleccionar términos potencialmente útiles es elegirlos de los $k$ mejores documentos recuperados por una consulta. Algunos métodos también utilizan los $k'$ peores documentos recuperados como una fuente de información para los términos irrelevantes. Las distintas variantes existentes han surgido porque distintos autores han propuesto diversas técnicas para la asignación de los pesos de los términos.

En~\cite{soboroff01ranking} se construyó un modelo de distribución de los documentos en el pool de TREC para seleccionar documentos aleatoriamente y crear un conjunto de documentos pseudorelevantes. Esto se hizo para tratar de probar la hipótesis de que ``los rankings de los sistemas de IR son robustos, aún si los documentos relevantes se eligen al \textit{azar} del conjunto de documentos recuperados'', aludiendo al desacuerdo existente en los juicios de los jueces de la mencionada competencia. Los resultados obtenidos mostraron cierta correlación con los rankings de algunos equipos participantes, pero poca con los mejores sistemas.

Otros sistemas utilizan funciones de similitud entre los documentos recuperados y la consulta~\cite{shang02precision}. Estas funciones tienen el propósito de encontrar términos que coocurran con los términos de la consulta en el mismo documento, en el mismo párrafo, en la misma sentencia o en una determinada ventana de palabras. La efectividad del sistema de IR se verá fuertemente afectada por la elección de la función de similitud~\cite{kim99comparison}. Algunas medidas de similitud fueron mencionadas en la \autoref{sec:similitud}.

\subsection{Fuentes externas}
En este grupo se encuentran los métodos que utilizan fuentes de información externas para ayudar en la tarea de inferencia. Algunos utilizan la información de los clicks del usuario (tuplas que consisten de una consulta y un resultado elegido por el usuario) para evaluaciones automáticas. Sin embargo, en estos datos existe un sesgo inherente: los usuarios tienden a hacer click en los documentos mejor rankeados, sin importar su calidad~\cite{boyan96machine}. De cualquier manera, en~\cite{joachims05accurately} se encontró que estos datos pueden utilizarse exitosamente para inferir preferencias relativas entre documentos. Para realizar evaluaciones automáticas, otras investigaciones han hecho uso de taxonomías tales como el ODP (referido también como DMOZ), el directorio de Yahoo! y Looksmart~\cite{haveliwala02evaluating, srinivasan04general}. Estas taxonomías dividen a una porción de la Web en una jerarquía de categorías, con páginas en ellas. Cada categoría tiene un título y un camino que representa su lugar en la jerarquía. Típicamente también existen títulos en las páginas, ingresados por editores de la taxonomía, que no se corresponden necesariamente con los títulos originales de las páginas.

La aparición de taxonomías online mantenidas por personas, como ODP, permiten una forma de evaluación automática. Se basa en tomar una gran muestra de las consultas de la Web actual y para cada una de ellas recolectar conjuntos de documentos pseudorelevantes en la taxonomía. En~\cite{beitzel03using} se analizan métodos para la evaluación de motores de búsqueda web de forma automática utilizando ODP. Se asume que los editores humanos aseguran un contenido de alta calidad y relevante dentro del directorio. Se tiene un conjunto de consultas que se mapean a documentos de la colección con algún criterio, en particular se proponen dos técnicas, la primera exige una coincidencia exacta de la consulta con el título de la página y la segunda una coincidencia exacta con el título de la categoría a la que pertenece la página, permitiendo esto último la existencia de más documentos relevantes. Luego esas consultas se ejecutan en los motores de búsqueda y se registra la ubicación del documento en los resultados. Como métrica de comparación se utilizó la media de esa ubicación para todas las consultas. 
Mostraron que el directorio utilizado no introduce un sesgo significativo en la evaluación y que las distintas consultas no afectan a la estabilidad de la métrica. Aunque la comparación con juicios de relevancia manuales mostró una correlación moderada, este tipo de métodos automáticos tiene la ventaja de poder evaluar muchas más consultas que los métodos manuales, abarcando una gran diversidad de tópicos.

En~\cite{haveliwala02evaluating} utilizaron ODP como una variante de estudio de usuarios precompilado para desarrollar un método automático de evaluación de estrategias de representación de documentos.
Se tuvieron en cuenta tres características de una página web $u$ a la hora de representarla en un espacio vectorial:
\begin{enumerate}
  \item su contenido, es decir, las palabras que contiene $u$;
  \item los enlaces que lo apuntan, es decir, los identificadores de los documentos $v$ que apuntan a $u$; y
  \item el texto en dichos enlaces, las palabras en los enlaces de $v$, o cercanas a estos, cuando apuntan a $u$.
\end{enumerate}
Se probaron distintas combinaciones y configuraciones de parámetros, mostrando la flexibilidad del método y el ahorro que se tiene en términos de costos y tiempo.

\section{Métricas}\label{sec:metricas}
La eficiencia de los sistemas de recuperación de información se mide comparando su rendimiento en un conjunto común de consultas y documentos. Para evaluar la confianza de tales comparaciones suelen utilizarse los test de significancia (pruebas \textit{t} o de Wilcoxon), que muestran que las diferencias observadas no son por azar.

Desde Cranfield 
 se han utilizado seis métricas principales:
\begin{enumerate}
  \item \label{it:mets1} La \textit{cobertura} de la colección, esto es, qué proporción de documentos relevantes incluye la colección;
  \item la \textit{demora de tiempo}, esto es, el intervalo promedio entre el momento en el que el pedido es hecho y el momento en el que se da la respuesta;
  \item la \textit{forma de presentación} de la salida;
  \item \label{it:mets4} el \textit{esfuerzo} que hace el usuario para obtener los resultados de su búsqueda;
  \item la \textit{cobertura} del sistema, esto es, la proporción de los documentos relevantes que aparecen en los resultados de la búsqueda; y
  \item la \textit{precisión} del sistema, esto es, la proporción de los documentos recuperados que son relevantes.
\end{enumerate}
Se podría decir que los puntos~\ref{it:mets1} a~\ref{it:mets4} son fácilmente cuantificables. La \textsc{Cobertura} y la \textsc{Precisión} son las métricas que se conocen como la \textit{efectividad} de un sistema de recuperación. En otras palabras es una medida de la habilidad del sistema para recuperar documentos relevantes y a la vez evitar los irrelevantes. Se asume que cuanto más eficiente es un sistema más satisfará al usuario, y también se asume que la precisión y la cobertura son métricas suficientes para cuantificar esa efectividad.

\begin{figure}[!ht]
\centerline{
\includegraphics[scale=0.5]{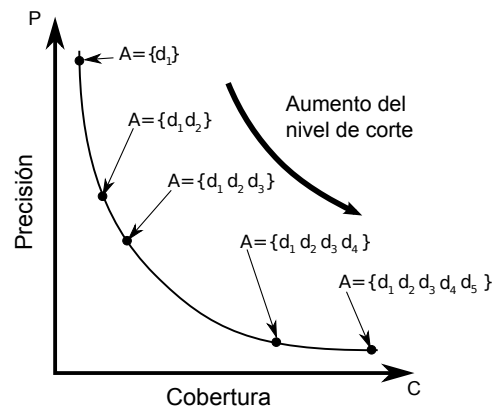}
}%
\caption{Curva de Precisión-Cobertura.}%
\label{fig:p-r-graph}
\end{figure}

Ha habido mucho debate sobre si de hecho la precisión y la cobertura son cantidades apropiadas de usar como métricas de la efectividad. Una alternativa han sido la cobertura y el \textit{fall-out} (la proporción de los documentos irrelevantes recuperados). Sin embargo, todas las alternativas siguen necesitando el cálculo de alguna forma de relevancia. La relación entre las distintas medidas y su dependencia de la relevancia serán explicadas luego. Las ventajas de basar la evaluación en la precisión y la cobertura son:
\begin{enumerate}
 \item es el par que más comúnmente se utiliza; y
 \item son métricas muy estudiadas y comprendidas.
\end{enumerate}
Muchas veces la técnica utilizada para medir la efectividad de un sistema de recuperación se ve influenciada por la estrategia de recuperación que se adopta y por la forma que tiene su salida. Por ejemplo, cuando la salida es un conjunto ordenado de documentos un parámetro obvio es la posición de cada uno de los documentos recuperados. Utilizando esta posición como un límite de \textit{corte} pueden calcularse los valores de precisión y cobertura para cada valor de corte. Luego, los resultados pueden resumirse en una curva suave que una dichos puntos en un gráfico de precisión-cobertura (P-C, \autoref{fig:p-r-graph}). El camino a lo largo de la curva sería interpretable como la variación de la efectividad en función del valor de corte. Desafortunadamente siguen existiendo preguntas que esta métrica no es capaz de responder, como ser, ¿cuántas consultas son mejores que la media y cuántas peores? A pesar de esto, explicaremos este tipo de medidas de la efectividad dado que son las más comunes.

Antes de continuar es necesario examinar el concepto de relevancia, ya que está detrás de las medidas de efectividad.
\subsection{Relevancia}\label{sub:relevancia}
La relevancia es un concepto \textit{subjetivo}, ya que distintos usuarios pueden diferir acerca de cuáles documentos son relevantes y cuáles no para ciertas consultas. Sin embargo estas diferencias no son lo suficientemente grandes como para invalidar experimentos que se han hecho con colecciones de documentos estándar. Las consultas se crean generalmente a partir de la necesidad de información de usuarios con conocimientos en el área. Los juicios se realizan por un grupo de expertos en esa disciplina. Entonces ya tenemos cierto número de consultas para las cuales se conocen las respuestas `correctas'. Una de las hipótesis generales dentro del campo de la IR es que si una estrategia de recuperación se comporta bastante bien para un gran número de experimentos entonces es probable que se desempeñe bien en una situación real, en donde la relevancia no se conoce a priori.

La relevancia es uno de los conceptos más importantes, sino el ``fundamental'', en la teoría de la IR~\cite{crestani01logic}. El concepto surge de considerar que si un usuario de un sistema de IR tiene una necesidad de información, entonces alguna información almacenada en algunos documentos en una colección pueden ser ``relevantes'' a esta necesidad. En otras palabras, la información que se considerará relevante para la necesidad de información del usuario es aquella que puede ayudarlo a satisfacerla. Cualquier otra información que no se considere relevante se considera ``irrelevante'' a esa misma necesidad de información. Esto es una consecuencia de aceptar que el concepto de relevancia tiene dos posibles valores.

Existe también un concepto de relevancia que puede decirse ser ``objetivo'' y merece mención.
Una definición lógica de relevancia fue introducida por Cooper en 1971~\cite{cooper71definition}. La definición de \textit{relevancia lógica} para IR se hizo por analogía con el mismo problema en los sistemas de respuesta a consultas. Esta analogía no llegó más allá de considerar consultas del tipo si-no (verdadero-falso). La relevancia fue definida en términos de \textit{consecuencia lógica} y para hacer esto las consultas y los documentos fueron definidos por conjuntos de sentencias declarativas. Una consulta si-no se representa con dos declaraciones formales, llamadas \textit{declaraciones de componentes}, de la forma $p$ y $\neg p$. Por ejemplo, si la consulta es `¿El nitrógeno es un gas noble?' las declaraciones en lenguaje formal serían `El nitrógeno es un gas noble' y `El nitrógeno no es un gas noble'. Las consultas más complicadas, como las del tipo \textit{¿cuál?} o \textit{¿qué pasa si?} pueden transformarse a esta forma~\cite{belnap63analysis, belnap76logic}.
%
Aunque inicialmente la relevancia lógica se definió sólo para sentencias, puede ser fácilmente extendida para su aplicación en documentos. Un documento es relevante para una necesidad de información si y sólo si contiene al menos una sentencia que sea relevante para esa necesidad. El autor también intentó abordar el problema de los \textit{grados de relevancia}, o como el dijera: ``los tonos de grises en lugar del blanco y negro''.

\vspace{-5mm}
\subsection{Precisión y Cobertura}\label{sub:prec_and_rec}
\vspace{-3mm}
Ahora dejamos las especulaciones acerca de la relevancia y volvemos a la medición de la efectividad. Asumimos una vez más que la relevancia tiene su significado usual de `pertenencia a un tópico' y `aptitud', esto es, el usuario es el que determinará si un documento es relevante o no. La efectividad es estrictamente una medida de la habilidad del sistema para satisfacer al usuario en términos de la relevancia de los documentos recuperados. Inicialmente, veremos cómo medir la efectividad en términos de la precisión y la cobertura.

\begin{table}[ht!]
\setlength{\abovecaptionskip}{0pt}   
\setlength{\belowcaptionskip}{0pt}   
\renewcommand{\arraystretch}{1.3} 
\begin{center}
 \begin{tabular}{r|c|c|l}
  \multicolumn{1}{l}{}	& \multicolumn{1}{c}{relevante} & \multicolumn{1}{c}{irrelevante} \\
  \cline{2-3}
  recuperado	& $R \cap A$ & $\bar{R} \cap A$ & $A$ \\
  \cline{2-3}
  no recuperado	& $R \cap \bar{A}$ & $\bar{R} \cap \bar{A}$ & $\bar{A}$ \\
  \cline{2-3}
  \multicolumn{1}{l}{} & \multicolumn{1}{c}{$R$} & \multicolumn{1}{c}{$\bar{R}$} & $N$\\
  \end{tabular}
\caption{Clasificación de los objetos en un sistema de recuperación.}
\label{tbl:contingency}
\end{center}
\end{table}

En la \autoref{tbl:contingency} se muestran los distintos conjuntos en los que se puede dividir la salida de un sistema de IR. A partir de ella pueden derivarse un gran número de medidas de efectividad a partir de esta tabla. Algunas son:
\vspace{-2mm}
\begin{align} 
   {\rm Precisi\acute{o}n} = P & =  \frac{\abs{R \cap A} }{\abs{A}}\text{,}\label{eqn:p}\\
   {\rm Cobertura} = C & =  \frac{\abs{R \cap A}}{\abs{R}}\text{,}\label{eqn:r}\\
   {\rm Cutoff}    & =  \frac{\abs{A}}{N}\text{,}\label{eqn:cu}\\
   {\rm Fallout}   & =  \frac{\abs{\bar R \cap A}}{\abs{\bar R}}\text{.}\label{eqn:fo}
\end{align} 

Se pueden construir tablas a partir de cada pedido $q$ que se envía al sistema de recuperación y a partir de estas se calculan los valores de precisión-cobertura. Si la salida del sistema depende de algún parámetro (digamos $k$), como puede ser la posición de los documentos o el \textit{nivel de coordinación} (el número de términos que una consulta tiene en común con un documento), pueden construirse tablas para cada valor de ese parámetro y obtener un valor distinto de P-C que puede notarse por el par ordenado ($P(k)$, $C(k)$). A partir de estos pares ordenados se construye un gráfico de P-C, en el cual los puntos generados para un $k$ dado se unen para formar una curva de P-C. Todas las curvas se promedian para medir el rendimiento conjunto del sistema.

\begin{figure}[!ht]
\centerline{%
 \includegraphics[scale=0.5]{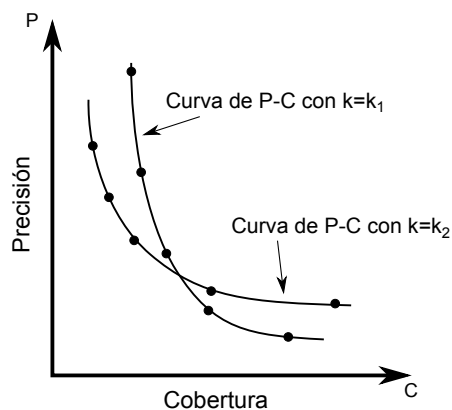}%
}%
\caption{Curva de rendimiento de un sistema para múltiples valores del parámetro $k$.}%
\label{fig:average-performance}%
\end{figure}

\subsection{Técnicas basadas en promedios}
Además de la técnica explicada más arriba, existen dos alternativas de promediado de curvas. Cuando se hace una recuperación por \textit{nivel de coordinación} se adopta la \textit{microevaluación}~\cite{salton66evaluation,rocchio66thesis,rocchio71evaluation}. Sea $Q$ el conjunto de pedidos de búsqueda, entonces el número de documentos relevantes recuperados para $Q$ es:
\begin{align*}
{{\tilde R}(Q)} = \sum\limits_{q \in Q} {|{R_q}|} \text{,}
\end{align*}
en donde ${R_q}$ es el conjunto de documentos relevantes recuperados por el pedido $q$. Si el nivel de coordinación es $k$, entonces el número de documentos recuperados por $Q$ a un nivel $k$ es:
\begin{align*}
{\tilde A}(k, Q) = \sum\limits_{q \in Q} {|{A(k, q)}|}  \text{,}
\end{align*}
en donde $A(k, q)$ es el conjunto de documentos recuperados por el sistema para $q$ a un nivel de coordinación $k$ o superior. Luego, los puntos $P(k)$, $C(k)$ se calculan de la siguiente manera:
\begin{align*}
 P_{\mu}(k, Q) &= \sum\limits_{q \in Q} {\frac{|{R_q \cap A(k, q)}|}{{\tilde A}(k, Q)}}\text{,}\\
C_{\mu}(k, Q) &= \sum\limits_{q \in Q} {\frac{|{R_q \cap A(k, q)}|}{{\tilde R}(Q)}}\text{.}
\end{align*}

\begin{figure}[!ht]
\centerline{
\includegraphics[scale=0.5]{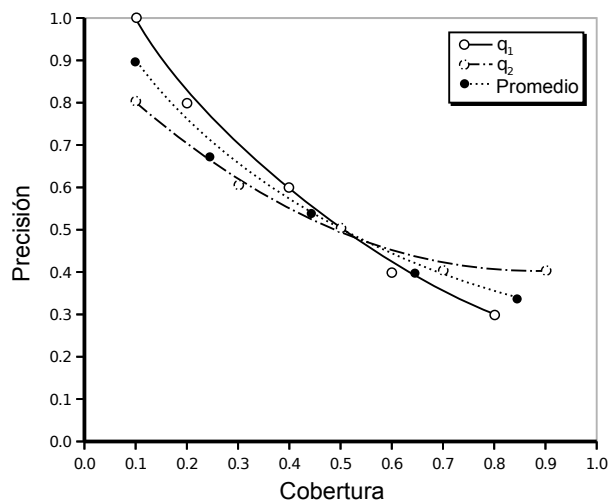}
} \caption{Curva interpolada promedio calculada utilizando la técnica de microevaluación.}
\label{fig:microevaluation}
\end{figure}

La \autoref{fig:microevaluation} muestra gráficamente que sucede cuando se combinan dos curvas individuales de P-C de esta manera. Los datos a partir de los que se generaron estas curvas pueden verse en la \autoref{tab:p-r-combinadas}.

\begin{figure}[!ht]
\centerline{
\includegraphics[scale=0.5]{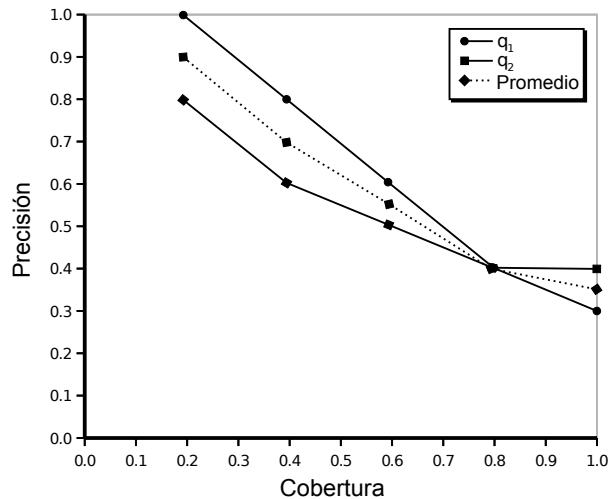}
} \caption{Curva promedio calculada utilizando la técnica de macroevaluación.}
\label{fig:macroevaluation}
\end{figure}

Una alternativa de promediado también es la \textit{macroevaluación}~\cite{salton66evaluation,rocchio66thesis,rocchio71evaluation}, la cual es independiente de cualquier parámetro del sistema. En este caso, se miden para cada consulta, los valores de precisión (cobertura) para determinados valores fijos o \textit{estándar} de cobertura (precisión), como por ejemplo, los valores de precisión alcanzados cuando se logra una cobertura de $\{0.2, 0.4, 0.6, 0.8, 1\}$. En la \autoref{fig:macroevaluation} se muestra un ejemplo. El valor promedio de precisión (cobertura) se calcula simplemente como el promedio de los valores de precisión (cobertura) de cada consulta para cada valor de cobertura (precisión).

Como puede verse, el promediado de valores es necesario, pero también trae problemas, porque puede existir una gran varianza en los puntos, lo que obliga a un análisis de significatividad estadística.

\begin{table}[ht!]
\begin{center}
{\sffamily
\scriptsize

\begin{tabular}{c|c|c|c}
 & $q_1$  & $q_2$ & ${{\tilde R}(Q)}$ \\
\hline
$R_q$ & 100 & 80 &  180\\
\end{tabular}

\begin{tabular}{c|cc|cc}
 & \multicolumn{2}{c|}{$q_1$} & \multicolumn{2}{c}{$q_2$}\\
$k$ & $C(k)$ & $P(k)$ & $C(k)$ & $P(k)$\\
\hline
1&0.1&1.0&0.1&0.8\\
2&0.2&0.8&0.3&0.6\\
3&0.4&0.6&0.5&0.5\\
4&0.6&0.4&0.7&0.4\\
5&0.8&0.3&0.9&0.4\\
\end{tabular}

\begin{tabular}{c|cc|cc|c}
$k$ & $\abs{A(k, q_1)}$ & $\abs{{R_{q_1} \cap A(k, q_1)}}$ & $\abs{A(k, q_2)}$ & $\abs{{R_{q_2} \cap A(k, q_2)}}$ & ${\tilde A}(k, Q)$ \\
\hline
1& 10&10& 10& 8& 20\\
2& 25&20& 40&24& 65\\
3& 67&40& 80&40&147\\
4&150&60&140&56&290\\
5&267&80&180&72&447\\
\end{tabular}

\begin{tabular}{|c|cc|cc|c|}
\hline
$k$ & $C_{\mu}(k,Q)$	& $P_{\mu}(k,Q)$\\
\hline
1&0.100&0.900\\
2&0.244&0.677\\
3&0.444&0.545\\
4&0.644&0.400\\
5&0.844&0.340\\
\hline
\end{tabular}
}
\caption{Datos para la generación de la \autoref{fig:microevaluation}.}
\label{tab:p-r-combinadas}
\end{center}
\end{table}

%
La precisión también puede calcularse en determinados puntos de corte de la lista ordenada de documentos que devuelve un sistema de recuperación. La Precisión a 10 documentos recuperados ($P@10$)~\cite{hawking01webtrack} se obtiene calculando la precisión de un conjunto de resultados considerando solamente los 10 primeros resultados de la lista ordenada. Esta es una métrica popular porque refleja el número de respuestas por defecto que devuelven la mayoría de los motores de búsqueda web. En forma más general, puede hablarse de \textit{Precisión a un rango k}. Dado un pedido $q$, el sistema devuelve una lista de documentos $D_q = \{d_1, \dots, d_m\}$, entonces se define la Precisión a un rango como:
\begin{align}
 P(k, D_q) = \frac{1}{k}\sum_{i=1}^{k}{\abs{{R_q \cap A(k, q)}}}\label{eqn:p_at_k}\text{,}
\end{align}
en donde $R_q$ y $A(k, q)$ son el conjunto de documentos relevantes para el pedido $q$ y el conjunto de documentos recuperados a un nivel de coordinación $k$ o superior, respectivamente, por el sistema para el pedido $q$. En adelante, para simplificar la notación, se utilizará $P(k)$ en lugar de $P(k, D_q)$.
\subsubsection{MAP}\label{subsub:map}
La Precisión Promedio Media (MAP\footnote{del inglés, Mean Average Precision.}) es una de las métricas más ampliamente usadas y nos da una representación numérica de la efectividad de un sistema~\cite{buckley05retrieval}. Geométricamente esta métrica es equivalente al área debajo de la curva de P-C. Para calcular la precisión promedio media sobre una única consulta se promedian los valores de precisión que se obtienen luego de recuperar cada documento relevante. Luego la MAP es la media sobre un conjunto de consultas. Esta métrica ha mostrado ser estable al variar el tamaño del conjunto de consultas~\cite{buckley00evaluating} y al ruido introducido por las variaciones de los juicios de relevancia~\cite{voorhees00variations}.

Formalmente, si se tiene un conjunto de consultas $Q$, y el resultado que devuelve el sistema de recuperación para $q_i \in Q$ es la lista ordenada de documentos $D_{q_i} = \{d_1, d_2, \dots, d_{m}\}$, entonces se define la MAP para una consulta como:

\begin{align*}
 \mathit{MAP}(q) &= \frac{1}{m}\sum\limits_{k=1}^{m}{P(k)},
\intertext{en donde $P(k)$ está dada por la \autoref{eqn:p_at_k} considerando la evaluación sobre el conjunto de resultados $D_{q_i}$. Finalmente se define la MAP para un conjunto de consultas como:}
 \mathit{MAP}(Q) &= \frac{1}{|Q|}\sum\limits_{i=1}^{|Q|}\mathit{MAP}(q_i).
\end{align*}

Esta métrica necesita tener disponible la información de relevancia de forma completa, o sea, haber evaluado para cada consulta si cada documento en la colección es relevante o no. 
El principal problema de MAP es que no puede interpretarse fácilmente, ya que si por ejemplo tenemos un $\mathit{MAP}=0.4$ no sabremos qué significa con exactitud, pero si tenemos una $P@10 = 0.4$ sabremos que obtuvimos cuatro documentos relevantes dentro de los diez primeros recuperados.

Como se dijo en la \autoref{sec:juicios_man}, es un problema obtener juicios de relevancia completos a medida que el tamaño de las colecciones de prueba crece. Al utilizar pooling aumenta el número de documentos que no son evaluados por los jueces (considerados irrelevantes por la métrica). Por esto en~\cite{buckley04retrieval} se define la métrica $\mathit{bpref}$, que utiliza información sobre los documentos no relevantes juzgados para generar una función que indica qué tan frecuentemente se recuperan documentos relevantes antes que los irrelevantes.

\begin{align*}
\mathit{bpref}(q) = \frac{1}{\abs{R_q}}\sum\limits_{d_j\in R_q}{\left(1-\frac{\text{\# de docs irrelevantes encima de }d_j}{\abs{R_q}}\right)}\text{.}
\end{align*}

Si se cuenta con la información de relevancia de toda la colección, esta métrica y $\mathit{MAP}$ muestran una alta correlación~\cite{buckley04retrieval}. Sin embargo, $\mathit{bpref}$ se comporta mejor en los casos de juicios incompletos. Recientemente han surgido algunas críticas a esta métrica y en~\cite{sakai07alternatives} se proponen otras alternativas.

\subsection{Medidas compuestas}\label{sub:comp_mes}
En muchas ocasiones es útil y necesario contar con una medida que represente la efectividad de un sistema y que posea una sola dimensión. Ya en 1967 Swets enunciaba, entre las propiedades de una métrica de rendimiento de recuperación, la necesidad de un único número en lugar de par de números que pueden llegar a covariar de formas poco claras, o en lugar de una curva construida a partir de una tabla~\cite{swets67effectiveness}. Esto permite el ordenamiento y la comparación de sistemas en términos absolutos.

En 1979 van Rijsbergen deriva una métrica compuesta de efectividad $E$ que tiene en cuenta las métricas de precisión y cobertura~\cite{vanrijsbergen79information}. La métrica incorpora el hecho de que los usuarios pudieran darle más importancia relativa a la precisión o a la cobertura. Esto se obtiene agregando un parámetro $\beta$ que indique que el usuario quiere darle $\beta$ veces más importancia a la cobertura que a la precisión y, por lo tanto, reflejará la relación $C/P$ para la cual el usuario está dispuesto a negociar un incremento en la precisión con un pérdida equivalente en cobertura:
\begin{align*}
 E = 1 - \frac{1}{\frac{\alpha}{P} + \frac{1 - \alpha}{C}}\text{.}
\end{align*}
Reemplazando $\alpha$ por $1/(\beta^2 + 1)$ se puede calcular la derivada parcial de $E$ respecto de la precisión e igualarla a la derivada parcial respecto de la cobertura para ver que $\beta = C / P$, como se quería. El segundo término de la ecuación es la conocida métrica $F$:
\begin{align*}
 F(P, C) &= \frac{1}{\frac{\alpha}{P} + \frac{1 - \alpha}{C}}\text{,}\\
 F(P, C) &= \frac{1}{\frac{1}{(\beta^2 + 1)P}+\frac{\beta^2}{(\beta^2+1)C}}\text{,}\\
 F_\beta(P, C) &= \frac{(\beta^2+1)PC}{C+\beta^{2}P}\text{.}
\end{align*}
Esta es una función $F:\mathbb{R}\times\mathbb{R}\rightarrow\mathbb{R}$ que contiene dos funciones de escala $\Phi_1$ y $\Phi_2$ para $P$ y $C$ y las combina de forma que la métrica resultante conserva el orden cualitativo de la efectividad.

Como casos particulares de esta función tenemos la \textit{media armónica} de precisión y cobertura, que le asigna igual importancia a ambos factores ($\beta=1$):
\begin{align*}
 F_1(P,C) = \frac{2PC}{C+P}\text{.}
\end{align*}
Otro caso particular ocurre si $\beta\to\infty$, cuando el usuario no desea asignarle importancia a la precisión, resultando en $F=C$. Por el contrario si $\beta\to 0$ entonces $F=P$.

C. J. van Rijsbergen mostró también que otras métricas que ya existían en la literatura~\cite{cleverdon66factors,heine73distance}, eran casos especiales de la medida $E$~\cite{vanrijsbergen79information}.
\subsection{Posición Rec\'{i}proca Media}\label{sub:mrr1}
La Posición Recíproca Media (MRR\footnote{del inglés, Mean Reciprocal Rank.}) de la respuesta que produce un sistema de IR, es la inversa multiplicativa de la posición de la primera respuesta relevante en esa salida~\cite{voorhees99qa}. Teniendo un conjunto de consultas, $Q = \{q_1, q_2, \dots, q_n\}$, se calcula el promedio de los valores obtenidos sobre sus resultados. En otras palabras, se tiene una lista ordenada de documentos, que son el resultado del sistema de IR para una consulta $q_i$ dada, se busca la posición, $\mathit{rank}$, de la primer respuesta correcta y se calcula la recíproca de ese valor. Finalmente se promedian los valores obtenidos al haber analizado todas las consultas. Esta medida se basa en el hecho de que se espera que las respuestas correctas se encuentren lo más arriba posible en la lista de resultados. Sea un función $\mathit{rank}(q)$ que devuelve la posición del primer documento relevante para $q$ devuelto por el sistema, entonces:
\begin{align*}
 \mathit{MRR}(Q) = \frac{1}{n}\sum\limits_{i=1}^{n}{\frac{1}{\mathit{rank}(q_i)}}\mbox{.}
\end{align*}
Si el sistema no logra recuperar ningún documento útil se asume que la recíproca es cero. Puede notarse que esta técnica no tiene en cuenta la cantidad de documentos relevantes recuperados.

Se la ha utilizado en general en el track \textit{Question Answering} de TREC en donde existen dos variantes de la métrica, de acuerdo a si se tiene en cuenta que los documentos relevantes son sólo los que responden perfectamente a la pregunta o si también se tienen en cuenta a aquellos que son acertados pero que no responden completamente a la pregunta. La primera suele llamarse MRR \textit{estricta} y la segunda versión \textit{indulgente}.

\section{Resumen}
En este capítulo se presentaron los fundamentos metodológicos de la evaluación de sistemas de recuperación de información. Se mostraron diversas métricas de comparación, distintas colecciones de prueba y varias formas de obtener los juicios de relevancia, tres componentes fundamentales para la comparación de nuevos sistemas y nuevas técnicas de IR.  
\chapter[Método incremental de recuperación de información basado en contexto]{%
Método incremental de\newline
\hspace{1em}recuperación de información\newline
basado en contexto\vspace*{-25mm}
}\label{chp4:metodo_incremental}
La calidad del material recuperado por los sistemas de búsqueda web basados en contexto es altamente dependiente del vocabulario que se usa para generar las consultas. Este trabajo propone aplicar un algoritmo semisupervisado para aprender incrementalmente nuevos términos.
El objetivo es contrarrestar la diferencia de terminología entre las palabras que el usuario pueda utilizar para expresar una consulta y el vocabulario empleado en los documentos relevantes. La estrategia de aprendizaje utiliza una selección de documentos web recuperados incrementalmente, dependientes del tópico, para ajustar los pesos que se les asignan a los términos, de manera de reflejar su aptitud como descriptores y discriminadores del tópico del contexto del usuario. Este algoritmo novedoso aprende nuevos descriptores buscando términos que tienden a aparecer en documentos relevantes, y aprende buenos discriminadores identificando términos que tienden a ocurrir únicamente en el contexto de un tópico dado. El vocabulario enriquecido de esta forma permite la creación de consultas de búsqueda que son más efectivas que aquellas generadas directamente con los términos de la descripción inicial del tópico, y que son en general los que emplea un usuario promedio. Una evaluación sobre una gran colección de tópicos y utilizando métricas de rendimiento estándar y ad~hoc sugiere que la técnica propuesta es superior a otras existentes en la literatura.
\section{Introducción}\label{sec:chp_metinc_introduccion}
Los primeros sistemas de IR fueron diseñados hace décadas para que los utilicen sólo algunos operadores entrenados. Este adiestramiento les permitía conocer de antemano: el lenguaje de consulta particular del sistema, el vocabulario particular del conjunto de documentos almacenados y la clase de documentos en la que estaban indexados. El acceso al sistema era a través de terminales de texto totalmente independientes y desconectadas de las computadoras que esos usuarios utilizaban para sus actividades laborales de todos los días.

El contexto en el cual se utilizan los sistemas de IR ha cambiado enormemente desde esos días. La computadora personal invadió la vida cotidiana y, tan importante como este punto o más, la disponibilidad masiva del acceso a la Internet permite utilizar los sistemas de IR desde cualquier lugar y desde una gran cantidad de dispositivos. Esta nueva situación provocó también un cambio en la clase de usuarios que utilizan tales sistemas y las actitudes de dichos usuarios, así como también un cambio en la información con la que cuentan los sistemas a la hora de procesar un pedido. Los usuarios acceden a los repositorios de información desde la misma máquina sobre la cual escriben sus artículos, leen las noticias y navegan los sitios web que les interesan. Este nuevo escenario trae aparejado nuevos desafíos y nuevas oportunidades para los diseñadores de los sistemas de acceso a la información, para que creen sistemas más amigables al usuario, que a su vez saquen mayor provecho de este nuevo ámbito colmado de información. 
\section{El problema del contexto en los sistemas de IR}\label{sec:context_problem}
El aumento de la disponibilidad y del uso de los sistemas no ha traído aparejado un crecimiento en la cantidad de información que el usuario le comunica al sistema, ni tampoco ha habido demasiados cambios en el diseño de los sistemas de interacción humano computadora. Los usuarios continúan manifestando sus pedidos a través de términos dentro de una consulta que luego se envía a un sistema que está aislado del contexto que produjo ese pedido.


Si miramos el problema desde el punto de vista de la interfaz de usuario, el lenguaje natural parece ser el medio ideal para la comunicación entre el sistema y el usuario. Desafortunadamente el sistema está separado de toda la información crítica que se necesita para su comprensión, 
incluso si el mismo intenta entender y representar los conceptos que están contenidos en los documentos indexados o en las consultas que procesa.
En otras palabras, esto quiere decir que los sistemas de información están aislados del \textit{contexto} en el cual se producen los pedidos. 

Cuando las personas recurren a un sistema de IR, en general, es porque quieren resolver algún problema, o lograr algún objetivo, para el cual su estado de conocimiento es insuficiente. Esto sugiere que no saben qué sería útil para esto y, por lo tanto, usualmente no son capaces de especificar de forma precisa las características más importantes de los objetos que son potencialmente útiles. Sin embargo, las consultas que se ingresan suelen basarse en un contexto que puede ayudar a interpretarlas y procesarlas. 
Por ejemplo, si el usuario está editando o leyendo un documento sobre un tópico específico, quizás quiera explorar material nuevo relacionado con ese tópico.
Si no se tiene acceso a este contexto las consultas se vuelven ambiguas, con los consecuentes resultados incoherentes y usuarios descontentos. Veamos el siguiente ejemplo que clarifica esta situación.

Consideremos un pedido de ``información acerca de java'', uno de los ejemplos más comunes en la literatura, y observemos cómo cambia drásticamente la utilidad de los resultados si manipulamos el contexto de la búsqueda en los siguiente escenarios.  
\begin{enumerate}
\renewcommand{\labelenumi}{Escenario \arabic{enumi}:}
\setlength{\leftskip}{12.5mm}
 \item \textit{Un importador de café buscando ampliar las variedades que distribuye}. En este caso, probablemente los recursos más apropiados son los que se refieran a la variedad de café que lleva ese nombre, su sabor, cualidades y las posibles compañías exportadoras que pudieran satisfacer su demanda.
 \item \textit{Un turista que está planeando sus próximas vacaciones}. Es muy probable que el turista esté buscando información sobre la isla de Indonesia, incluyendo paquetes turísticos, sitios de interés e historia.
%
 \item\label{it:lenguaje} \textit{Un estudiante de computación escribiendo un informe sobre Lenguajes de Programación}. En este caso, quisiéramos ver información acerca del lenguaje Java, con ejemplos, sintaxis y comparación con otros lenguajes.

\end{enumerate}
Estos escenarios muestran tres clases de problemas asociados con la interpretación de una consulta sin la consideración de un contexto~\cite{budzik01information}.
\begin{enumerate}
\renewcommand{\labelenumi}{Problema \arabic{enumi}:}
\setlength{\leftskip}{12.5mm}
 \item \textit{Relevancia de los objetivos actuales}. Los objetivos actuales del usuario influyen en gran medida en la interpretación de la consulta y en el criterio de evaluación de cuáles son resultados relevantes para ella.
 \item \textit{Ambigüedad del significado de las palabras}. El sentido del término ``java'' es distinto en el escenario~\ref{it:lenguaje} respecto de los otros. El contexto en el que se realiza una consulta proporciona una noción clara del sentido de la palabra.
 \item \textit{Perfil del usuario}. El público en cada uno de los escenarios también permite restringir los resultados. Una fuente de información que probablemente sea útil para un importador, no sea apropiada para un turista.
\end{enumerate}
Los ejemplos mencionados nos ayudan a ver algunos de los grandes problemas con los que se encuentran los sistemas actuales de IR al intentar responder a consultas sin tener en cuenta el contexto en el cual éstas se producen. Además, para empeorar la situación, se ha mostrado que las consultas enviadas a los motores de búsqueda tienen un promedio de 2.21 palabras~\cite{jansen98searchers} y, en una consulta de dos términos, es muy fácil ver que 
no existe la información suficiente para discernir los objetivos del usuario, o incluso para inferir el sentido correcto de las palabras.

Por lo tanto, vemos que la Búsqueda Basada en Contexto es el proceso de buscar información relacionada con el contexto temático del usuario~\cite{budzik01information,maguitman05suggesting,kraft06searching,ramirez06semantic}. Las búsquedas automáticas basadas en contexto sólo tienen éxito si las consultas reflejan la semántica que tienen los términos en el contexto bajo análisis. Desde un punto de vista pragmático, los términos toman significado de acuerdo a la forma en la que se los utilice y de acuerdo con su co-ocurrencia con otros términos. Luego, explotar grandes conjuntos de datos (como la Web) guiados por el contexto del usuario puede ayudar a descubrir el significado de un pedido de información por parte del usuario y a identificar buenos términos para refinar consultas incrementalmente.

Intentar encontrar buenos subconjuntos de términos para crear las consultas apropiadas es un problema combinatorial, y la situación empeora cuando atacamos un espacio de búsqueda abierto, como por ejemplo, cuando las consultas pueden incorporar términos que no pertenecen al contexto actual. Esta no es una situación inusual cuando intentamos refinar consultas basadas en una descripción de un contexto que contiene pocos términos y, a su vez, poseemos un corpus externo de gran tamaño.

Sin embargo, éste no es un problema nuevo en el área. La siguiente sección analiza algunos trabajos previos.

\section{Antecedentes} \label{sec:antecedentes}
Para acceder a información relevante es necesario generar las consultas apropiadas. En la búsqueda web basada en texto, los pedidos de información de los usuarios y los posibles recursos de información se caracterizan con términos. Para medir la importancia relativa de estos términos, en los métodos tradicionales de IR, se utilizan pesos (ver \autoref{sec:fundamentos}). 
Sin embargo, como lo han mostrado varias investigaciones, surgen problemas a la hora de intentar aplicar los esquemas convencionales de IR para medir la importancia de los términos en sistemas de búsqueda de datos en la Web~\cite{kobayashi00information,belkin00helping}. Una dificultad que aparece es que estos métodos no tienen acceso a una colección predefinida de documentos, generando dudas acerca de la aplicabilidad de tales esquemas en la medición de la calidad de los términos en las búsquedas web. Además, esta calidad dependerá de la tarea que se esté realizando; la noción de importancia tiene distintas interpretaciones dependiendo de si el término se necesita para la construcción de una consulta, para la generación de índices, para resumir un documento o para determinar similitud (\autoref{sec:similitud}). Por ejemplo, un término que es útil como descriptor del contenido de un documento y, por lo tanto útil para juzgar similitud, puede fallar si se lo utiliza como discriminante y ser ineficaz como término en una consulta, ya que lograría muy baja precisión en los resultados. Pero, podría darse el caso que se mejoren los resultados de esa búsqueda si se combina con otros términos que discriminen entre resultados buenos y malos.

En particular, la mayoría de los métodos que emplean realimentación de relevancia\linebreak (\autoref{sub:feedback}) eligen los términos más frecuentes de los documentos que se recuperaron en la primera pasada, con la esperanza de que éstos ayuden al proceso de recuperación. Este criterio de selección algunas veces brinda términos útiles, pero otras veces términos inútiles o incluso, términos que deterioran la calidad de los resultados. En~\cite{cao08selecting} se propone agregar términos en función de su impacto en la efectividad de los documentos recuperados, mostrando que tratar de separar los buenos términos	 de los malos basándose en su frecuencia es una tarea muy difícil. Proponen como alternativa el uso de un clasificador que se alimenta con un conjunto novedoso de propiedades, tales como la coocurrencia y la proximidad con los términos de la consulta inicial.

La comunidad de IR ha investigado desde hace décadas los roles de los términos como descriptores y discriminadores. Desde el trabajo de Spärck Jones sobre la interpretación estadística de la especificidad de un término~\cite{jones72statistical}, el poder discriminante de un término ha sido explicado estadísticamente, como una función del uso de dicho término. De la misma manera, la importancia de los términos como descriptores de un contenido ha sido tradicionalmente estimada midiendo su frecuencia dentro de un documento. La combinación de descriptores y discriminadores da lugar a esquemas para la medición de la relevancia de los términos como el conocido modelo de evaluación TF-IDF (\autoref{sub:mod_vectorial}).

Por otro lado, se ha avanzado mucho en el problema del cálculo del grado de información\footnote{del inglés, informativeness.} que tiene un término a lo largo de un corpus~\cite{amati02probabilistic,rennie05using,cai09learning}. Una vez que ese grado se calcula sobre un conjunto de términos, se pueden formular mejores consultas.

El principal problema de los métodos de refinamiento de consultas es que su efectividad, como se dijo, está correlacionada con la calidad de los documentos mejor clasificados recuperados en la primera pasada. Por otro lado, si se cuenta con un contexto temático, el proceso de refinamiento de consultas puede guiarse calculando una estimación de la calidad de los documentos recuperados. Esta estimación puede utilizarse para predecir cuáles términos pueden ayudar a refinar las consultas subsiguientes.

Durante los últimos años se han propuesto muchas técnicas para generar consultas desde el contexto del usuario~\cite{budzik01information,kraft06searching}. Otros métodos realizan el proceso de expansión y refinamiento de consultas con la explícita intervención del usuario~\cite{scholer02query,billerbeck03query}. Sin embargo poco se ha hecho en el campo de los métodos semisupervisados que saquen ventaja simultáneamente del contexto del usuario y de los resultados que obtienen del proceso de búsqueda.

\subsection*{Trabajo relacionado}\label{sub:trabajo_relacionado}
Los sistemas existentes en la literatura pueden clasificarse en cuatro categorías: \textit{realimentación de relevancia}, sistemas que hacen uso de \textit{perfiles de usuario}, métodos basados en técnicas implícitas y explícitas de \textit{desambiguación de términos} y métodos de \textit{modelado simbólico de usuarios}.

\begin{enumerate}
 \item \textbf{Realimentación de relevancia}. Esta técnica, explicada en la \autoref{sub:feedback}, tiene poco soporte en los buscadores comerciales de la actualidad. La razón puede verse en el hecho de que, si bien en muchas ocasiones la primer ronda de resultados no es satisfactoria para los usuarios, los mismos no tienen deseos de invertir tiempo y esfuerzo en el uso de técnicas manuales o semiautomáticas de refinamiento. Esto fue corroborado por estudios sobre el motor de búsqueda Excite\footnote{http://www.excite.com}~\cite{spink98users,jansen98searchers} y es un argumento a favor de los métodos que realizan refinamiento automático.
 \item \textbf{Perfiles de usuario}. Los métodos que construyen perfiles de usuario, como una forma de representar sus intereses, pueden ser vistos como métodos de recolección de información contextual. Como en el caso anterior, los perfiles de usuario pueden crearse recolectando términos a partir de los documentos clasificados por el usuario. La diferencia con el método anterior radica en que la información persiste a través de las distintas sesiones de búsqueda~\cite{chen98webmate}.

 Otros sistemas aplican algoritmos de aprendizaje automatizado para inducir un clasificador de documentos que se entrena con los documentos que el usuario califica~\cite{pazzani96syskill,billsus98personal}.
Esta técnica implícita parece ser más prometedora que la Realimentación de relevancia, sin embargo, los sistemas que crean un perfil del usuario, al concentrarse en la captura de intereses generales y a largo plazo, presentan la desventaja de restar importancia a la relevancia de los objetivos actuales.
 \item \textbf{Desambiguación de términos}. Para realizar la desambiguación de los términos, de entre sus múltiples significados, algunos sistemas requieren de la intervención explícita de los usuarios~\cite{cheng97experiment}, mientras que otros utilizan la información subyacente en los documentos para estimar su significado más común~\cite{bradshaw99mining}.

 Desafortunadamente, desambiguar palabras sólo resuelve parte del problema de comprender el contexto en un entorno aislado, quedando sin solucionar los objetivos actuales y el perfil del usuario. Además, algunos sistemas siguen teniendo la necesidad de que los usuarios intervengan de forma explícita, lo que nos lleva nuevamente a los problemas existentes en la Realimentación de relevancia.

 Los sistemas que utilizan un texto de referencia a la hora de ordenar los documentos tienen la ventaja de que no requieren de la intervención por parte del usuario, sin embargo como métrica de ordenamiento utilizan la frecuencia de los términos y no hacen un análisis más profundo del contexto del usuario, lo que limita su efectividad. Por ejemplo, un sistema que procesa la consulta ``monitores de plasma'' podría devolver resultados relacionados con el plasma sanguíneo si la mayoría de los documentos que contienen la palabra ``plasma'' tratan acerca de ese tema.

 \item \textbf{Modelado de la semántica de la tarea}. Se han desarrollado sistemas que crean y utilizan un modelo explícito del comportamiento del usuario en una aplicación dada y asocian las consultas, o las acciones de los usuarios, en un estado particular de la tarea con los recursos que le son útiles~\cite{horvitz98lumiere,johnson99integrating}. Por ejemplo, Argus~\cite{johnson99integrating} observa a los usuarios interactuar con un Sistema de Apoyo al Desempeño\footnote{del inglés, Performance Support System.} y utiliza un modelo de tareas explícito para detectar el mejor momento para recuperar artículos de un sistema de memoria organizacional.

 Aunque el rendimiento de estos sistemas puede ser muy alto, suelen ser difíciles de construir y normalmente tienen un alcance limitado. Esencialmente sufren de los mismos problemas que aparecen a la hora de diseñar un buen documento HTML: el diseñador debe tener un conocimiento previo de los documentos existentes para así poder crear los enlaces apropiados. En entornos en los cuales la colección de documentos es muy grande y dinámica (como lo es la Internet) se torna realmente poco práctico.
\end{enumerate}

A continuación se mencionarán algunos sistemas, desarrollados en la literatura, que asisten al usuario en la tarea de recuperación de información.
\subsubsection*{Asistentes para el Manejo de Contenido}\label{sub:imas}

En~\cite{budzik99watson} se desarrollaron una clase de sistemas llamados Asistentes para el Manejo de Contenido (IMA\footnote{del inglés, Information Management Assistants.}). Los IMAs descubren material relacionado en representación del usuario, actuando como un intermediario inteligente entre él y los sistemas de recuperación de información. Un IMA observa la interacción del usuario con las aplicaciones diarias (p. ej. un editor de texto o un navegador web), y usa estas observaciones para anticipar sus necesidades de información. Entonces intenta satisfacer estas necesidades accediendo a los sistemas de IR tradicionales (p. ej. motores de búsqueda en Internet, resúmenes de artículos, entre otros), filtrando los resultados y presentándolos al usuario. Además, la arquitectura de los IMAs proporciona una plataforma potente para contextualizar las consultas ingresadas explícitamente por un usuario conectándolas con las tareas que está realizando. 
\subsubsection*{Watson}\label{sub:watson}
Watson utiliza información contextual extraída de los documentos que los usuarios están manipulando para generar consultas web~\cite{budzik01information}. Para la selección de los mejores términos para conformar las consultas utiliza diversas técnicas de extracción y evaluación de términos. Luego filtra los resultados, agrupa las páginas similares y las presenta como sugerencias al usuario.

\subsubsection*{Remembrance Agent}\label{sub:rem_agent}
El Agente de la Memoria (RA\footnote{del inglés, Remembrance Agent.}) es un programa que ``aumenta la memoria humana'' mostrando una lista de documentos que pudieran ser relevantes para el contexto actual del usuario~\cite{rhodes96remembrance}. A diferencia de muchos otros sistemas de esa época, el RA se ejecutaba sin la intervención del usuario. Esta interfase no invasiva permitía al usuario tomar las sugerencias que le parecían interesantes o directamente ignorarlas.

Su filosofía de diseño se basa en que, como un programa que se está ejecutando y actualizando continuamente, nunca debe distraer al usuario de su tarea principal, sólo debe mejorarla. Sugiere fuentes de información que pueden llegar a ser relevantes para la tarea actual del usuario en la forma de resúmenes de una línea al final de la pantalla. En ese lugar la información puede ser fácilmente monitoreada, pero no distrae del trabajo que se está realizando. En caso de aceptar la sugerencia, el usuario puede visualizar el texto completo presionando una tecla.

\subsubsection*{Letizia}\label{sub:letizia}
Letizia es un agente intermediario que asiste al usuario que está navegando la Web~\cite{lieberman95letizia}. A medida que el usuario trabaja en la forma habitual con un explorador web, el agente registra el comportamiento del usuario y trata de traer por adelantado ítems de interés explorando de forma concurrente y autónoma los enlaces de la página que el usuario está viendo en ese momento. Automatiza una estrategia de navegación que consiste en una búsqueda mejorada, con la técnica del primero-mejor, infiriendo los intereses del usuario a partir de su comportamiento.
Utiliza con conjunto de heurísticas simples para modelar cuál podría ser el comportamiento de navegación del usuario. 
A pedido, puede visualizar una página conteniendo sus recomendaciones actuales, en la cual se puede elegir seguir una recomendación o regresar a las actividades convencionales de navegación.

\subsubsection*{WebWatcher}\label{sub:webwatcher}
WebWatcher es un agente que ayuda interactivamente a los usuarios a localizar la información que desean, utilizando conocimiento aprendido acerca de cuáles son los hipervínculos que mejor pueden guiar al usuario a la información que está buscando~\cite{armstrong95webwatcher}.

WebWatcher es un agente de búsqueda de información que se ``invoca'' ingresando en su página web e indicando en un formulario qué información se necesita (p. ej., una publicación de algún autor). Entonces el agente lleva al usuario a una página nueva que contiene una copia de la página recomendada más algunos agregados que asisten al usuario a medida que sigue los hipervínculos hacia la información buscada. Mientras el usuario navega la Web, WebWatcher va aprendiendo y usa ese conocimiento para recomendar hipervínculos prometedores, resaltándolos en las páginas.

\subsubsection*{SenseMaker}\label{sub:sensemaker}


SenseMaker es una interfase de exploración de fuentes de información (principalmente referencias a artículos) heterogéneas, como p. ej. motores de búsqueda web y de bibliografía de distintos proveedores~\cite{baldonado97sensemaker}. El sistema puede ``unir'' (en forma de clusters) datos que muestran cierto grado de similitud, de acuerdo a algún criterio definido por el usuario. Por ejemplo, para páginas web, un criterio puede ser agrupar aquellos sitios que se refieran a la misma página, o en el caso de artículos, por título o autor. El sistema permite que el usuario elija alguno de varios criterios para visualizar los resultados recuperados, así como también para eliminar elementos duplicados.
Permite a los usuarios experimentar de forma iterativa con distintas vistas de los resultados. Dentro de una vista se puede reducir la complejidad de la visualización filtrando los campos mostrados y agrupando resultados similares. También es posible evolucionar consultas por medio de la expansión o la sustitución de términos.


\subsubsection*{Fab}\label{sub:fab}
Fab es un sistema recomendador colaborativo híbrido basado en contenido, que aprende a navegar por la Web en representación de un usuario~\cite{balabanovic97fab}. Genera recomendaciones utilizando un conjunto de agentes de recolección (que buscan las páginas de un determinado tema) y agentes de selección (que buscan páginas para un usuario en particular). Las valuaciones explícitas de los usuarios de las páginas que el sistema le recomendó se combinan con algunas heurísticas para: actualizar los perfiles de los agentes personales, eliminar agentes que no tuvieron éxito, y duplicar a los exitosos.

Es un sistema híbrido porque propone eliminar las debilidades y aprovechar las ventajas de los recomendadores colaborativos y de los basados en contenidos al combinarlos en uno solo.
Los sistemas de recomendación basados en \textit{contenido} intentan recomendar objetos similares a aquellos a los cuales el usuario a seleccionado en el pasado, mientras que los sistemas recomendadores \textit{colaborativos} identifican otros usuarios cuyos gustos sean similares a un usuario dado y le recomienda objetos que otros seleccionaron.

\subsubsection*{Broadway}\label{sub:broadway}
El sistema \textit{Broadway}\footnote{del inglés, BROwsing ADvisor reusing pathWAYs.} es un sistema de razonamiento basado en casos (CBR\footnote{del inglés, Case-Based Reasoning.}), que monitorea la actividad de navegación del usuario y da consejos reutilizando casos que se extraen del historial de navegación de otros usuarios~\cite{jaczynski97broadway}. Los consejos se basan en una lista ordenada en base a la similitud con los documentos visitados.
El paradigma CBR se utiliza para aprender el conjunto de casos relevantes de los historiales de navegación de los usuarios, los cuales son utilizados para mejorar y mantener actualizado el proceso de asesoramiento. El sistema se basa en la siguiente hipótesis: si dos usuarios han recorrido la misma secuencia de páginas, pueden tener objetivos similares y, por lo tanto, un usuario puede aprovechar los documentos que el otro usuario consideró relevantes.

\subsubsection*{SiteSeer}\label{sub:siteseer}
Siteseer es un sistema de recomendación de páginas web que utiliza la estructura de los \textit{Favoritos} de un navegador web, para predecir y recomendar páginas relevantes~\cite{rucker97siteseer}. El sistema interpreta que las páginas contenidas en los Favoritos son una declaración implícita de interés en su contenido y que las carpetas en las que se agrupan esas páginas son una indicación de la coherencia semántica entre los elementos. Además trata a las carpetas como un sistema de clasificación personal, lo cual posibilita contextualizar las recomendaciones en clases definidas por el usuario. Siteseer aprende con el tiempo las preferencias y categorías de cada usuario, y a su vez aprende de cada página web a qué comunidades pertenecen sus usuarios. Luego ofrece recomendaciones personalizadas organizadas dentro de las carpetas propias de cada usuario.

Siteseer emplea los descubrimientos de un usuario como recomendaciones implícitas para otros, basándose en los Favoritos que pueden encontrarse en un conjunto de revisores confiables. Un usuario es confiable si el \textit{grado de solapamiento} de sus carpetas es suficientemente alto.
La similitud se calcula a través del contenido de las carpetas, sin derivar ningún valor semántico del contenido de las páginas ni del título de las carpetas. De esta manera crea dinámicamente comunidades virtuales de intereses, particulares a cada usuario y específicas para cada categoría de interés de éste.
Al calcular la pertenencia de forma relativa a cada carpeta, el sistema no impone rigidez a las categorías y logra ser útil a usuarios con intereses relativamente poco frecuentes o muy específicos.

\subsubsection*{Syskill \& Webert}\label{sub:syskill-webert}
Syskill \& Webert es un agente que aprende a evaluar páginas en la Web y decide qué páginas pueden interesarle a un usuario~\cite{pazzani96syskill}. Los usuarios califican las páginas que visitan en una escala de tres puntos y el agente aprende el perfil del usuario analizando la información que extrae de cada página. Este perfil se utiliza para sugerirle al usuario enlaces que podrían interesarle, y para construir consultas que se envían a un motor de búsqueda para descubrir nuevas páginas de interés.

El sistema aprende un perfil separado para cada tópico de interés del usuario. Esto ayuda a la precisión del aprendizaje de cada tópico. Cada uno de estos posee una página web con los URLs que se indexaron del tópico. De esta manera el sistema permite a un usuario usar esta página como un punto de partida para la exploración.
La utilización del sistema de esta forma tiene la limitación de que debe existir una página con un buen conjunto de URLs sobre un tópico. Si esto no ocurriera, el agente envía consultas, basadas en el perfil del usuario, a un motor de búsqueda para recolectar páginas nuevas.

\subsubsection*{Extender}\label{sub:extender}

El sistema Extender aplica técnicas de búsqueda incremental para construir descripciones más completas del contexto del usuario y las utiliza para la identificación de tópicos relacionados con ese contexto~\cite{maguitman05suggesting}. El objetivo del sistema es ayudar a expertos en la construcción de modelos de conocimiento, dándole sugerencias sobre nuevos tópicos. Estas sugerencias son pequeños conjuntos de términos que tratan de conducirlo al significado del tópico (p. ej., una etiqueta de la forma ``\textit{lunar}, \textit{luna}, \textit{explorador}'' se utiliza para describir el tópico \textit{Misión de exploración a la Luna}).

El sistema comienza a partir de un mapa conceptual~\cite{ausubel63psychology} e iterativamente hace búsquedas en la Web para encontrar material novedoso relacionado con todo el mapa. También permite que el usuario resalte algún tema para forzar la búsqueda hacia ese concepto. En cada iteración, los documentos recuperados se representan como una matriz de documentos-términos, se hace clustering para identificar los tópicos dentro del conjunto y el material que se considera irrelevante se descarta. Este proceso continúa hasta alcanzar algún criterio de convergencia del tópico o un límite en la cantidad de iteraciones.

\subsubsection*{Suitor}\label{sub:suitor}
Suitor\footnote{del inglés, Simple User Interest Tracker.} es una colección de ``agentes atentos'' que recolectan información de los usuarios monitoreando su comportamiento y su contexto desde múltiples canales, incluyendo la mirada, las teclas presionadas en el teclado, el movimiento del mouse, los sitios visitados y las aplicaciones que se están ejecutando~\cite{maglio00suitor}. Esta información se utiliza para recuperar de la Web y otras bases de datos, material relevante al contexto del usuario.

Suitor observa al usuario, lo modela y se anticipa a sus necesidades. Esta clase de sistemas incluye a los mayordomos, los dispositivos comerciales como el TiVo\footnote{http://www.tivo.com/}, que graba automáticamente programas de televisión que le gustan al usuario, y sitios como Amazon.com que monitorean las costumbres de compra y el comportamiento de navegación y le sugieren al usuario nuevos libros que podrían interesarle.\\

En la siguiente sección se presentará una propuesta para refinar consultas temáticas, basada en el análisis de los términos que se encuentran en el contexto del usuario y en los resultados recuperados incrementalmente.

\section{Una Plataforma novedosa para la selección de términos}
Este trabajo presenta técnicas generales para aprender incrementalmente términos relevantes asociados a un contexto temático. Específicamente se estudian tres preguntas:
\begin{enumerate}
\item ¿Puede el contexto del usuario explotarse satisfactoriamente para acceder a material relevante en la Web?
\item ¿Puede un conjunto de términos específicos de un contexto ser refinado incrementalmente basándose en el análisis de los resultados de una búsqueda?
\item ¿Los términos específicos de un contexto aprendidos mediante métodos incrementales, son mejores para generar consultas comparados con aquellos encontrados por técnicas clásicas de IR o métodos clásicos de reformulación de consultas?
\end{enumerate}
La contribución de este trabajo es un algoritmo semisupervisado que aprende incrementalmente nuevo vocabulario con el propósito de mejorar consultas. El objetivo es que las consultas reflejen la información contextual y así puedan recuperar efectivamente material relacionado semánticamente. En este trabajo se utilizó una métrica estándar de evaluación del rendimiento y dos métricas ad~hoc para descubrir si estas consultas son mejores que las generadas utilizando otros métodos. Estas nuevas métricas se presentarán en detalle en la \autoref{sub:metricas_propuestas}.

La pregunta principal que guió este trabajo es cómo aprender términos específicos a un contexto basándonos en la tarea del usuario y en una colección abierta de documentos web recuperados incrementalmente. De ahora en más, asumiremos que la tarea del usuario está representada como un conjunto de términos cohesivos que resumen el tópico del contexto del usuario. Consideremos un ejemplo que involucra la \textit{Máquina Virtual de Java}, descripto por los siguientes términos:{\small {\tt
\begin{center}
\begin{tabular*}{0.90\textwidth}{@{\extracolsep{\fill}}lllll}
java       & virtual       & machine &programming   & language\\
computers & netbeans & applets      & ruby  & code\\
sun  & technology     & source      & jvm     & jdk\\
\end{tabular*}
\end{center}}}
\noindent Los términos específicos a un contexto juegan distintos roles. Por ejemplo, el término \textit{java} es un buen descriptor del tópico para el común de las personas. Por otro lado, términos como \textit{jvm} y \textit{jdk} (acrónimos de ``Java Virtual Machine'' y ``Java Development Kit'' respectivamente) pueden no ser buenos descriptores del tópico para esas mismas personas, pero son efectivos recuperando información similar al tópico cuando se los utiliza en una consulta. Luego, \textit{jvm} y \textit{jdk} son buenos discriminadores del tópico.

En~\cite{maguitman04dynamic} se propone estudiar el poder descriptivo y discriminante de un término basándose en su distribución a través de los tópicos de las páginas recuperadas por un motor de búsqueda. Allí, el espacio de búsqueda es la Web completa y el análisis del poder descriptivo o discriminante de un término está limitado a una pequeña colección de documentos que se va construyendo incrementalmente y que varía en el tiempo. A diferencia de los esquemas de IR tradicionales, los cuales analizan una colección predefinida de documentos y buscan en ella, los métodos propuestos utilizan una cantidad limitada de información para medir la importancia de los términos y documentos así como también para la toma de decisiones acerca de cuáles términos conservar para análisis futuros, cuáles descartar, y qué consultas adicionales generar.

Para distinguir entre descriptores y discriminadores de tópicos se argumenta que \textit{buenos descriptores de tópicos} pueden encontrarse buscando aquellos términos que aparecen en la \underline{mayoría} de los documentos relacionados con el tópico deseado. Por otro lado, \textit{buenos discriminadores de tópicos} pueden hallarse buscando términos que \underline{sólo} aparecen en documentos relacionados con el tópico deseado. Ambos tipos de términos son importantes a la hora de generar consultas. Utilizar términos descriptores del tópico mejora el problema de los resultados falso-negativos porque aparecen frecuentemente en páginas relevantes. De la misma manera, los buenos discriminadores de tópicos ayudan a reducir el problema de los falsos-positivos, ya que aparecen principalmente en páginas relevantes.

\subsection*{Cálculo del Poder Descriptivo y del Poder Discriminante}\label{sec:computing}

Como una primera aproximación al cálculo del poder descriptivo y discriminante se comienza con un conjunto de $m$ documentos y $n$ términos. En primera medida se construye una matriz de $m\times n$ elementos, de forma tal que ${\bf \M}[i,j]=p$, donde $p$ es el número de apariciones del término $k_i$ en el documento $d_j$. En particular se puede asumir que uno de los documentos (p. ej., $d_0$) corresponde al contexto inicial. El ejemplo siguiente muestra esta situación:{
\begin{spacing}{1}
\small
\begin{center}
\begin{math}\M = \bordermatrix{          & d_0 & d_1 & d_2 & d_3 & d_4 \cr
{\tt \ \ \ \ \ \ \ \ \ \ \ \ \ \ java}   & 4  & 2 &  5 &  5 &  2 \cr
{\tt \ \ \ \ \ \ \ \ machine}            & 2 & 6 & 3 & 2 &0\cr
{\tt \ \ \ \ \ \ \ \ virtual}            & 1 & 0 & 1 & 1 & 0\cr
{\tt \ \ \ \ \ \ language}               &1 & 0 & 2 & 1 & 1 \cr
{\tt programming}                        & 3 & 0 & 2 & 2 & 0 \cr
{\tt \ \ \ \ \ \ \ \ \ \ coffee}         & 0 & 3 & 0 & 0 &3 \cr
{\tt \ \ \ \ \ \ \ \ \ \ island}         & 0 & 4 &0 &0 &2 \cr
{\tt \ \ \ \ \ \ province}               &0 &4 &0 &0 & 1\cr
{\tt \ \ \ \ \ \ \ \ \ \ \ \ \ \ \ \ jvm}&0 &0 &2 &1 &0\cr
{\tt \ \ \ \ \ \ \ \ \ \ \ \ \ \ \ \ jdk}&0 &0 &3 &3 &0\cr }
\end{math} \ \ \ \ \ \ \ \ \ \ \ \ \
    \begin{tabular}{rl}
   \multicolumn{2}{l}{{\bf Documents:}}\\
       $d_0$: & \textit{contexto inicial del usuario}\\
    $d_1$: & {espressotec.com}\\
    $d_2$: & {netbeans.org}\\
    $d_3$: & {sun.com}\\
    $d_4$: & {wikitravel.org}\\
    \end{tabular}
\end{center}
\end{spacing}
}

La matriz $\M$ permite formalizar las nociones de buenos descriptores y buenos discriminadores. 

\begin{definition}\textnormal{\cite{maguitman04dynamic}}
Se define el \textnormal{poder descriptivo de un término en un documento} como la función $\descD:\{d_0,\ldots,d_{m-1}\}\times \{k_0,\ldots,k_{n-1}\}
\rightarrow [0,1]$:
\begin{align*}
\descD(d_j,k_i) = \frac{\M[i,j]}
                     {\sqrt{\sum_{h=0}^{n-1} (\M[h,j])^2}}.
\end{align*} 
\end{definition}
\begin{definition}\textnormal{\cite{maguitman04dynamic}}
Se define una función $\step(p)$ tal que $\step(p)=1$ cuando $p>0$ y $\step(p)=0$ en cualquier otro caso. Luego se define el \textnormal{poder discriminante de un término en un documento} como la función $\discD: \{d_0,\ldots,d_{m-1}\}\times
\{k_0,\ldots,k_{n-1}\} \rightarrow [0,1]$:
\begin{align*}
\discD(d_j, k_i) = \frac{\step(\M[i,j])}
                     {\sqrt{\sum_{h=0}^{m-1} \step(\M[i,h])}}.
\end{align*}
\end{definition}
\ \noindent Se puede ver que $\descD$ y $\discD$ satisfacen las siguientes condiciones 
\begin{align*}
\sum_{h}{(\descD(d_j,k_h))^2}=1\mbox{ \ \ y \ \  }\sum_{h}{(\discD(d_h,k_i))^2}=1.
\end{align*}
\noindent Dado el término $k_i$ en el documento $d_j$, el término tendrá un poder descriptivo alto en $d_j$ si aparece frecuentemente en él, mientras que tendrá un alto poder discriminante si tiende a aparecer sólo en él (o sea, aparece esporádicamente en otros documentos). Para los términos del ejemplo, el poder descriptivo y el poder discriminante serían los siguientes:
\begin{spacing}{1}
\small
\begin{center}
\begin{math}
\bordermatrix{& \descD(d_0,k_i) \cr
  {\tt \ \ \ \ \ \ \ \ \ \ \ \ \ \ java} & 0.718 \cr
  {\tt \ \ \ \ \ \ \ \ machine} & 0.359 \cr
  {\tt \ \ \ \ \ \ \ \ virtual} & 0.180 \cr
  {\tt \ \ \ \ \ \ language} & 0.180 \cr
  {\tt programming} & 0.539 \cr
  {\tt \ \ \ \ \ \ \ \ \ \ coffee}& 0.000 \cr
  {\tt \ \ \ \ \ \ \ \ \ \ island} & 0.000 \cr
  {\tt \ \ \ \ \ \ province}& 0.000 \cr
  {\tt \ \ \ \ \ \ \ \ \ \ \ \ \ \ \ \ jvm}& 0.000 \cr
  {\tt \ \ \ \ \ \ \ \ \ \ \ \ \ \ \ \ jdk}& 0.000 \cr }
\end{math}
\begin{math}
\bordermatrix{ & \discD(d_0,k_i) \cr  &  0.447 \cr  & 0.500 \cr  &
0.577 \cr  &0.500 \cr  & 0.577 \cr & 0.000 \cr & 0.000 \cr  & 0.000
\cr  & 0.000 \cr  & 0.000\cr }
\end{math}
\end{center}
\end{spacing}
Los pesos anteriores reflejan algunas de las limitaciones de esta primera aproximación. Por ejemplo, los pesos asociados con los términos {\tt jvm} y {\tt jdk} no reflejan su importancia como discriminadores del tópico analizado. De la misma manera que las populares medidas TF e IDF (explicadas en la \autoref{sub:mod_vectorial}), las funciones $\descD$ y $\discD$ permiten descubrir términos que son buenos descriptores y buenos discriminadores de un documento, en lugar de buenos descriptores y discriminadores del \textit{tópico} del documento.

El objetivo es desarrollar nociones de descriptores y discriminadores de tópicos que sean aplicables a la Web. En lugar de extraer descriptores y discriminadores directamente del contexto del usuario, se busca extraerlos del \textit{tópico} de ese contexto. Para realizar esta tarea es necesario un método incremental que caracterice el tópico del contexto del usuario, lo cual se lleva a cabo identificando los documentos que son similares a ese contexto. Asumiendo que el contexto y los documentos recuperados de la Web están representados como vectores en un espacio de términos, para determinar qué tan similares son dos documentos $d_j$ y $d_k$ se adopta la medida de similitud por coseno (\autoref{def:vector_model}, \autoref{eqn:cosine_similarity}), 
que en términos de $\descD$ queda expresada como:
\begin{align*}
\simi(d_j,d_k) = \sum_{h=0}^{n-1}[\descD(d_j,k_h)\cdot\descD(d_k,k_h)].
\end{align*}
\noindent Los valores de similitud entre el contexto del usuario ($d_0$) y los otros documentos del ejemplo son los siguientes:
\begin{center}
\begin{math}
\bordermatrix{& d_1 & d_2 & d_3 & d_4 \cr
 \simi(d_0,d_j) = & 0.399 &    0.840 &   0.857 &  0.371\cr}
 \end{math}
\end{center}
%
La noción de descriptor de tópico se definió informalmente con anterioridad como ``los términos que ocurren \textit{frecuentemente} en el contexto de un tópico''. 
\begin{definition}\textnormal{\cite{maguitman04dynamic}}
Se define el \textnormal{poder descriptivo de un término en el tópico de un documento} como la función $\descT : \{d_0,\ldots, d_{m-1}\} \times \{k_0, \ldots, k_{n-1}\} \to [0,1]$. \newline Si $\sum\nolimits_{%
  \begin{subarray}{l}%
    h=0 \\ %
    h\neq j%
  \end{subarray}%
    }^{m-1}{\simi(d_j,d_h)}= 0$ se define $\descT(d_j,k_i)=0$. En los otros casos se define $\descT(d_j,k_i)$ de la siguiente manera:
\begin{align*}
\descT(d_j,k_i) = \frac{\sum\nolimits_{%
  \begin{subarray}{l}%
    h=0 \\ %
    h\neq j%
  \end{subarray}%
    }^{m-1}{[\simi(d_j,d_h)\cdot [\descD(d_h,k_i)]^2]}}{\sum\nolimits_{%
  \begin{subarray}{l}%
    h=0 \\ %
    h\neq j%
  \end{subarray}%
    }^{m-1}{\simi(d_j,d_h)}}.%
\end{align*}
De esta manera el poder descriptivo de un término $k_i$ en el tópico de un documento $d_j$ es la medida de la calidad de $k_i$ como descriptor de documentos similares a $d_j$. 
\end{definition}
\noindent Como se definió informalmente antes, un término es un buen discriminador de un tópico si ``tiende a aparecer \textit{sólo} en documentos asociados con ese tópico''.
\begin{definition}\textnormal{\cite{maguitman04dynamic}}
Se define el \textnormal{poder discriminante de un término en el tópico de un documento} como la función $\discT:\{d_0,\ldots,d_{m-1}\}\times\{k_0,\ldots,k_{n-1}\} \rightarrow [0,1]$ que se calcula:
\begin{align*}
\begin{array} {rl}
\discT(d_j,k_i) = &  \sum\nolimits_{%
  \begin{subarray}{l}%
    h=0 \\ %
    h\neq j%
  \end{subarray}%
    }^{m-1}{[\simi(d_j,d_h) \cdot [\discD(d_h,k_i)]^2]}.
\end{array}
\end{align*}
Así el poder discriminante de un término $k_i$ en el tópico del documento $d_j$ es un promedio ponderado de la similitud entre $d_j$ y otros documentos discriminados por $k_i$.
\end{definition}
Los siguientes valores son el poder descriptivo y discriminante de los términos del ejemplo:
\begin{spacing}{1}
\small
\begin{center}
\begin{math}
\bordermatrix{
 & \descT(d_0,k_i) \cr
 {\tt \ \ \ \ \ \ \ \ \ \ \ \ \ \ java}& 0.385 \cr
 {\tt \ \ \ \ \ \ \ \ machine}& 0.158 \cr
 {\tt \ \ \ \ \ \ \ \ virtual}& 0.014 \cr
 {\tt \ \ \ \ \ \ language}& 0.040 \cr
 {\tt programming}& 0.055 \cr
 {\tt \ \ \ \ \ \ \ \ \ \ coffee}& 0.089 \cr
 {\tt \ \ \ \ \ \ \ \ \ \ island}& 0.064 \cr
 {\tt \ \ \ \ \ \ province} & 0.040 \cr
 {\tt \ \ \ \ \ \ \ \ \ \ \ \ \ \ \ \ jvm}& 0.032 \cr
 {\tt \ \ \ \ \ \ \ \ \ \ \ \ \ \ \ \ jdk} & 0.124 \cr
 }
\end{math}
\begin{math} \bordermatrix{
 & \discT(d_0,k_i) \cr
 & 0.493 \cr
 & 0.524 \cr
 & 0.566 \cr
 & 0.517 \cr
 & 0.566 \cr
 & 0.385 \cr
 & 0.385 \cr
 & 0.385 \cr 
 & 0.848 \cr 
 & 0.848 \cr }
\end{math}
\end{center}
\end{spacing}

Con las nociones de descriptores y discriminadores de tópicos es posible aprender términos nuevos y específicos a un contexto y reajustar los pesos de los existentes. Esto produce una mejor representación del contexto de búsqueda del usuario, mejorando el proceso de refinamiento de consultas y el filtrado basado en ese contexto.
Podemos apreciar en el ejemplo, que los términos {\tt jvm} y {\tt jdk}, que no pertenecían al contexto inicial del usuario, resultaron ser excelentes discriminadores del tópico.

\section{Mecanismo incremental para refinar consultas temáticas}\label{sec:incremental_method}
En esta sección se abordarán los detalles del mecanismo incremental para el refinado de consultas temáticas propuesto. Ésta es una de las contribuciones de esta tesis.

\begin{figure}[!ht]
\centerline{\includegraphics[scale=0.5]{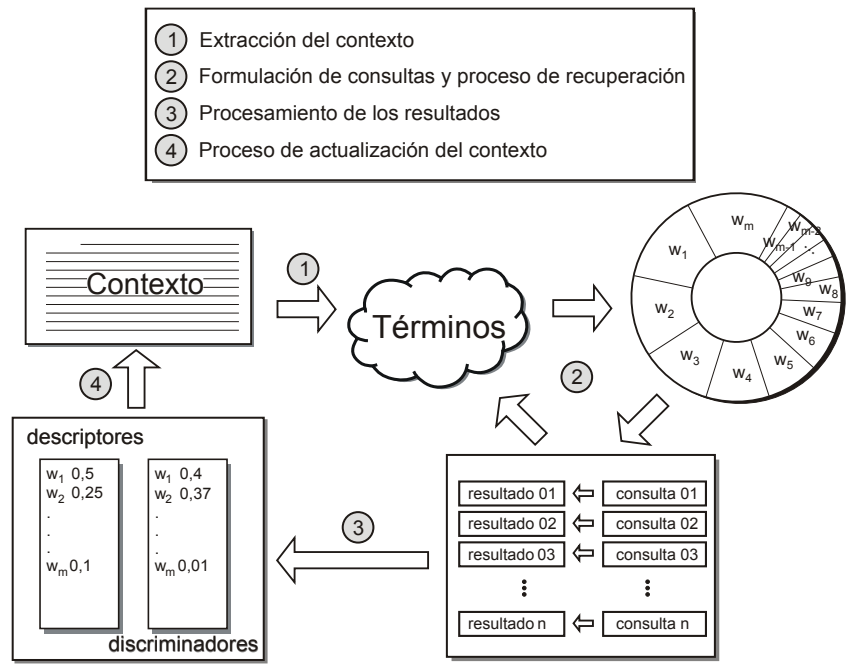}} \caption[Representación esquemática del método incremental.]{Una representación esquemática del método incremental para el refinamiento de consultas temáticas.} \label{fig:diagram}
\end{figure}

La propuesta es aproximar el poder descriptivo y discriminativo de los términos del contexto bajo análisis con el propósito de generar buenas consultas. Esta aproximación adapta el mecanismo típico de realimentación de relevancia para que considere un contexto temático en evolución $\mathcal{C}_i$. Un esquema del método incremental para el refinamiento de consultas basado en un contexto temático se muestra en la \autoref{fig:diagram} y se resume en el \autoref{alg:main}.

\begin{algorithm} \caption{Principal} \label{alg:main}
\begin{algorithmic}
\STATE $i \Leftarrow 0$ \STATE $\mathcal{C}_i \Leftarrow InitialContext$
\REPEAT[evolución de la fase \ensuremath{\mathcal{P}_i}]
    \STATE $i \Leftarrow i+1$
    \STATE Calcular $\mathcal{C}_i$
    \STATE Actualizar $\mathcal{C}_i$
\UNTIL{ $(i > v) \wedge FinalConvergence$}
\end{algorithmic} 
\end{algorithm}

\begin{algorithm}
\caption{Calcular $\mathcal{C}_i$} \label{alg:compute}
\begin{algorithmic}
\REQUIRE $\alpha + \beta = 1$
\STATE $j \Leftarrow 0$
\STATE $\descT_j \Leftarrow \emptyset$
\STATE $\discT_j \Leftarrow \emptyset$
\REPEAT[evolución de las pruebas]
    \STATE $j \Leftarrow j+1$
    \STATE Crear consultas a partir de $\mathcal{C}_i$ y realizar la Búsqueda
    \STATE Calcular $\descT'$ y $\discT'$ basándose en los resultados de la búsqueda
    \STATE $\{\descT_{j}|\discT_{j}\} = \alpha\{\descT_{j-1}|\discT_{j-1}\} + \beta\{\descT'|\discT'\}$
    \STATE Prueba-de-convergencia
\UNTIL{$(j > u) \wedge PhaseConvergence$}
\end{algorithmic}
\end{algorithm}

\begin{algorithm}
\caption{Prueba-de-convergencia} \label{alg:converg}
\begin{algorithmic}
\REQUIRE $\mu > \nu$
\STATE $PhaseConvergence \Leftarrow max(\mathit{sim}(Results, \mathcal{C}_i)) < \mu$
\STATE $FinalConvergence \Leftarrow max(\mathit{sim}(Results, \mathcal{C}_i)) < \nu$
\end{algorithmic}
\end{algorithm}

El sistema lleva a cabo una serie de \textit{fases} con el objetivo de aprender mejores descripciones de un contexto temático. En la figura esto está representado por el ciclo de pasos que van desde el paso 1 al paso 4. Al final de cada fase se actualiza la descripción del contexto con el nuevo material aprendido (paso 4). En una fase $\mathcal{P}_i$ dada, se representa el contexto con un conjunto de términos con sus respectivos pesos. La estimación de la importancia del término $k$ en el contexto $\mathcal{C}$ durante la fase $i$ se representa con $w^{\mathcal{P}_i}(k,\mathcal{C})$. Si $k$ aparece en el contexto inicial, entonces el valor de $w^{{\mathcal{P}_0}}(k,\mathcal{C})$ se inicializa con la frecuencia normalizada del término $k$ en $\mathcal{C}$, mientras que los términos que no son parte del contexto $\mathcal{C}$ se asumen en 0. Cada fase a su vez evoluciona a través de lo que se denominó una serie \textit{pruebas} (paso 2); en ellas se formulan una serie de consultas, se analizan los resultados obtenidos y se calcula el poder descriptivo y discriminante de los nuevos términos descubiertos. 

La generación de las consultas en cada una de las pruebas se implementó a través de un mecanismo de selección por ruleta, en donde la probabilidad de elegir un término en particular $k$, para que forme parte de una consulta, es proporcional a $w^{\mathcal{P}_i}(k,\mathcal{C})$. La técnica de selección por ruleta se utiliza comúnmente en los Algoritmos Genéticos~\cite{holland75adaptation} para la elección de las soluciones potencialmente útiles que luego se recombinarán; la probabilidad de selección está dada por el nivel de aptitud del individuo. Este método produce una exploración no determinística del espacio de términos que favorece a los términos más aptos.

La estimación del poder descriptivo y discriminate de un término $k$ para un contexto $\mathcal{C}$ en una fase $i$ en una prueba $j$ se calcula incrementalmente de la siguiente manera:
\begin{align*}
w^{(i,j+1)}_\descT(k,\mathcal{C}) &= \alpha .  w^{(i,j)}_\descT(k,\mathcal{C}) + \beta . \descT^{(i,j)}(k,\mathcal{C})\text{,}\\
w^{(i,j+1)}_\discT(k,\mathcal{C}) &= \alpha . w^{(i,j)}_\discT(k,\mathcal{C}) + \beta . \discT^{(i,j)}(k,\mathcal{C})\text{.}
\end{align*}
Se asume que los valores iniciales de cada prueba son nulos, $w^{(i,0)}_\descT(k,\mathcal{C})=w^{(i,0)}_\discT(k,\mathcal{C})=0$ y se utilizan los resultados recuperados durante cada prueba $j$ para calcular el poder descriptivo, $\descT^{(i,j)}(k,\mathcal{C})$, y discriminante, $\discT^{(i,j)}(k,\mathcal{C})$, del término $k$ para el tópico $\mathcal{C}$ respectivamente. Las constantes $\alpha$ y $\beta$ determinan la cantidad de información que el sistema recordará de una prueba a la siguiente. Cuanto mayor sea el valor asignado a la constante $\alpha$ más tiempo perdurará el conocimiento aprendido y menos influencia tendrán los nuevos documentos recuperados en el cálculo del poder descriptivo y discriminante de un término.

\subsection*{Monitoreo de la efectividad}
Con el objetivo de medir la efectividad que el sistema alcanza en cada iteración, se calculan un conjunto de métricas sobre el material recuperado. Las medidas que pueden aplicarse para guiar al algoritmo son las clásicas del área de IR (como las vistas en la \autoref{sec:metricas}) u otras nuevas, que serán definidas en el capítulo siguiente.

Si luego de una ventana de $u$ intentos la efectividad no supera un determinado nivel $\mu$ (o sea, que no se han observado mejoras importantes luego de cierto número de intentos), el sistema provoca un cambio de fase para explorar nuevas regiones del vocabulario que sean potencialmente útiles. Un cambio de fase puede verse como un salto en el vocabulario, lo cual puede pensarse como una transformación importante de la representación del contexto (típicamente una mejora). Si se realiza un cambio de fase durante la prueba $j$, el valor de $w^{{\mathcal{P}_i}}_\descT(k,\mathcal{C})$ queda determinado como 
$w^{(i,j)}_\descT(k,\mathcal{C})$ y
$w^{{\mathcal{P}_i}}_\discT(k,\mathcal{C})$ como 
$w^{(i,j)}_\discT(k,\mathcal{C})$. 
\begin{align*}
w^{{\mathcal{P}_{i+1}}}(k,\mathcal{C}) = \gamma . w^{{\mathcal{P}_i}}(k,\mathcal{C}) +
\zeta . w^{{\mathcal{P}_i}}_\descT(k,\mathcal{C}) + \xi .
w^{{\mathcal{P}_i}}_\discT(k,\mathcal{C}).
\end{align*}
Los términos con los pesos modificados de esta forma se utilizan para generar nuevas consultas durante la prueba $i+1$. La convergencia final del algoritmo se alcanza luego de al menos $v$ cambios de fase si la efectividad no supera un cierto límite $\nu$ ($\nu < \mu$). Para evitar que los pesos de los términos crezcan indefinidamente es necesario que $\gamma + \zeta + \xi = 1$. El valor de $\gamma$ indica la cantidad de memoria que tendrá el sistema de una fase a la siguiente. Los valores de $\zeta$ y $\xi$ indican la cantidad de conocimiento que se conservará de una fase a la siguiente. Estas constantes afectan la velocidad de aprendizaje del sistema y su habilidad para conservar el foco de las búsquedas.


El método propuesto es prometedor y la eficacia del mismo se evaluará en el \autoref{chp5:evals} utilizando una plataforma de evaluación especialmente diseñada para el análisis de sistemas de IR temáticos. 

\section{Alcances y aplicaciones}
La generación automática de consultas a partir de un contexto temático necesita de técnicas que tengan la habilidad de unir el contexto dado con las fuentes de  material relevante. El problema de las coincidencias falso-negativas es una situación común que surge cuando el texto contiene un tópico similar pero el vocabulario de términos no coincide. Un problema complementario son las coincidencias falso-positivas, que aparecen cuando coinciden los términos del vocabulario pero los documentos pertenecen a tópicos diferentes. Estas dos situaciones problemáticas han sido identificadas desde hace mucho en la comunidad de IR como uno de los desafíos principales, y muchas propuestas han tratado de superar estas cuestiones con distintos grados de éxito.

En este capítulo se ha propuesto un método que muestra un avance en el intento de solucionar los problemas mencionados anteriormente aprendiendo nuevos vocabularios. Se propuso utilizar los descriptores de tópicos para identificar aquellos términos que aparecen más frecuentemente en documentos asociados con el tópico dado. Estos términos no son necesariamente parte de la especificación de las necesidades de información del usuario, sin embargo, pueden encontrarse iterativamente analizando los conjuntos de documentos que se recuperan incrementalmente de la Web. Por otro lado, también se propuso utilizar los discriminadores de tópicos para identificar a aquellos términos que tienden a aparecer sólo en los documentos del tópico y muy pocas veces en documentos que no pertenecen a él. En el siguiente capítulo se presentará una Plataforma de evaluación sobre la cual se evaluará el método incremental propuesto, comparándolo con otros existentes en la literatura.

El desarrollo de métodos que evolucionen consultas de alta calidad y recuperen recursos relevantes al contexto puede tener consecuencias importantes en la manera en la que los usuarios interactúan con la Web. Estos métodos pueden ayudar a construir sistemas para un amplio espectro de servicios de información:\newpage
\begin{itemize}
\item {\textbf{Búsqueda basada en la tarea del usuario}.} Los sistemas de búsqueda basada en la tarea del usuario explotan la interacción del usuario con las aplicaciones en su computadora para determinar la tarea actual del usuario y poner en contexto sus necesidades de información~\cite{leake00capture,budzik01information}. Las búsquedas basadas en palabras clave podrían muy fácilmente fallar en el intento de hallar páginas relevantes a esa tarea. Un sistema de búsqueda basado en la tarea del usuario que evolucione consultas de alta calidad, puede generar automáticamente sugerencias que estén contextualizadas con esa tarea.

\item {\textbf{Recuperación de recursos para portales web temáticos}.}
Los portales web temáticos tienen el propósito de reunir recursos sobre temas específicos. El material recolectado se utiliza para construir directorios y sitios de búsqueda especializados. Típicamente, la recolección del material de estos portales está a cargo de crawlers enfocados~\cite{chakrabarti99focused,menczer04topic}. Una alternativa a éstos puede hallarse formulando consultas temáticas en un motor de búsqueda y eligiendo del conjunto de resultados aquellos recursos que están relacionados con el tópico en cuestión.
\item {\textbf{Búsqueda en la Web oculta}.} La mayor parte de la información de la Web puede encontrarse en la forma de páginas generadas dinámicamente, y constituyen lo que se conoce como la Web oculta~\cite{kautz97hiddenweb}. Las páginas que la componen no existen hasta que son creadas dinámicamente, como el resultado de una consulta presentada a un formulario de búsqueda en sitios específicos (p. ej., PubMed Central\footnote{http://www.ncbi.nlm.nih.gov/pmc/} o Amazon\footnote{http://www.amazon.com/}). Por lo tanto, la generación de consultas de alta calidad es de gran importancia al momento de querer acceder a recursos de la Web oculta, y los métodos de búsqueda basada en contexto pueden aportar soluciones útiles a la hora de acceder a su contenido.

\item {\textbf{Soporte para la administración y modelado de conocimiento}.} El modelado del conocimiento es el proceso mediante el cual se representa un cuerpo de conocimiento para facilitar su posterior acceso. La administración efectiva del conocimiento puede necesitar ir más allá de la captura inicial del mismo~\cite{leake03aiding,maguitman05suggesting}. La Web proporciona una fuente rica en información en donde buscar nuevo material para incluir en los modelos de conocimiento. De esta manera, el material puede accederse por medio de consultas contextualizadas que se presentan a un motor de búsqueda convencional, en donde el contexto está dado por el modelo de conocimiento que se está construyendo. 
Las técnicas discutidas aquí, utilizando la Web como un gran repositorio de la memoria colectiva y partiendo de un modelo de conocimiento en construcción, pueden facilitar el proceso de captura de conocimiento y ayudan a aumentar la memoria organizacional.
\end{itemize}

\section{Resumen}
%
En este capítulo se presentó una de las contribuciones de esta tesis, un Método Incremental de Recuperación de Información basada en Contexto. Se comenzó con una breve introducción del método y de los problemas que tienen los sistemas de IR al no tener en cuenta el contexto en el cual se desarrollan las actividades de los usuarios. A continuación se hizo un análisis de los trabajos existentes en la literatura y se clasificaron los sistema de IR basados en contexto. Luego se introdujeron las nociones de descriptores y discriminadores de tópicos, que son los principios sobre los que se basaron los algoritmos presentados en esta tesis. Con ellos es posible mejorar la valoración de los pesos asignados a los términos que representan el contexto de un usuario y aumentar así el rendimiento de un sistema de recuperación. Luego se describió el método incremental propuesto, que es capaz de refinar consultas temáticas a partir de la técnica recién mencionada, adaptando el mecanismo de realimentación de relevancia para que considere el contexto temático del usuario. Finalmente, se presentó un análisis de los alcances y aplicaciones de la herramienta de recuperación propuesta.

\chapter{Plataforma de evaluación}\label{chp5:evals}
El método descripto en el capítulo anterior fue implementado en el contexto de una plataforma general de IR. Esta se desarrolló con el propósito de proponer y evaluar nuevos algoritmos en el área de IR. El objetivo final es publicar la plataforma bajo una licencia de Código Abierto, de modo que otros grupos de investigación puedan realizar comparaciones y/o mejoras sobre los algoritmos implementados en la misma. Los resultados de las evaluaciones del algoritmo incremental presentado en el capítulo anterior y de otros algoritmos serán mostrados a continuación.

\section{Estructura}\label{sec:framework_structure}
La plataforma incluye actualmente una colección local de documentos que fueron indexados con la plataforma de código abierto Terrier\footnote{http://www.terrier.org}, desarrollada por la Universidad de Glasgow\footnote{http://www.gla.ac.uk/}. El acceso a este índice se realiza a través de una interfaz que es capaz de aceptar otros tipos de índices e incluso, motores de búsqueda web. En un comienzo, se implementó una interfaz para el servicio web SOAP de Google\footnote{http://www.google.com}, que luego fuera reemplazado por la empresa por una API AJAX. La utilización del servicio web permitió el desarrollo de las primeras versiones de los algoritmos presentados en esta tesis. Finalmente se optó por un índice local de documentos web debido a las limitaciones que se encontraron en cuanto a los tiempos de ejecución de los algoritmos y al límite impuesto por Google a la cantidad de consultas que se podían realizar por día.

\begin{figure}[!ht]
\begin{center}
\includegraphics[width=\textwidth]{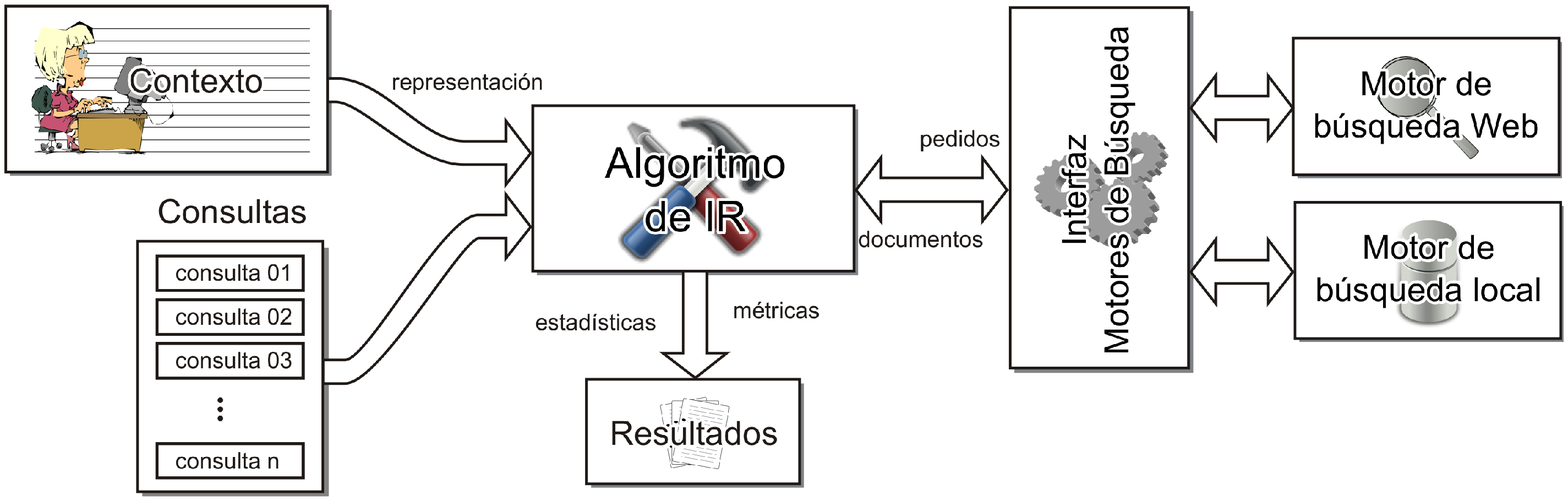}
\caption{Representación esquemática de la plataforma de evaluación.}
\label{fig:framework_structure}
\end{center}
\end{figure}

Una representación esquemática de la Plataforma de Evaluación se muestra en la \autoref{fig:framework_structure}. Como se puede observar existe una primera parte que se encarga de la representación de las consultas. Estas pueden ingresarse como un conjunto o como un documento, a partir del cual el sistema generará las consultas necesarias. Por otro lado, la plataforma ofrece una interfaz de comunicación con los distintos motores de búsqueda. Como se dijo más arriba, una de las posibilidades es contar con un motor de búsqueda web. También existe un componente dedicado al cálculo de las métricas que guiarán los algoritmos de búsqueda y que también servirán para su evaluación.

A continuación se explicarán con más detalles los componentes de la plataforma.
\section{Generación de consultas}
Este componente del sistema se encarga del proceso de generación de las consultas que iniciarán o que utilizarán los algoritmos que se evaluarán con la plataforma. Los algoritmos evaluados y propuestos en esta tesis caen en la categoría algoritmos de recuperación basada en contexto (\autoref{sec:context_problem}), por lo que en todos los casos se cuenta con el contexto del usuario.

La generación de consultas puede llevarse a cabo con distintas técnicas. La primera es de forma aleatoria, en donde se seleccionan al azar palabras del contexto del usuario, todas con la misma probabilidad, y es la que se utilizó en~\cite{cecchini08using} y en~\cite{cecchini10MOEAforContext}. Otra técnica es el mecanismo de selección por ruleta, en donde la probabilidad de selección de un término está dada por el peso que tiene asignado. Esto provoca una exploración no determinística del espacio de términos que favorece a los más aptos. Este método fue el que se utilizó en~\cite{lorenzetti09semisupervised}.

\section{Motores de búsqueda}\label{sec:framework_search_engines}
Este componente del sistema se encarga de realizar los pedidos de información a los distintos motores de búsqueda con los que cuenta la plataforma. También lleva a cabo la tarea de preprocesar y convertir los resultados a un formato uniforme a todos los motores. Uno de los motores de búsqueda con los que el sistema se comunica es un motor web, en particular se implementó la comunicación con el servicio de búsqueda web de Google. Una forma de acceder al motor de búsqueda web Google es hacer uso de su formulario web y luego aplicar un analizador sintáctico al código HTML que obtenemos. Esta técnica se conoce en inglés como \textit{web scraping} y no es recomendada por la compañía, pudiendo dar lugar a que filtren el acceso de la computadora que origina los pedidos. Para evitar este tipo de prácticas y para favorecer la investigación sobre los recursos indexados por su buscador, desarrollaron un servicio web SOAP\footnote{http://code.google.com/apis/soapsearch/} que permitía hacer uso del motor de búsqueda. Este servicio tenía una limitación importante para su uso en cuanto a que imponía un límite máximo a la cantidad de consultas diarias (1000 en un principio); por otra parte, no se garantizaba que los servidores que atendían los pedidos estuvieran todos actualizados con la misma versión del índice, lo que provocó dificultades a la hora de repetir los experimentos. Hoy en día este servicio sigue vigente pero no tiene más soporte. En su lugar existe una API AJAX\footnote{http://code.google.com/apis/ajaxsearch/} más orientada a su uso en páginas web.

El formato de la consulta del servicio web es similar al formato de la consulta que se ingresa por medio del formulario web, excepto que la cantidad máxima de palabras estaba limitada a 10 (luego ampliada a 32). Los resultados devueltos por el servicio web incluyen, entre otros datos, el URL de cada página y un pequeño fragmento del documento\footnote{en inglés este fragmento se denomina \textit{snippet}.}. 

Otro de los motores de búsqueda que se utilizan es la plataforma de recuperación de información Terrier\footnote{del inglés, TERabyte retRIevER}~\cite{ounis06terrier}. Esta plataforma implementa varios algoritmos del estado del arte en el área de IR y ofrece técnicas eficientes y efectivas en una forma modular y extensible. El sistema incluye modelos de recuperación basados en la Divergencia de la Aleatoriedad, que es un modelo de teoría de la información para expansión de consultas y ordenamiento de documentos (un ejemplo se presentó en la \autoref{sub:feedback}). También incluye una gran variedad de modelos, como algunas variantes del esquema clásico TF-IDF y el método de modelado de lenguaje basado en la fórmula de ordenamiento estadístico BM25~\cite{robertson94approximations}.

En esta tesis se utilizó principalmente el componente de indexación de Terrier para crear una colección local de documentos. Terrier ofrece varias APIs para indexación y consulta que fueron adaptadas para su uso en la plataforma propuesta. Al igual que otras plataformas de código abierto disponibles, incluye analizadores sintácticos para indexar documentos en formato de texto plano, HTML, PDF y otros, así como también varios algoritmos de \textit{stemming}\footnote{del inglés, método para reducir una palabra a su término raíz o \textit{stem}.}. Otro componente utilizado de esa plataforma fue el de recuperación, el cual admite una serie de operadores en el formato de las consultas, como los más comunes de inclusión y exclusión de términos, pero también permite la asignación de pesos a cada término de la consulta, la expansión de consultas y la búsqueda de términos en determinados campos de un documento.
\subsection*{Colección local de documentos}
Como se mencionó anteriormente, la plataforma desarrollada para el soporte y la evaluación de los algoritmos propuestos en esta tesis incluye una colección local de documentos, indexados con el propósito de acelerar el proceso de evaluación de algoritmos, principalmente frente a las limitaciones de velocidad que presenta un servicio web. Otras limitaciones que se superaron con el uso de una colección local se mencionarán en la sección siguiente. Como se mencionó en la \autoref{sec:juicios_man}, al crear una colección de documentos uno de los mayores desafíos es evaluar la relevancia de cada uno de los documentos del índice, entonces al crear la colección se optó por indexar un conjunto de documentos de la ontología ODP. Esto permitió la evaluación de los algoritmos propuestos con métricas que hacen uso de relaciones de similitud semántica (vista en la \autoref{sec:similitud_semantica}) y que se detallarán en la sección siguiente. 

Para realizar los experimentos se utilizaron $\abs{V}=448$ tópicos del ODP. Fueron elegidos los tópicos que se encuentran en el tercer nivel de la jerarquía. Se impusieron un número de restricciones a esta selección para asegurar la calidad del conjunto de datos. El tamaño de cada tópico seleccionado fue de 100 URLs como mínimo y se limitaron los tópicos a aquellos que pertenecen al lenguaje inglés. Para cada uno de estos tópicos se recuperaron todos sus URLs así como también los pertenecientes a sus subtópicos. Todas las páginas pertenecientes a un tópico o a alguno de sus descendientes se consideraron, a los efectos de las métricas, como el mismo tópico. El número total de páginas resultó ser más de 350 mil.
\section{Métricas de evaluación}\label{sub:metricas_propuestas}
Con el objetivo de medir la efectividad que alcanza el sistema, este componente se encarga de calcular distintas métricas sobre los resultados que le entrega el algoritmo en evaluación.
Las medidas que pueden aplicarse para guiar al algoritmo son las clásicas del área de IR (como las vistas en la \autoref{sec:metricas}) u otras nuevas.

Como se mencionó en la sección anterior, las evaluaciones comenzaron utilizando un motor de búsqueda web que, como parte de los resultados, devolvía un pequeño fragmento del documento recuperado. Por lo tanto, una de las primeras métricas con las que se evaluaron los algoritmos fue la medida clásica de IR, similitud por coseno (discutida en la \autoref{sub:mod_vectorial}). Por cuestiones de eficiencia, la métrica sólo se utilizó para comparar el contexto del usuario con cada fragmento de los documentos recuperados por el motor de búsqueda. En particular, puede notarse que al utilizar un motor web no es posible calcular otras métricas como la cobertura, dado que no hay una forma de saber de antemano qué páginas pertenecen al conjunto de documentos relevantes para la consulta que se está haciendo. Otro problema que aparece es que, en general, este fragmento contiene porciones del documento cercanas a las palabras de la consulta y, por lo tanto, es muy probable que éstas estén contenidas dentro de ese fragmento. Puede verse que es sencillo alcanzar valores altos de similitud, perjudicando incluso a aquellos documentos que pudieran ser relevantes y que no emplean el mismo vocabulario que el usuario, desfavoreciendo la exploración de material novedoso.

\subsection{Similitud Novedosa}
En vista de las consideraciones expuestas arriba, en esta tesis se propone una nueva métrica llamada \textit{Similitud Novedosa}. La similitud novedosa es una medida de similitud ad~hoc que está basada en la \autoref{eqn:cosine_similarity} definida en el \autoref{chp2:state}.
Esta nueva medida descarta los términos que forman parte de la consulta al momento de hacer los cálculos, reduciendo el sesgo introducido por esos términos y favoreciendo la exploración de nuevos documentos. 

\begin{definition}\label{def:new_similarity}
Dado un vector de consulta $\overrightarrow{q}=(w(k_1,q), \dots, w(k_t,q))$ y dos documentos\linebreak $\overrightarrow{d_j}=(w(k_1,d_j), \dots, w(k_t,d_j))$ y $\overrightarrow{d_k}=(w(k_1,d_k), \dots, w(k_t,d_k))$, representados en el modelo vectorial de acuerdo con la forma enunciada en la \autoref{def:vector_model}. Sean $k_i$ con $i \in [1\dots t]$, en donde $t$ es el número total del términos en el sistema, cada uno de los términos que pudieran contener $\overrightarrow{q}$, $\overrightarrow{d_j}$ o $\overrightarrow{d_k}$. Entonces, la similitud novedosa se define como: 
\begin{align}
  \mathit{sim}^{N}(q,d_j,d_k) &= \mathit{sim}(\overrightarrow{d_j - q}, \overrightarrow{d_k - q}) \label{eqn:new_similarity} \text{,}\\
\intertext{La notación $\overrightarrow{d_j - q}$ es la representación del documento $d_j$ con valores nulos en todos los términos correspondientes a la consulta ${q}$. Lo mismo se aplica a $\overrightarrow{d_k - q}$. Por lo tanto:}
  &= \frac{\sum\nolimits_{%
  \begin{subarray}{l}%
    i = 1 \\ %
    \forall {k_i} \notin q%
  \end{subarray}%
    }^{t}{w(k_i,d_j) . w(k_i,d_k)}}{\sqrt {\sum\nolimits_{%
  \begin{subarray}{l}%
    i = 1 \\ %
    \forall {k_i} \notin q%
  \end{subarray}%
    }^t w(k_i,d_j)^{2}} .\sqrt {\sum\nolimits_{%
  \begin{subarray}{l}%
    i = 1 \\ %
    \forall {k_i} \notin q%
  \end{subarray}%
    }^t w(k_i,d_k)^{2}}}\text{.}\nonumber
\end{align}
\end{definition}

\subsection{Precisión Semántica}
Otras métricas que se implementaron dentro de la plataforma de evaluación fueron, la Precisión, la Cobertura y la Precisión a un rango \textit{k}. Por otro lado, en esta tesis también se propone otra métrica, la \textit{Precisión Semántica}. De acuerdo a la visto en el \autoref{chp3:evaluacion}, la Precisión se define como la porción de documentos recuperados que son relevantes (\autoref{eqn:p}). Si creamos una función de similitud a partir de esta métrica, $\mathit{sim}_{rel}$, que le asigne un peso de 1 a un documento, $d_j$, si pertenece al conjunto de los documentos relevantes, $R$, y un peso nulo en otro caso, tenemos
\begin{align*}
 \mathit{sim}_{rel}(d_j) = \left\{ {\begin{array}{*{20}l}
   1	& \mbox{si }d_j\in {R}\mbox{,}\\
   0	& \mbox{si no.} \\
\end{array}} \right.
\end{align*}
Vemos que esta función es binaria e indica si un documento pertenece al conjunto de documentos relevantes. Luego, recordando que $A$ representa el conjunto de documentos recuperados, la ecuación de Precisión, $P$, puede reescribirse de la siguiente manera:
\begin{align*}
 P = \frac{\sum_{d_j\in A}\mathit{sim}_{rel}(d_j)}{|A|}\text{.}
\end{align*}

El objetivo de los algoritmos desarrollados en esta tesis es encontrar documentos que sean relevantes para el tópico del contexto del usuario, $\mathcal{C}$, por lo tanto, utilizar una métrica de relevancia binaria descarta la posibilidad de recuperar muchos documentos parcialmente relevantes y por lo tanto potencialmente útiles. Entonces, se propone una medida de precisión semántica, basada en las nociones de similitud vista en la \autoref{sec:similitud_semantica}.
\newpage
\begin{definition}\label{def:semantic_precision}
Sea $\mathfrak{T}(d_j)$ una función que devuelve el tópico al que pertenece un documento. Sea $\mathcal{C}$ el contexto del usuario el cual pertenece a un tópico $\tau_0 = \mathfrak{T}	(\mathcal{C})$. Sea $\semsimilarity(\tau_1,\tau_2)$ la similitud semántica entre estos dos tópicos, definida en la \autoref{eqn:sim_sem_grafos}. La precisión semántica, $P^S$, se define como: 
\begin{align}
P^S&= \frac{\sum_{d_j\in A}\semsimilarity(\mathfrak{T}(\mathcal{C}),\mathfrak{T}(d_j))}{|A|}\text{.}\label{eqn:semantic_precision}
\end{align}
\end{definition}

Las métricas de Precisión, Precisión Semántica y Cobertura se utilizaron en conjunto con la colección local de documentos, por lo tanto, la función $\mathfrak{T}(d_j)$ simplemente devuelve el tópico dentro de la ontología ODP al que pertenece $d_j$.

\section{Evaluación de los Algoritmos}\label{sec:evaluation}
En esta sección ilustraremos la aplicación de la plataforma de evaluación propuesta. Con tal fin se presentarán los resultados obtenidos en las evaluaciones del Algoritmo Incremental para el Refinamiento de Consultas propuesto en la \autoref{sec:incremental_method} y otros algoritmos de búsqueda basados en contexto propuestos por el grupo de investigación del autor de la presente tesis, publicados en~\cite{cecchini10thesis}. En las siguientes dos subsecciones se compararán los métodos propuestos con otros existentes en la literatura.
\subsection{Algoritmos Incrementales}\label{sec:incremental_algorithms}
El objetivo de esta sección es comparar el método propuesto con otros dos métodos. El primero es un método base que genera las consultas directamente del contexto temático y no aplica ningún mecanismo de refinamiento. El segundo método es el Bo1-DFR descripto en la \autoref{sub:feedback}.

Para la creación del contexto inicial $\mathcal{C}$ utilizado en los experimentos, se hizo uso de la descripción que contiene cada tópico en ODP. El algoritmo propuesto se ejecutó en cada tópico por al menos $v=10$ iteraciones, con 10 consultas por iteración y recuperando 10 resultados por consulta. Las listas de descriptores y discriminadores en cada iteración se limitaron a 100 términos cada una. Otros parámetros del método fueron ajustados de la siguiente manera: 
el número de iteraciones por prueba $u=10$, 
los coeficientes de conservación y de actualización del poder descriptivo y discriminante durante una prueba $\alpha$=0.5 y $\beta$=0.5, 
el coeficiente de conservación del peso de los términos en el contexto de un usuario $\gamma$=0.33, 
los coeficientes de actualización del peso de un término respecto del poder descriptivo y discriminante aprendido en la fase anterior $\zeta$=0.33 y $\xi$=0.33, 
el nivel de efectividad mínimo esperado durante una prueba $\mu$=0.2 y 
el nivel de efectividad mínimo esperado durante una fase $\nu$=0.1. 
Además se utilizó la lista de palabras frecuentes proporcionada con Terrier, se aplicó el algoritmo de stemming de Porter~\cite{porter80stripping} en todos los términos y no se aplicaron ninguno de los métodos de expansión de consultas provistos por la plataforma.

\begin{figure}[!ht]
\begin{center}
\includegraphics[scale=0.6]{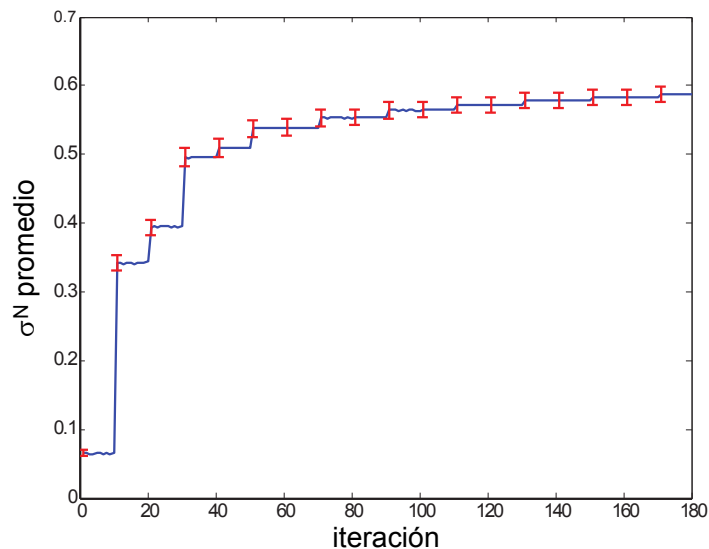}
\caption{Evolución de la similitud novedosa máxima.}
 \label{fig:averagesimilarity}
 \end{center}
\end{figure}

Se calculó la métrica de similitud novedosa $\nsimilarity$ entre el contexto inicial (la descripción del tópico) y los resultados recuperados. La \autoref{fig:averagesimilarity} muestra la evolución de la similitud novedosa promedio sobre todos los tópicos evaluados. Además, el gráfico muestra las barras de error cada 10 iteraciones (las cuales normalmente coinciden con un cambio de fase). Las mejoras observadas, especialmente durante el primer cambio de fase, proporcionan evidencias de que el algoritmo propuesto puede ayudar a mejorar el vocabulario del tópico.

\begin{figure}[!ht]
\begin{center}
  \includegraphics[width=100mm]{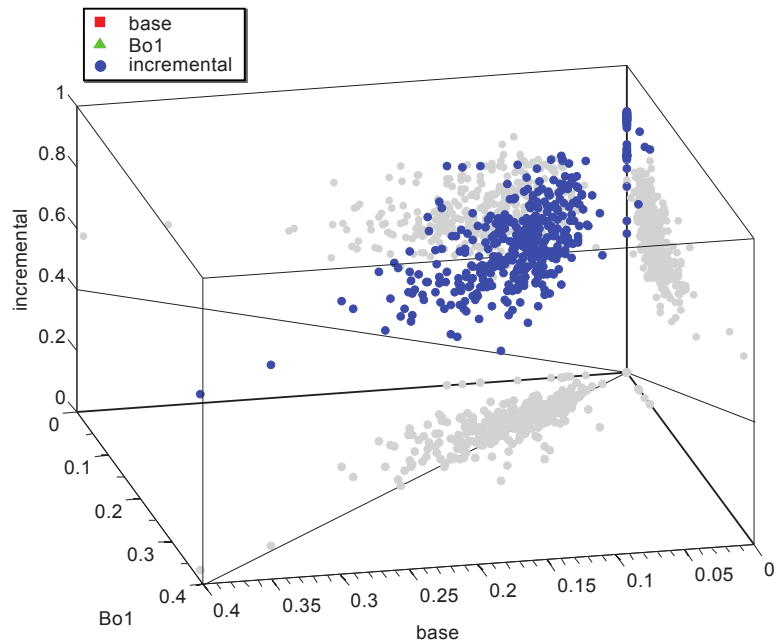}
\end{center}
  \caption{Comparación de los tres métodos basada en Similitud Novedosa.}
  \label{fig:noveltydrivensimilarity}
\end{figure}

\begin{figure}[!ht]
\begin{center}
  \includegraphics[width=100mm]{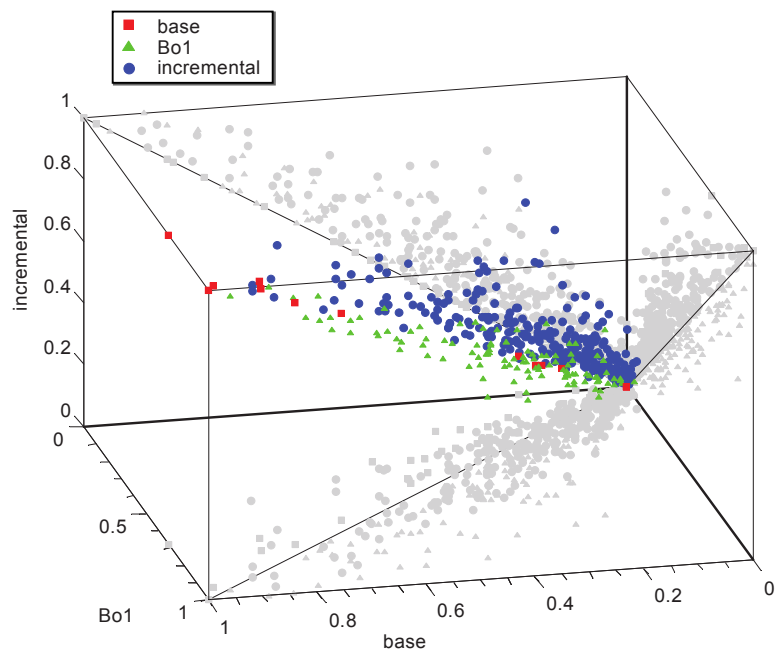}
\end{center}
  \caption{Comparación de los tres métodos basada en Precisión.}
  \label{fig:precision}
\end{figure}

\begin{figure}[!ht]
\begin{center}
  \includegraphics[width=100mm]{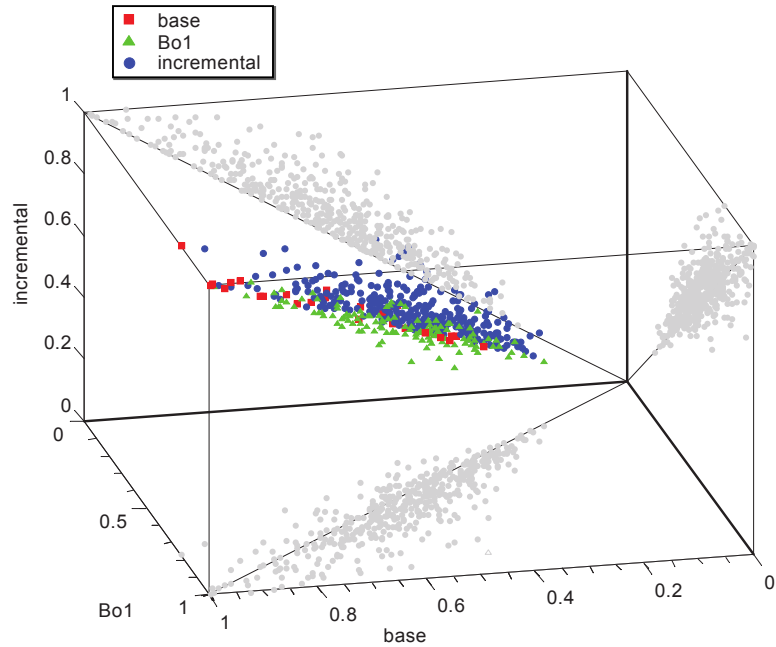}
\end{center}
  \caption{Comparación de los tres métodos basada en Precisión Semántica.}
  \label{fig:semanticprecision}
\end{figure}

Los diagramas en las Figuras~\ref{fig:noveltydrivensimilarity}, \ref{fig:precision}, \ref{fig:semanticprecision} comparan el rendimiento de las consultas para cada método evaluado utilizando similitud novedosa, precisión y precisión semántica.
Se tomaron cada uno de los 448 tópicos indexados y se ejecutó el algoritmo incremental con el propósito de recuperar la mayor cantidad de documentos de ese tópico. Cada tópico corresponde a un experimento y está representado en las figuras por un punto. La coordenada vertical del punto (z) corresponde al rendimiento del método incremental, mientras que las otras dos (x e y) corresponden al método base y al método Bo1-DFR. Se puede observar además la proyección de cada punto sobre los planos x-y, x-z e y-z. Sobre el plano x-z, los puntos por encima de la diagonal corresponden a los casos en los que el método incremental es mejor que el base. De la misma manera, para el plano y-z, los puntos por encima de la diagonal corresponden a los casos en los que el método incremental es mejor que el Bo1-DFR. El plano x-y compara el rendimiento del método base con el Bo1-DFR. Puede notarse que se utilizaron distintas marcas para ilustrar los casos en los que un método se comporta mejor que los otros dos.

Es interesante notar que para todos los casos evaluados el método incremental fue superior al método base y al método Bo1-DFR en términos de similitud novedosa. Esto muestra la utilidad de evolucionar el vocabulario del contexto para descubrir buenos términos para las consultas. En la métrica precisión, el método incremental fue estrictamente superior a los otros dos métodos en el 66.96\% de los tópicos evaluados. El método Bo1-DFR fue el mejor método en el 24.33\% de los tópicos y el método base se comportó tan bien como alguno de los otros dos en el 8.70\% de los tópicos. Finalmente, en la métrica precisión semántica el método incremental fue estrictamente superior a los otros dos métodos en el 65.18\% de los tópicos, Bo1-DFR fue superior en el 27.90\% de los tópicos y el método base se comportó tan bien como alguno de los otros dos en el 6.92\% de los tópicos.

La \autoref{fig:ci} presenta las medias y los intervalos de confianza del rendimiento de los métodos basados en $\nsimilarity$, Precisión y $P^S$. Estas tablas de comparación muestran que las mejoras alcanzadas por el método incremental con respecto a los otros métodos son estadísticamente significativas.
\begin{figure}
\begin{center}
  \begin{tabular}{|l|c|cc|}
    \hline  
  \multicolumn{4}{|c|}{$\nsimilarity$} \\ \hline \hline	
    método & $\abs{V}$& media& 95\% I.C. \\
    \hline
    Baseline    & 448 &0.087 & [0.0822;0.0924]\\
    Bo1-DFR & 448 & 0.075&[0.0710;0.0803]\\
    Incremental & 448 & 0.597 & [0.5866;0.6073]\\
    \hline
  \end{tabular}
\end{center}
\begin{center}
  \begin{tabular}{|l|c|cc|}
    \hline  
      \multicolumn{4}{|c|}{Precisión} \\ \hline \hline
      método & $\abs{V}$& media& 95\% I.C. \\
    \hline
    Baseline    & 448&  0.266 & [0.2461;0.2863] \\
    Bo1-DFR & 448& 0.307 & [0.2859;0.3298] \\
    Incremental & 448 & 0.354& [0.3325;0.3764] \\
    \hline
  \end{tabular}
\end{center}
\begin{center}
  \begin{tabular}{|l|c|cc|}
    \hline  
      \multicolumn{4}{|c|}{$P^S$} \\ \hline \hline
    método & $\abs{V}$& media& 95\% I.C. \\
    \hline
    Baseline    & 448 & 0.553&[0.5383;0.5679]\\
    Bo1-DFR & 448 &  0.590&[0.5750;0.6066]\\
    Incremental & 448 &  0.622&[0.6068;0.6372]\\
    \hline
  \end{tabular}
\end{center}
\caption[Medias e ICs del rendimiento de los tres métodos basados en $\mathit{sim}_S$, $P$ y $P^S$.]{Medias e intervalos de confianza (I.C.) del rendimiento de los tres métodos basados en similitud semántica, precisión y precisión semántica.}\label{fig:ci}
\end{figure}

\subsection[AGs para la Búsqueda basada en Contextos Temáticos]{\fontsize{14pt}{16.8pt}\selectfont 
Algoritmos Genéticos para la Búsqueda basada en Contextos Temáticos}
\begin{figure}[!ht]
\begin{center}
\begin{tabular}{cc}
\includegraphics[width=7cm]{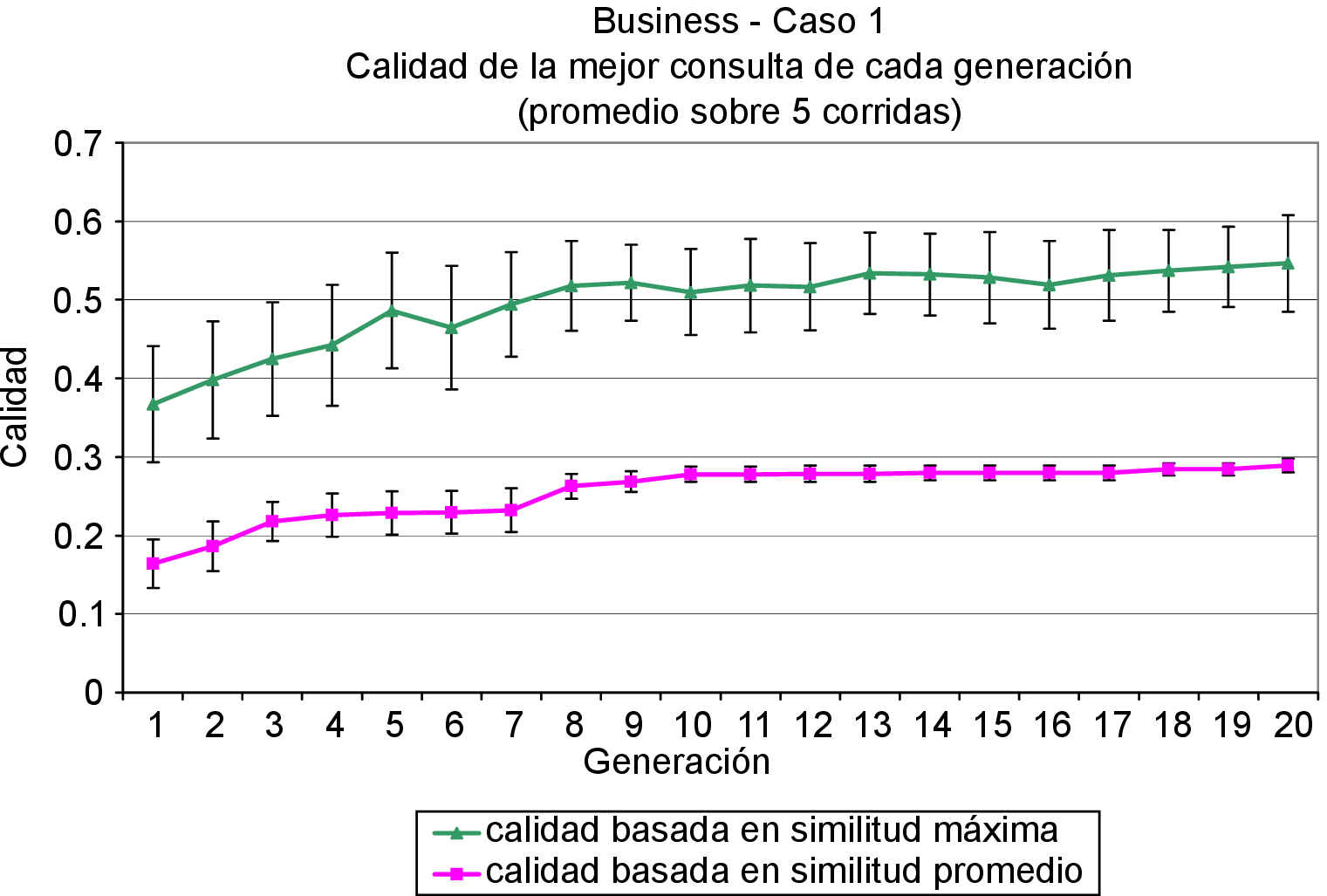} & \includegraphics[width=7cm]{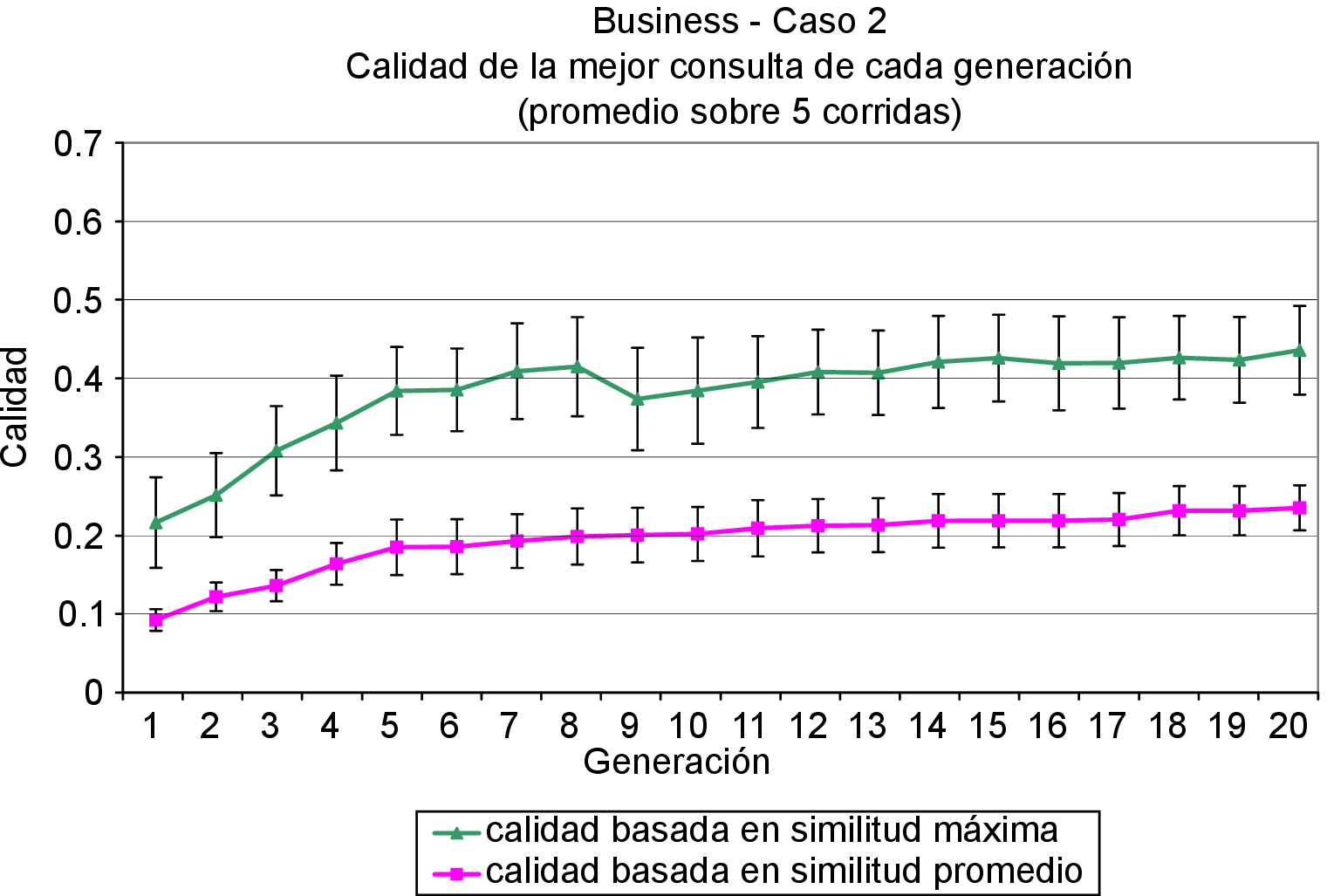} \\\hline
\includegraphics[width=7cm]{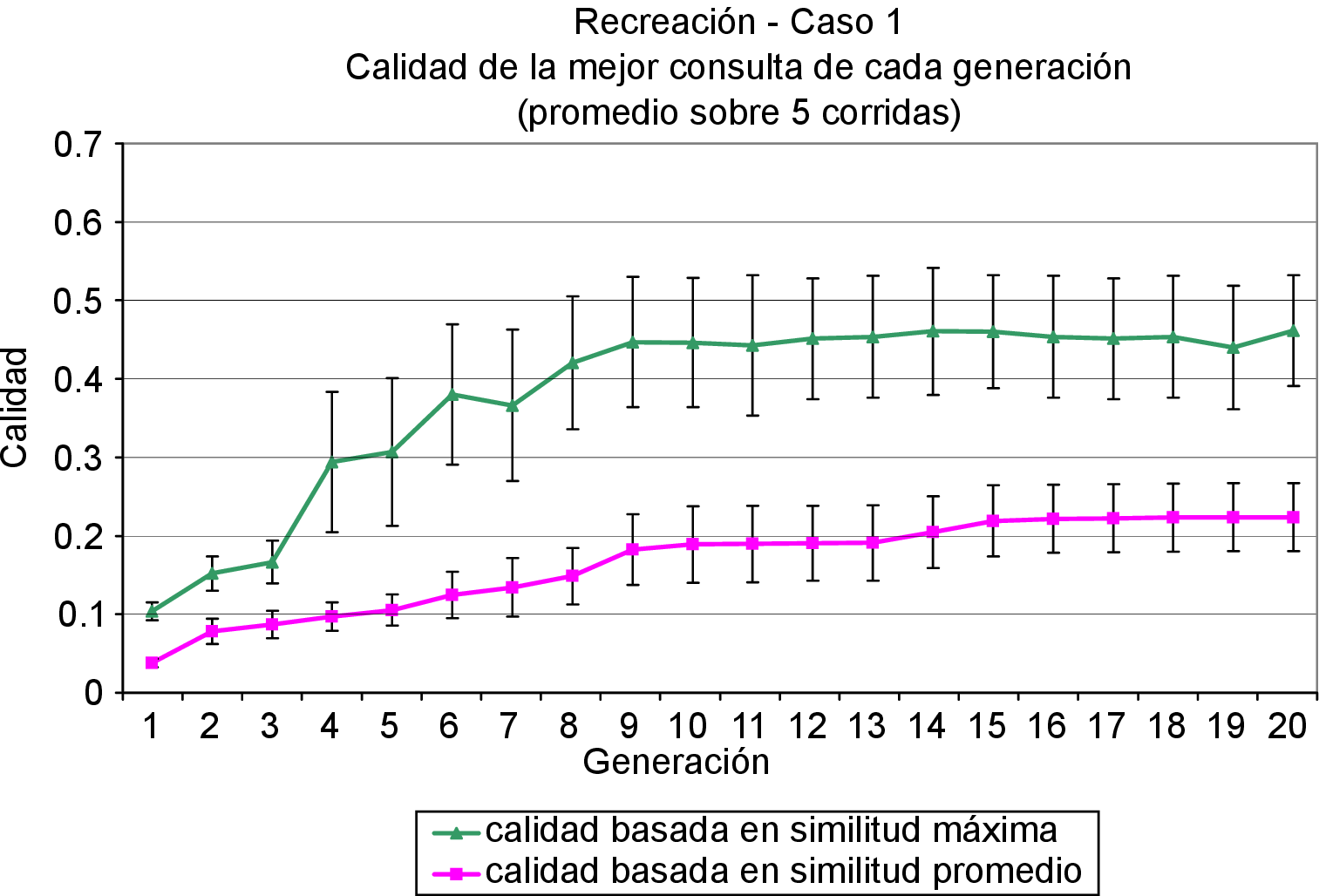} & \includegraphics[width=7cm]{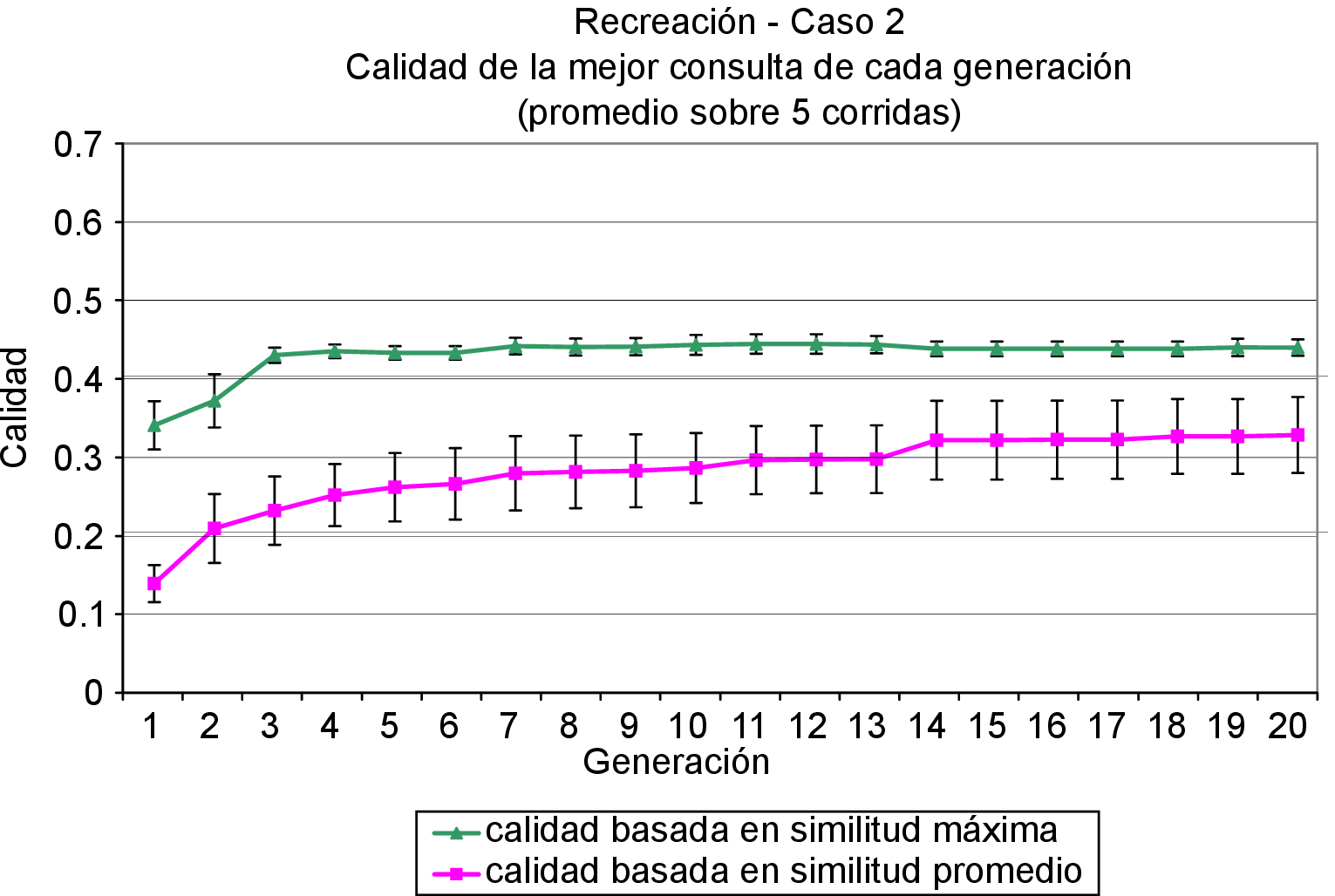} \\\hline
\includegraphics[width=7cm]{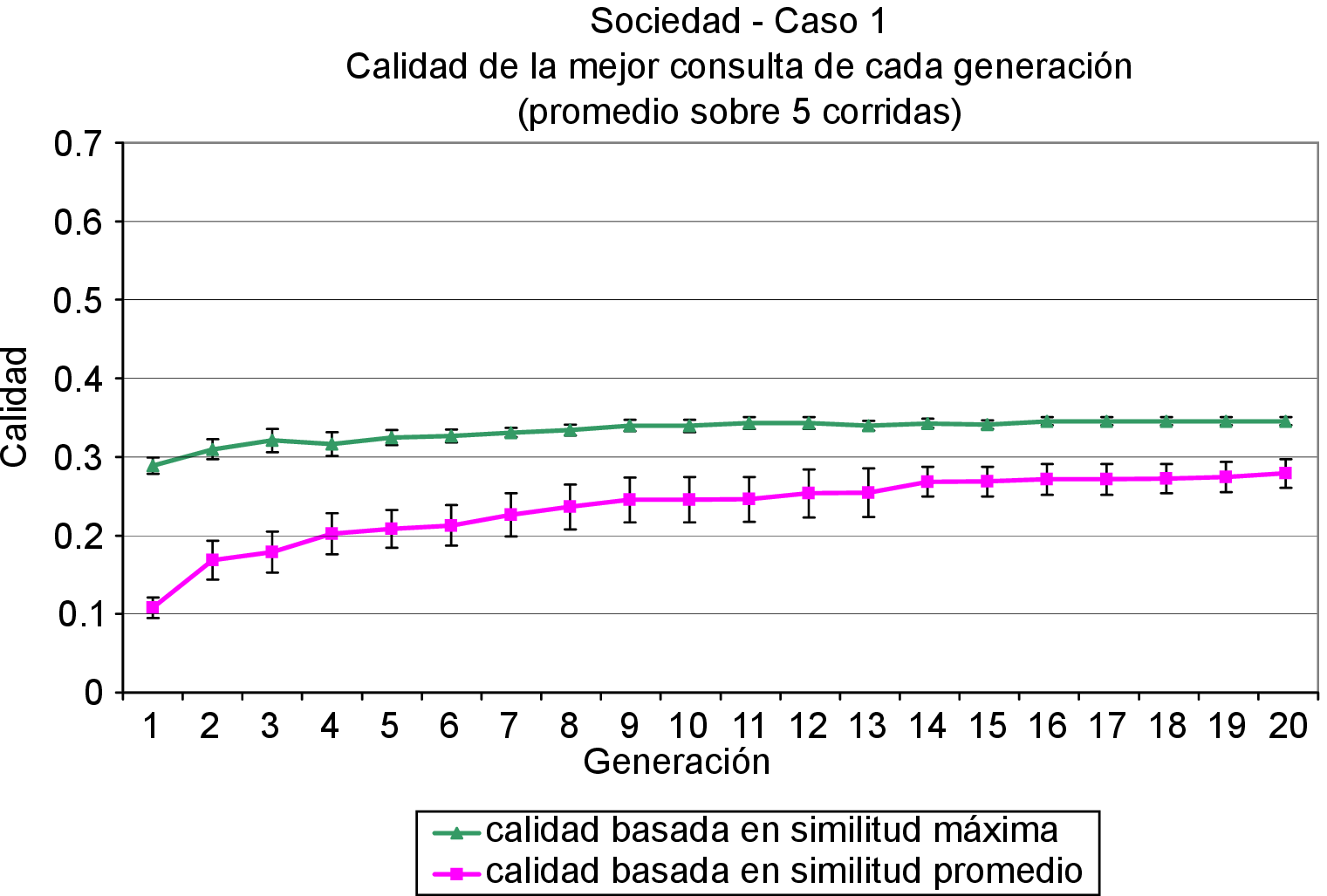} & \includegraphics[width=7cm]{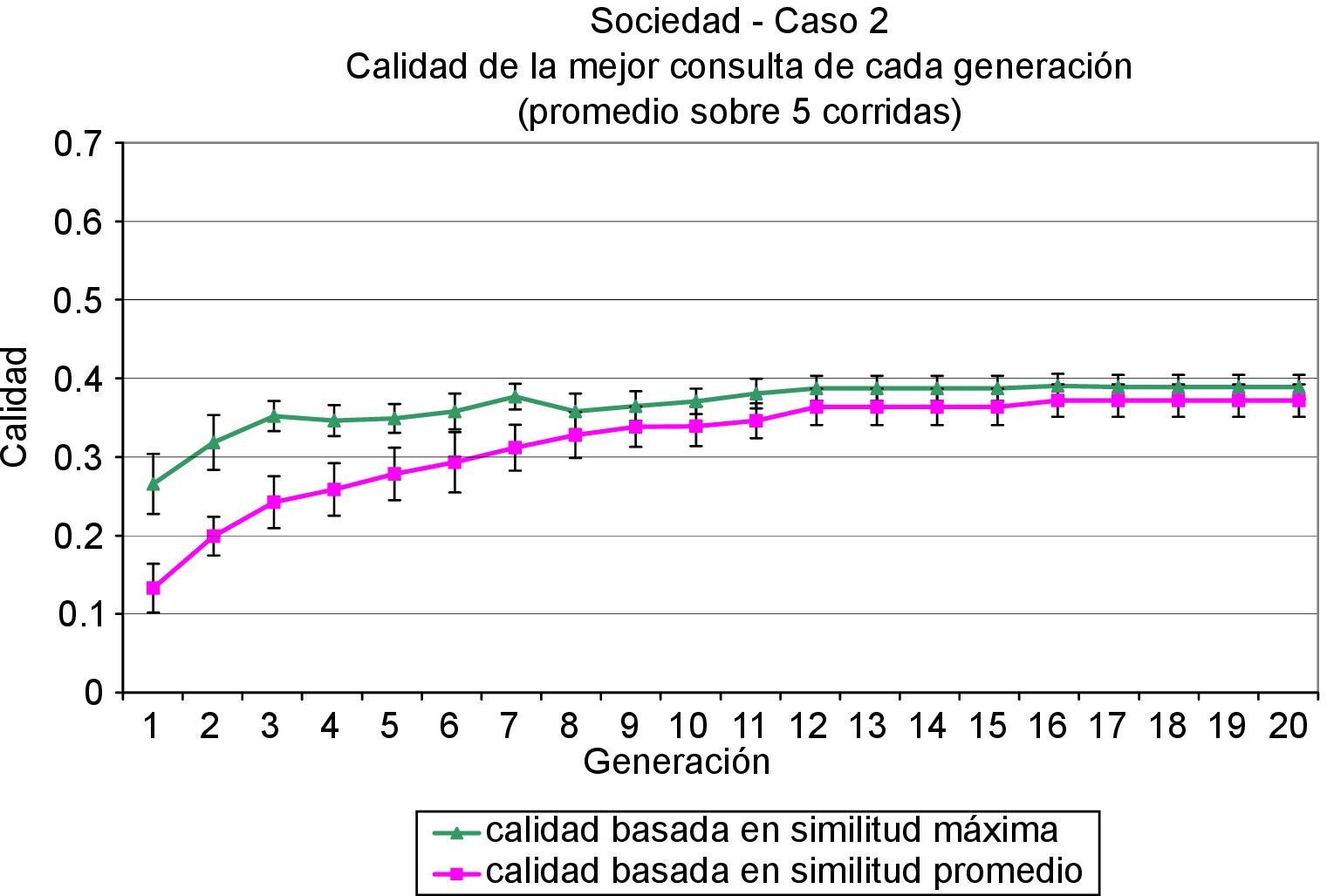} \\
\end{tabular}
\caption[Evaluación del AG mono-objetivo utilizando la métrica $\mathit{sim}_{cos}$.]{Resultados de la evaluación del Algoritmo Genético mono-objetivo utilizando la métrica similitud por coseno para tres tópicos del ODP.}
\label{fig:genetic_mono_results}
\end{center}
\end{figure}

La selección de buenos términos para una consulta puede verse como un problema de optimización, en donde la función objetivo que quiere optimizarse está relacionada directamente con la efectividad de la consulta para recuperar material relevante. Algunas características de este problema de optimización son:
\begin{enumerate}
 \item la alta dimensionalidad del espacio de búsqueda, en donde las soluciones candidatas son consultas, dentro de las cuales cada término que las compone se corresponde con una dimensión de ese espacio,
 \item la existencia de soluciones subóptimas aceptables, 
 \item la posibilidad de hallar múltiples soluciones, y,
 \item la búsqueda de material novedoso.
\end{enumerate}
En~\cite{cecchini08using} planteamos técnicas de optimización basadas en Algoritmos Genéticos~\cite{goldberg89genetic} para evolucionar ``buenos términos para consultas'' en el contexto de un tópico dado. Las técnicas propuestas hacen hincapié en la búsqueda de material novedoso que esté relacionado con el contexto de la búsqueda.
Para evaluar el rendimiento de estos algoritmos, los mismos fueron implementados en su totalidad dentro de la plataforma propuesta y evaluados utilizando el motor de búsqueda web explicado en la \autoref{sec:framework_search_engines}. Más detalles de los parámetros propios del algoritmo genético pueden encontrarse en~\cite{cecchini08using}. Las métricas aplicadas fueron la similitud por coseno (\autoref{eqn:cosine_similarity}) y la similitud novedosa (\autoref{eqn:new_similarity}). En la \autoref{fig:genetic_mono_results} se resumen los resultados obtenidos para los tópicos del ODP \textit{Negocios}, \textit{Recreación} y \textit{Sociedad}. En cada uno de ellos se eligieron dos páginas al azar y se las utilizó como contexto inicial de búsqueda.

La formulación de consultas de alta calidad es un aspecto fundamental en la búsqueda basada en contexto. Sin embargo, el cálculo de la efectividad de una consulta es un desafío importante porque normalmente hay varios objetivos involucrados, como pueden ser la Precisión y la Cobertura. Los algoritmos genéticos presentados arriba son, en particular, algoritmos Mono-objetivo, lo que quiere decir que no son capaces de evaluar múltiples objetivos de forma simultánea. En~\cite{cecchini10MOEAforContext} proponemos técnicas que pueden aplicarse para evolucionar consultas contextualizadas cuya calidad se basa en múltiples objetivos. 
Estos algoritmos fueron implementados utilizando la plataforma PISA~\cite{pisa03bleuler} y evaluados utilizando la plataforma de evaluación propuesta en esta tesis.
La plataforma PISA está dedicada principalmente a la búsqueda multi-objetivo y divide el proceso de optimización en dos módulos. El primero contiene todo lo específico al problema que se está abordando, como son la evaluación de las soluciones y la representación del problema; este módulo fue el que se implementó dentro de la plataforma propuesta. El otro módulo contiene las partes que son independientes al problema, siendo la más importante el proceso de selección de los mejores individuos. En este módulo fueron utilizados los algoritmos de selección más utilizados en la literatura, y que ya están implementados dentro de PISA\footnote{http://www.tik.ee.ethz.ch/pisa/}.

\begin{figure}[!ht]
\begin{center}
\begin{tabular}{c}
\includegraphics[width=14cm]{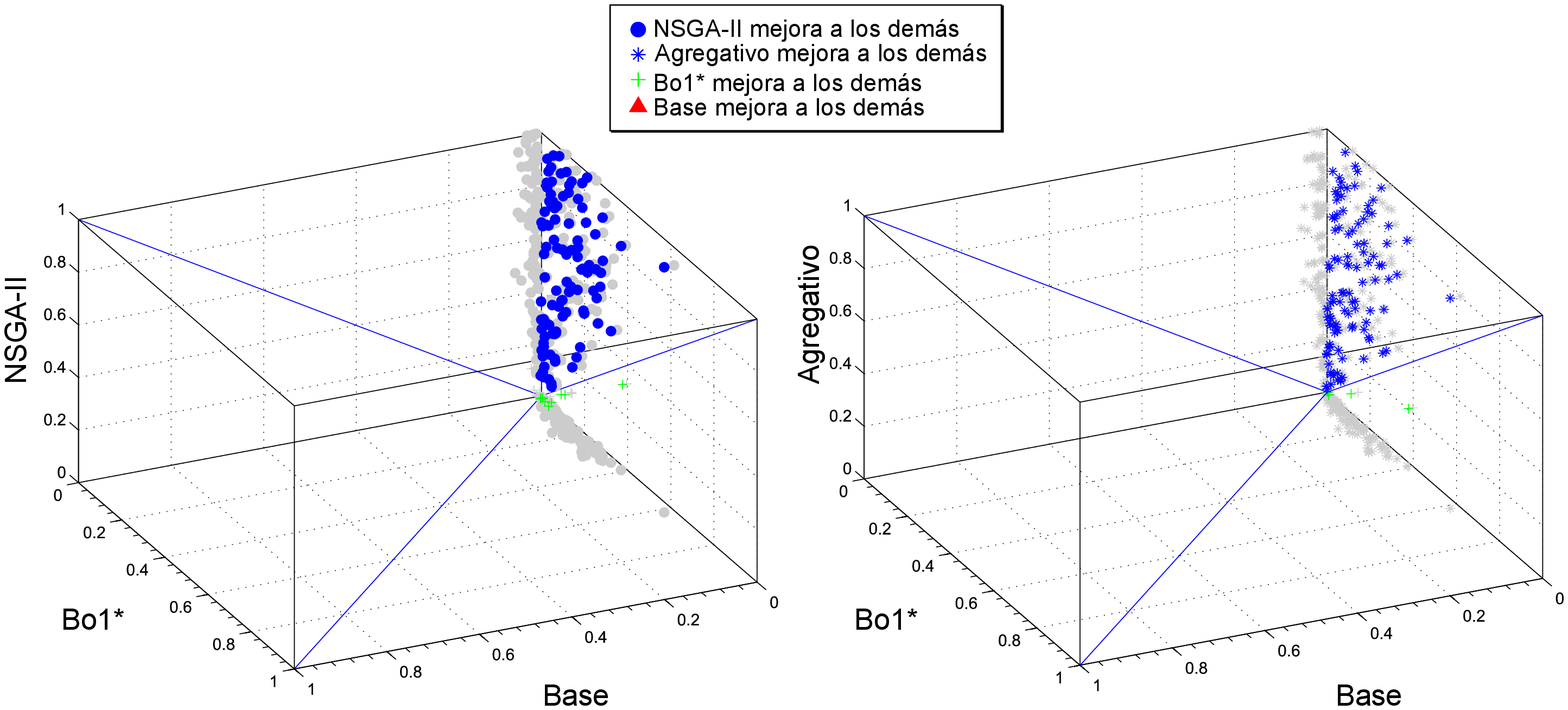}\\
\hline
\\
\includegraphics[width=14cm]{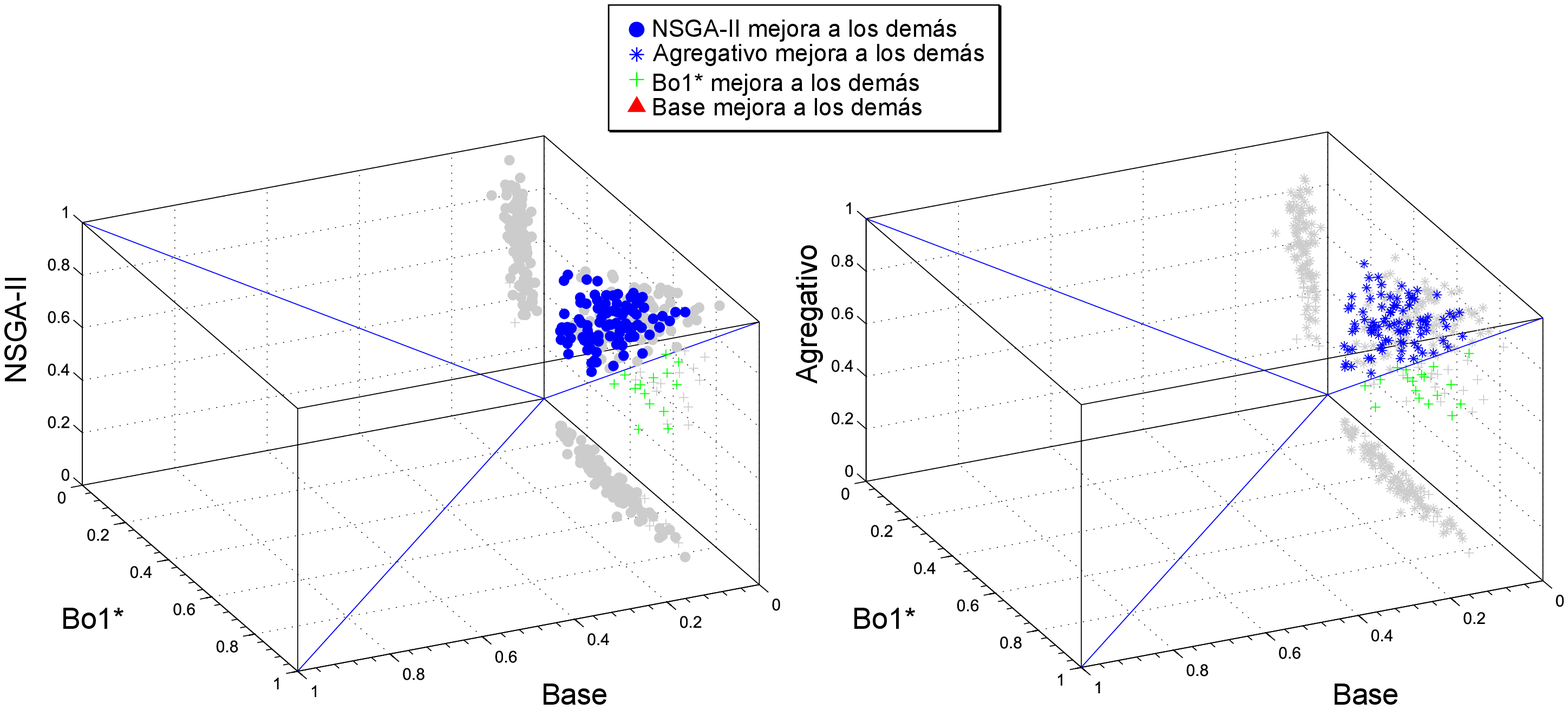}
\end{tabular}
\caption[Evaluación de los AGs multi-objetivo para las métricas $P$ y $C$.]{Resultados de la evaluación de los Algoritmos Genéticos multi-objetivo para las métricas Precisión (arriba) y Cobertura (abajo).}
\label{fig:genetic_multi_results}
\end{center}
\end{figure}

Las métricas que se utilizaron fueron la Precisión a un rango de 10 resultados ($P@10$) y la Cobertura. Además se definió una nueva métrica, $F^\star$, que es la media armónica de la $P@10$ y la Cobertura. En la comparación se estudió la efectividad de 3 algoritmos base contra 2 algoritmos propuestos. Estos últimos evolucionaron siguiendo dos objetivos simultáneos, la $P@10$ y la Cobertura. Uno de ellos fue el algoritmo NSGA-II~\cite{deb02fast} y el otro un algoritmo multi-objetivo agregativo que combina ambos objetivos utilizando $F^\star$. Dentro de los algoritmos que se utilizaron como base de la comparación tenemos un algoritmo Base que sólo genera consultas a partir del contexto inicial del usuario y no realiza ninguna evolución. El segundo es el ya mencionado algoritmo Bo1-DFR. Por último, dado que las comparaciones se realizaron utilizando el motor de búsqueda que cuenta con una colección local de documentos, se tenía información acerca de la relevancia de los resultados obtenidos por cada consulta. Es por esto que se implementó una versión supervisada del método Bo1-DFR, $Bo1^\star$, que a diferencia del algoritmo original, no crea las consultas con los mejores términos de los primeros resultados de la primera pasada (ver \autoref{sub:feedback}), sino que utiliza los mejores términos de los primeros documentos que se sabe que son relevantes (o sea, que pertenecen al tópico del contexto del usuario).

En la \autoref{fig:genetic_multi_results} se resumen los resultados obtenidos para las métricas Precisión y Cobertura de los algoritmos multi-objetivo. En este caso, para la creación del contexto inicial, se empleó una técnica similar a la mencionada en la \autoref{sec:incremental_algorithms}, pero en esta oportunidad con 110 tópicos de la ontología elegidos al azar. Para más detalles de los parámetros de los algoritmos puede consultarse~\cite{cecchini10MOEAforContext}.

\section{Resumen}
En este capítulo se presentaron las evaluaciones realizadas a partir de la implementación del \textit{algoritmo incremental} para la recuperación de información temática presentado en el capítulo anterior. Para esto se creó una \textit{plataforma general de evaluación de algoritmos de IR}, presentándose su estructura y sus componentes principales. La misma soporta distintas fuentes de información, como pueden ser motores de búsqueda web o motores locales y es capaz de comunicarse con otras plataformas, como la plataforma Terrier y la plataforma PISA. Para crear la colección local de documentos se indexaron una gran cantidad de tópicos de la ontología DMOZ, seleccionados de forma de asegurar la calidad del conjunto de datos. Respecto de la métricas necesarias para las evaluaciones, se implementaron varias técnicas existentes y se propusieron otras, como la \textit{similitud novedosa} que descarta a los términos presentes en las consultas, de modo de eliminar el sesgo introducido por ellos; y la \textit{precisión semántica}, que tiene en cuenta relaciones de similitud parcial entre los tópicos de una ontología.
La plataforma también mostró ser útil para la evaluación de otros algoritmos propuestos en el área de IR.

\chapter{Conclusiones y trabajo a futuro}\label{chp:conclusiones}
\section{Conclusiones}
A lo largo de esta tesis se desarrolló una herramienta de recuperación de información que ayuda al usuario en la tarea que está realizando, brindándole información relevante y basada en su contexto actual.
Para ello se propuso una solución al problema de la sensibilidad semántica, que es la limitación que surge cuando no se puede hallar una relación entre dos documentos similares semánticamente, porque contienen distintos términos en su vocabulario, resultando en un falso-negativo al intentar recuperar material relevante. Además, mediante la identificación de buenos discriminadores de tópicos, la propuesta presentada en esta tesis ayuda a mitigar el problema de falsos-positivos, que aparece cuando el mismo término (p. ej., java) aparece en dos tópicos diferentes. El método enunciado trabaja aprendiendo incrementalmente mejores vocabularios de un gran conjunto de datos como la Web.

A partir de este trabajo se concluye que la información contextual puede ser utilizada con éxito para acceder a material relevante. Sin embargo, los términos más frecuentes en ese contexto no son necesariamente los más útiles. Es por ello que se propone un método incremental para el refinamiento del contexto, que se basa en el análisis de los resultados de las búsquedas y que
mostró ser aplicable a cualquier dominio caracterizable por términos.

En este trabajo se demostró que al implementar un método incremental semisupervisado de refinamiento del contexto se puede mejorar el rendimiento alcanzado por un método base, el cual envía consultas generadas directamente a partir del contexto inicial, y mejorar también el rendimiento del método de refinamiento Bo1-DFR, el cual no refina las consultas basándose en un contexto. Esto muestra la utilidad de aprovechar simultáneamente los términos existentes en el contexto temático actual y los de un conjunto externo de datos a la hora de aprender mejores vocabularios y de refinar consultas automáticamente.

En esta tesis se implementó una plataforma de evaluación de métodos y técnicas para la recuperación de información. La misma permitió el desarrollo de los algoritmos presentados en este trabajo, proporcionando el soporte necesario para un análisis detallado de los resultados obtenidos. Dentro de esta plataforma también se implementaron las nuevas métricas propuestas en esta tesis.
Una de ellas es la Similitud novedosa, una medida de comparación entre documentos que descarta los términos que pudieran introducir un sesgo en la medición, favoreciendo la exploración de nuevo material.
La otra es la Precisión semántica, una métrica para la comparación de los resultados de un sistema de recuperación de información. Esta medida brinda una noción más rigurosa de la calidad de los documentos recuperados por un algoritmo de IR, al incorporar la noción de relevancia parcial entre tópicos.

En la literatura se han propuesto otros métodos basados en corpus para atacar el problema de la sensibilidad semántica. Por ejemplo, el análisis de la semántica latente visto en la \autoref{sub:lsa}. 
Otra técnica de este estilo que se aplicó para estimar la similitud semántica en PMI-IR\footnote{del inglés, Pointwise Mutual Information -- Information Retrieval}~\cite{turney01mining}. Este método de recuperación de información está basado en la información de polaridad mutua, que mide la relación entre dos elementos (p. ej., términos) comparando sus frecuencias observadas con respecto a las esperadas. Estas técnicas se diferencian de la que se propone en que no se basan en un proceso incremental de refinamiento de consultas, sino que utilizan una colección predefinida de documentos para identificar relaciones semánticas. Además, estas técnicas no distinguen las nociones de descriptores y discriminadores de tópicos. Las técnicas para la elección de los términos de las consultas propuestas en este trabajo están inspiradas y motivadas sobre la misma base de otros métodos de expansión y refinamiento de consultas~\cite{scholer02query,billerbeck03query}. Sin embargo, los sistemas que aplican estos métodos se diferencian de la plataforma propuesta en que el proceso se realiza a través de consultar o navegar en interfaces que necesitan la intervención explícita del usuario, en lugar de formular consultas automáticamente.

En los sistemas de recuperación proactivos, el uso del contexto juega un rol vital a la hora de seleccionar y filtrar información. Tales sistemas observan las interacciones del usuario e infieren necesidades adicionales de información, buscando documentos relevantes en la Web u otras librerías electrónicas. 
Aprender mejores vocabularios es una manera de aumentar la percepción y la accesibilidad del material útil. Se propuso un método prometedor para identificar la necesidad detrás de la consulta, lo cual es uno de los principales objetivos para muchos servicios y herramientas web actuales y futuras.

\section{Trabajo a futuro}
Dentro de las limitaciones encontradas durante el desarrollo de esta tesis, la más importante resultó ser el tiempo de ejecución de los algoritmos presentados. La velocidad es un obstáculo muy grande a la hora de realizar una evaluación con usuarios y es un aspecto a tener en cuenta a futuro. Por otro lado, el tiempo límite de ejecución podría incluirse como un parámetro a ser definido por el usuario, indicando qué tanto está dispuesto a esperar por resultados o si en cambio, desea un determinado número de documentos novedosos sin importar el tiempo de espera. 
Otro aspecto que no fue abordado dentro de los objetivos y contribuciones de estas tesis es la determinación del contexto actual del usuario, que también es de especial interés al momento de realizar las evaluaciones con usuarios. 
En lugar de esto, en las evaluaciones presentadas, se utilizó un conjunto de términos extraídos de una página de un tópico dado o la descripción de un tópico realizada por un editor de una ontología temática. En la literatura existen diversos trabajos que abordan el tema del reconocimiento automático del contexto actual de un usuario~\cite{balabanovic95adaptive, bharat00searchpad, bauer01wordsieve,budzik06context}.

%
Se está trabajando actualmente para aplicar el método propuesto para el aprendizaje de mejores vocabularios en otras tareas de IR, como la clasificación de texto. También se están analizando las distintas estrategias que ayudan a mantener al sistema enfocado en el contexto inicial, luego de que se han llevado a cabo varios pasos incrementales. 
Por otro lado, se espera adaptar la plataforma propuesta para evaluar otras aplicaciones de recuperación de información, tales como algoritmos de clasificación y clustering.

%
Se ampliará la plataforma de evaluación presentada en esta tesis con el propósito de ponerla a disponibilidad de la comunidad de IR, lo que resultará de gran utilidad a la comunidad científica del área, proveyéndola de una herramienta que permitirá analizar de manera objetiva la efectividad de nuevos métodos. Entonces, se diseñará un instrumento de evaluación para sistemas de IR basado en un gran número de tópicos y documentos obtenidos a partir de ontologías de tópicos, para luego integrarlo con métodos de evaluación existentes y novedosos. En tal sentido será importante el uso de las nociones de similitud semántica y relevancia parcial incorporadas a partir de esta tesis.

Como se mostró en los capítulos anteriores, la construcción de colecciones de prueba ha merecido especial atención del ámbito de la IR experimental,
ya que analizar grandes colecciones de documentos y juzgar su relevancia es una tarea sumamente costosa, especialmente cuando los documentos cubren tópicos diversos.
A la luz de estas necesidades y dificultades, y a partir de ontologías de tópicos editadas por humanos, tales como ODP, hemos desarrollado, y esperamos seguir refinando, un marco de experimentación para la evaluación automática y semi-automática de sistemas de IR, aprovechando 
el número masivo de relaciones disponibles entre tópicos y documentos.
\cleardoublepage
\phantomsection
\addcontentsline{toc}{chapter}{Índice de Figuras}
\label{chp:figures}
\listoffigures


\cleardoublepage
\phantomsection
\addcontentsline{toc}{chapter}{Bibliografía}

\begin{spacing}{0.9}
  \begin{small}
    \newcommand{\etalchar}[1]{$^{#1}$}

    \label{chp:biblio}
    \bibliographystyle{alpha}
  \end{small}
\end{spacing}

\end{document}